\documentclass[fleqn,usenatbib]{mnras}

\usepackage[T1]{fontenc}
\usepackage{hhline}
\usepackage{tabularx}
\usepackage{longtable}

\DeclareRobustCommand{\VAN}[3]{#2}
\let\VANthebibliography\thebibliography
\def\thebibliography{\DeclareRobustCommand{\VAN}[3]{##3}\VANthebibliography}

\def\thickhline{\noalign{\hrule height1.1pt}}
\newcommand{\cnc}{\texttt{cosmocnc}}

\usepackage{graphicx}	
\usepackage{amsmath}	

\usepackage{ctable}

\title[\cnc{}: a cluster number count likelihood code]{\cnc{}: A fast, flexible, and accurate galaxy cluster number count likelihood code for cosmology}

\author[\'{I}. Zubeldia \& B. Bolliet]{
\'{I}\~{n}igo Zubeldia$^{1,2}$\thanks{inigo.zubeldia@ast.cam.ac.uk} and
Boris Bolliet$^{2,3}$
\\
$^{1}$Institute of Astronomy, University of Cambridge, Madingley Road, Cambridge CB3 0HA\\
$^{2}$Kavli Institute for Cosmology, University of Cambridge, Madingley Road, Cambridge CB3 0HA\\
$^{3}$DAMTP, Centre for Mathematical Sciences, Wilberforce Road, Cambridge CB3 0WA, UK
}

\date{Accepted XXX. Received YYY; in original form ZZZ}

\pubyear{2024}

\begin{document}
\label{firstpage}
\pagerange{\pageref{firstpage}--\pageref{lastpage}}
\maketitle

\begin{abstract}
We introduce \cnc{}, a Python package for computing the number count likelihood of galaxy cluster catalogues in a fast, flexible and accurate way. \cnc{} offers three types of likelihoods: an unbinned, a binned, and an extreme value likelihood. It also supports the addition of stacked cluster data, which is modelled consistently with the cluster catalogue. The unbinned likelihood, which is the main focus of the code, can take an arbitrary number of mass observables as input and deal with several complexities in the data, such as variations in the properties of the cluster observable across the survey footprint, the possibility of different clusters having measurements for different combinations of mass observables, redshift measurement uncertainties, and the presence on unconfirmed detections in the catalogue. If there are more than one mass observables, the unbinned likelihood is computed with the backward convolutional approach, a novel approach that is first implemented in \cnc{}. After developing the likelihood formalism and describing its implementation, we validate the code with synthetic Simons-Observatory-like catalogues, finding excellent agreement between their properties and \cnc{}'s predictions and obtaining constraints on cosmological and scaling relation parameters featuring negligible biases. \cnc{} is publicly available at \href{https://github.com/inigozubeldia/cosmocnc}{\texttt{github.com/inigozubeldia/cosmocnc}}.
\end{abstract}




\section{Introduction}\label{sec:intro}


The abundance of galaxy clusters as a function of mass and redshift is a powerful cosmological probe, sensitive to cosmological parameters such as the matter density parameter $\Omega_{\mathrm{m}}$, the amplitude of matter clustering $\sigma_8$, the equation of state of dark energy, and the sum of the neutrino masses (e.g., \citealt{Allen2011,Weinberg13}). This has been demonstrated over the past two decades in a number of cosmological analyses using clusters detected in X-ray, optical and millimetre observations (e.g., \citealt{Rozo2010,  Hasselfield2013, Planck2014XX, Mantz2015, Bocquet2015, deHaan2016,  Ade2016,  Bocquet2018, Zubeldia2019, Bolliet2019, Costanzi2019, Abdullah2020, Abbot2020, To2021,  Garrel2022, Chaubal2022, Lesci2022, Chiu2023, Sunayama2023,  Fumagalli2024, Bocquet2024, Ghirardini2024}). With $10^4$--$10^5$ objects, cluster catalogues from current and upcoming observatories such as eROSITA \citep{Merloni2012}, \textit{Euclid} \citep{Euclid2011}, the Vera C. Rubin Observatory (Rubin/LSST; \citealt{LSST2012}), SPT-3G \citep{Benson2014}, the Simons Observatory (SO; \citealt{SO2019}), and CMB-S4 \citep{Abazajian2016} have the potential to improve significantly upon the constraints derived from their predecessors, taking advantage of their sheer statistical power and of the synergies between different observations (e.g., using galaxy weak lensing observations to calibrate X-ray and thermal Sunyaev-Zeldovich (tSZ) mass--observable scaling relations, as in, e.g., \citealt{Bocquet2018, Bocquet2024, Ghirardini2024}). 

A key factor for the success of these upcoming analyses is the development of likelihoods that are capable of dealing with the corresponding cluster catalogues, both in terms of efficiency and of accuracy. Indeed, a good cluster number count likelihood must be fast to evaluate and accurate enough for the statistical precision provided by the cluster catalogue to be analysed. In addition, it must be able to model all the cluster observables in a consistent way, and it must also be able to deal with various complexities in the input data, such as variations in the properties of the observables across the survey footprint (due to, e.g., inhomogeneous noise and/or foregrounds), missing data (e.g., only a subset of clusters having measurements for a given observable, or some clusters having no redshift measurements), cluster-dependent redshift measurement uncertainties, and the presence of unconfirmed detections in the catalogue. Furthermore, in some scenarios it may be useful to combine the cluster catalogue with stacked data, such as stacked cluster lensing profiles for mass calibration (e.g., \citealt{Costanzi2019,Lesci2022,Fumagalli2024}). In this case, a good likelihood ought to be able to model the cluster catalogue and the stacked data in a consistent way.

Although a number of cluster number count likelihoods of different types have been developed and implemented with success (see the number count analyses cited above), to the authors' knowledge these likelihoods have always been designed with a specific cluster catalogue in mind, being highly adapted to it at the expense of flexibility. In this work we introduce \cnc{}, a Python package for cluster number count likelihood computation that is designed to satisfy all our requirements for a good cluster number likelihood while being flexible enough so that it can be used in order to perform a cosmological analysis with most cluster catalogues with little to no modification. After developing the formalism underlying \cnc{}'s likelihoods and describing in detail their implementations, we validate the code with a set of synthetic SO-like cluster catalogues, performing several sets of cosmological analyses with them. In a follow-up paper, Zubeldia \& Bolliet (in prep.), we will demonstrate the application of \cnc{} to real data, obtaining cosmological constraints from the \textit{Planck} MMF3 \citep{Ade2016,Planck2016xvii} and the SPT2500d  \citep{Bocquet2018} cluster catalogues.

This paper is structured as follows. After a brief description of the main features of the code in Section\,\ref{sec:mainfeatures}, in Section\,\ref{sec:formalism} we lay out the formalism of \cnc{}'s three likelihoods. Next, in Section\,\ref{sec:implementation} we give a detailed account of how these likelihoods are implemented in the code, and in Section\,\ref{sec:validation} we test the performance of \cnc{} by applying it to a set of synthetic SO-like cluster catalogues, each containing about $16\,000$ clusters. We find excellent agreement between the synthetic catalogues and \cnc{}'s theoretical predictions and, most notably, we obtain constraints on cosmological and scaling relation parameters featuring negligible biases. We then conclude in Section\,\ref{sec:conclusion}. Finally, in Appendix\,\ref{sec:appendixa} we develop and implement a formalism to account for unconfirmed detections in the cluster catalogue, in Appendix\,\ref{appendix:b} we show two internal consistency test that \cnc{} has passed, and in Appendix\,\ref{appendix:c} we give a list all of \cnc{}'s input parameters.


\section{Main features of the code}\label{sec:mainfeatures}

\cnc{} is a Python package for evaluating the number count likelihood of galaxy cluster catalogues in a fast, flexible and accurate way. It is based on the use of Fast Fourier Transform (FFT) convolutions in order to evaluate some of the likelihood integrals. \cnc{} is fully written in Python and is publicly available at \href{https://github.com/inigozubeldia/cosmocnc}{\texttt{github.com/inigozubeldia/cosmocnc}}. Its main features, which are described in detail in the rest of this paper, are the following:

\begin{itemize}
    \item It supports three types of likelihoods: an unbinned likelihood, a binned likelihood, and an extreme value likelihood.
    \item It also supports the addition of stacked cluster data (e.g., stacked lensing profiles), which is modelled in a consistent way with the cluster catalogue.
    \item It links the cluster mass observables (also known as mass proxies, e.g., tSZ signal-to-noise, richness, lensing mass estimate, or X-ray flux) to the cluster mass and redshift through a hierarchical model with an arbitrary number of layers, allowing for correlated scatter between the different mass observables. In each layer, the mass--observable scaling relations and the scatter covariance matrix can be easily defined in a custom way, and can depend on sky location and redshift.
    \item It incorporates several widely-used halo mass functions, including that of \citet{Tinker2008}, which is computed in a fast way using the \texttt{cosmopower} power spectrum emulator \citep{Spurio2022,Bolliet2023}, as well as the Mira-Titan halo mass function emulator \citep{Bocquet2020}.
    \item The unbinned likelihood is the most general and flexible of the three. It supports an arbitrary number of cluster mass observables for each cluster in the sample and it allows for the set of mass observables to vary from cluster to cluster. It also allows for redshift measurement uncertainties. If some mass observables have completely uncorrelated scatter, \cnc{} takes advantage of this fact, boosting its computational performance significantly.
    \item The binned and extreme value likelihoods, on the other hand, consider the cluster abundance only across one mass observable and/or redshift, and do not allow for redshift measurement uncertainties.
    \item \cnc{} can produce mass estimates for each cluster in the sample, which are derived assuming the hierarchical model that is used to model the mass observables.
    \item It also allows for the generation of synthetic cluster catalogues for a given observational set-up.
    \item Several of \cnc{}'s computations can be accelerated with Python's \texttt{multiprocessing} module.
    \item The code is interfaced with the Markov chain Monte Carlo (MCMC) code \texttt{Cobaya}\footnote{  \texttt{\href{https://cobaya.readthedocs.io/en/latest/} {cobaya.readthedocs.io}}} \citep{Torrado2021}, allowing for easy-to-run MCMC parameter estimation.
    \item The code is also interfaced with \texttt{class\_sz}\footnote{\href{https://github.com/CLASS-SZ/class_sz}{ \texttt{github.com/CLASS-SZ/class\_sz}} }\label{footnote}\citep{Bolliet2023b}, allowing a wide range of cosmological models as well as enabling joint analyses with Cosmic Microwave Background (CMB) and Large Scale Structure (LSS) survey data.
\end{itemize}

It must be noted that little is hard-coded in \cnc{}. For example, the number of mass observables to be considered, their scaling relations, and the clusters for which each observable is available are all specified at a high level for any given cluster sample, \cnc{} using the same machinery regardless of the cluster sample. If a given cluster sample with a number $n_{\mathrm{obs}}$ of mass observables is ready to be used for a cosmological analysis, one or more mass observables can be dropped by just changing the value of one input parameter. The type of likelihood to be used (unbinned, binned, or extreme value) can also be specified by simply setting the value of an input parameter.

In their current implementation, the three types of likelihoods in \cnc{} assume that the clusters in the catalogue are statistically independent from each other, i.e., sample variance due to cluster clustering is neglected (see \citealt{Hu2003,Lima2004,Smith2011,Payerne2024}). Sample variance will be included in a later release of the code.

As noted in the Introduction, an exhaustive list of the input parameters to the code is given in Appendix\,\ref{appendix:c}.

\section{Likelihood formalism}\label{sec:formalism}

In this section we describe the mathematical formalism underlying \cnc{}'s calculations. First, we describe the code's input data (Section\,\ref{subsec:inputdata}) and the hierarchical model for the mass observables (Section\,\ref{sec:hierarchical}). We then derive the unbinned, binned, and extreme value likelihoods (Sections\,\ref{subsec:unbinned}, \ref{subsec:binned}, and \ref{subsec:extreme}, respectively), and how \cnc{} accounts for unconfirmed detections (Section\,\ref{sec:nonval}). Finally, we derive the stacked likelihood (Section\,\ref{subsec:stacked}).

\subsection{Input data}\label{subsec:inputdata}

\cnc{} takes as input data a cluster catalogue $\mathcal{C}$ with $N_{\mathrm{tot}}$ clusters, each with measured values for a set of $n_{\mathrm{obs},i}$ `mass observables' (also known as mass proxies, such as tSZ signal-to-noise, lensing mass estimate, X-ray flux or optical richness), denoted with $\bmath{\omega}_{\mathrm{obs},i}$, where $i$ denotes the $i$-th cluster in $\mathcal{C}$. For each cluster, $\bmath{\omega}_{\mathrm{obs},i}$ is a vector of dimension $n_{\mathrm{obs},i}$. 

There must be at least one mass observable for which a measurement is available for all the clusters in the catalogue. This is the `selection observable', denoted with $\zeta_{\mathrm{obs}}$, through which the catalogue $\mathcal{C}$ has been constructed by imposing a selection threshold $\zeta_{\mathrm{th}}$, selecting the clusters with $\zeta_{\mathrm{obs}} > \zeta_{\mathrm{th}}$. Optionally, a minimum and/or maximum redshift can be imposed in the construction of the catalogue. Furthermore, if there are other mass observables in the catalogue (in addition to the selection observable), a given cluster may have available measurements for any subset of them. Each cluster can also have a redshift measurement, $z_{\mathrm{obs},i}$, which, together with $\bmath{\omega}_{\mathrm{obs,i}}$, comprise the cluster data vector for cluster $i$, $\mathbfit{D}_i = \{ z_{\mathrm{obs},i}, \bmath{\omega}_{\mathrm{obs,i}} \}$. In addition, the catalogue may also contain the sky location of each cluster, $\hat{\mathbfit{n}}_i$, relevant if the properties of the mass observables vary across the sky (e.g., due to inhomogeneous noise or foregrounds). There is no limit as to how fine the dependency of the observables on sky location can be. It is, however, assumed that, for cluster detection, the survey sky footprint is tessellated into a set of $n_{\mathrm{tile}}$ `selection tiles', with the properties of the selection observable (namely, scaling relations and scatter) assumed to be constant across each selection tile.

In summary, the data in the catalogue can be written as $\mathcal{C} = \{ N_{\mathrm{tot}}, \underline{\hat{\mathbfit{n}}}, \underline{\mathbfit{D}} \}$, where $\underline{\hat{\mathbfit{n}}}$ is the vector whose $i$-th component is $\hat{\mathbfit{n}}_i$ and $\underline{\mathbfit{D}}$ is the vector whose $i$-th component is $\mathbfit{D}_i$. Let us now consider the likelihood $\mathcal{L}$ of such a cluster catalogue, i.e., the probability of the data given a set of model parameters (both cosmological and scaling-relation). Two different types of likelihood exist, both of which can be computed with \cnc{}: an `unbinned' and a `binned' likelihood. We derive them in Sections \ref{subsec:unbinned} and \ref{subsec:binned}, respectively. Furthermore, it is also possible to consider the likelihood of the cluster in the catalogue with the largest value of the selection observable. \cnc{} also includes an implementation of this `extreme value' likelihood, which is described in Section\,\ref{subsec:extreme}.  

In the unbinned case, \cnc{} can also take as input data a set of $n_{\mathrm{stacks}}$ stacked measurements, $\mathcal{S} = \{ \mathbfit{S}_r, r = 1,\dots,n_{\mathrm{stacks}} \}$, where each stacked data vector $\mathbfit{S}_r$ corresponds to a given observable (e.g., a lensing profile) stacked across a subset of $n_{\mathrm{stack},r}$ clusters in the input catalogue. The corresponding `stacked' likelihood, which, for a given catalogue, can be combined with the unbinned likelihood, is described in Section\,\ref{subsec:stacked}.

Finally, the catalogue can include unconfirmed objects, i.e., objects that are detected through the selection observable but which have not been confirmed to be real clusters (e.g., via complementary observations), and which therefore may be false detections. \cnc{} can support these detections in its unbinned likelihood implementation and in one subtype of the binned likelihood. This is described in Section\,\ref{sec:nonval}.

\subsection{Cluster mass observable hierarchichal model}\label{sec:hierarchical}

\subsubsection{Model description}

Consider cluster $i$, with measured values for $n_{\mathrm{obs},i}$ mass observables, $\bmath{\omega}_{\mathrm{obs},i}$. We recall that the set of available mass observables can vary from cluster to cluster: only a measurement for the selection observable must be available for all the clusters in the catalogue. 

In \cnc{}, $\bmath{\omega}_{\mathrm{obs},i}$ is related to the cluster mass $M_i$ and its true redshift $z_i$ through a hierarchical model with $n_{\mathrm{layer}}$ layers. We note that here and in the rest of Section\,\ref{sec:formalism}, we refer to the cluster mass in a generic way and simply denote it with $M$. The formalism developed here applies for any mass definition, as long as it is used consistently. On the other hand, by true redshift we mean the cluster's real redshift, as opposed to its measured value, which may have a non-negligible measurement uncertainty.

In \cnc{}'s hierarchical model, a given layer consists of two operations: (i) the application of a set of scaling relations and (ii) the application of Gaussian scatter. Consider the $j$-th layer of the model. In it, first, a set of $n_{\mathrm{obs},i}$ scaling relations relations, denoted with the function $\mathbfit{f}^{(j+1)}$, are applied to the layer's `input variables', which are the `output variables' of the $j-1$-th layer, $\bmath{\omega}^{(j-1)}_i$. This operation produces the `intermediate variables' of the $j$-th layer, $\bmath{\omega}^{(j)}_{\mathrm{in},i}$,

\begin{equation}
\bmath{\omega}_{\mathrm{in},i}^{(j)} \equiv \mathbfit{f}^{(j)} (\bmath{\omega}_i^{(j-1)},z_i, \hat{\mathbfit{n}}_i).    
\end{equation}
As we have made explicit, the scaling relations may also depend on $z_i$ and $\hat{\mathbfit{n}}_i$. They can also depend on the cosmological model and its parameters, as well as on other parameters (the `scaling relation parameters').

The second operation in the $j$-th layer is the application of correlated Gaussian scatter to its intermediate variables. This means that the probability density function (pdf) of the output variables of the the $j$-th layer, $\bmath{\omega}^{(j)}_i$, conditioned on the value of $\bmath{\omega}^{(j)}_{\mathrm{in},i}$, is an $n_{\mathrm{obs},i}$-dimensional Gaussian centred on $\bmath{\omega}^{(j)}_{\mathrm{in},i}$. The elements of its covariance matrix are considered input model parameters.

Finally, the input variables of the first layer are equal the cluster mass $M_i$, i.e., every component of  $\bmath{\omega}^{(0)}_i$ is $M_i$, and the output variable of the $n_{\mathrm{layer}}$-th layer is equal to the observed values of the mass observables, $\bmath{\omega}^{(n_{\mathrm{layer}})}_i = \bmath{\omega}_{\mathrm{obs},i}$.

With this hierarchical model in mind, if $n_{\mathrm{layer}} > 1$, the pdf of the cluster mass observables of cluster $i$, $\bmath{\omega}_{\mathrm{obs},i}$, conditioned on the cluster mass $M_i$, true redshift $z_i$, and sky location $\hat{\mathbfit{n}}_i$, can then be written as

\begin{multline}\label{eq:scatterintegral}
    P (\bmath{\omega}_{\mathrm{obs},i} | M_i, z_i, \hat{\mathbfit{n}}_i) = \\ \int \prod_{j=1}^{n_{\mathrm{layer}}-1} \left[ d\bmath{\omega}^{(j)}_i P (\bmath{\omega}^{(j+1)}_i | \bmath{\omega}^{(j)}_i, z_i, \hat{\mathbfit{n}}_i) \right] P (\bmath{\omega}^{(1)}_i | M_i, z_i, \hat{\mathbfit{n}}_i),
\end{multline}
where $ P (\bmath{\omega}^{(j+1)}_i | \bmath{\omega}^{(j)}_i, z_i, \hat{\mathbfit{n}}_i)$ is the pdf accounting for the scatter in the $j$-th layer, expressed in terms of the layer's input and output variables. We refer to the integral in Eq.\,(\ref{eq:scatterintegral}) as the `scatter integral'. It has has $(n_{\mathrm{layer}} - 1 ) \times n_{\mathrm{obs},i}$ dimensions. 

If $n_{\mathrm{layer}} = 1$, on the other hand, the scatter integral disappears, with Eq.\,(\ref{eq:scatterintegral}) reducing to $ P (\bmath{\omega}_{\mathrm{obs},i} | M_i, z_i, \hat{\mathbfit{n}}_i)  = P (\bmath{\omega}^{(1)}_i | M_i, z_i, \hat{\mathbfit{n}}_i)$.

\subsubsection{Some examples}

Consider a toy cluster catalogue featuring a single mass observable consisting of a noisy, unbiased measurement of each cluster's true mass $M_i$, the measurement noise assumed to be Gaussian. In this case, the hierarchical model consists of one single layer. For each cluster, the intermediate variable of the first (and only) layer is equal to the cluster mass, $\bmath{\omega}^{(1)}_{\mathrm{in},i} = \mathbfit{f}^{(1)} ( \bmath{\omega}^{(0)}_i , z_i , \hat{\mathbfit{n}}) = \bmath{\omega}^{(0)}_i  = (M_i)$, and the scatter accounts for the Gaussian measurement noise.

Suppose now that there is also some log-normal `intrinsic scatter' in the mass measurements, which is added before the observational noise. In this case, the number of layers increases to two, with the scatter in the first layer accounting for the intrinsic scatter and that in the second layer accounting for the measurement noise. In particular, the intermediate variable of the first layer is $\bmath{\omega}^{(1)}_{\mathrm{in},i} = \mathbfit{f}^{(1)} ( \bmath{\omega}^{(0)}_i , z_i , \hat{\mathbfit{n}}) = (\ln M_i)$, and that of the second layer is $\bmath{\omega}^{(2)}_{\mathrm{in},i} = \mathbfit{f}^{(2)} ( \bmath{\omega}^{(1)}_i , z_i , \hat{\mathbfit{n}}) = (\mathrm{exp} ( \omega^{(1)}_{i,1}))$, where $\omega^{(1)}_{i,1}$ denotes the first (and only) component of $\bmath{\omega}^{(1)}_{i}$. Note that the use of logarithmic variables (e.g., the logarithm of the mass) is necessary in order to be able to accommodate the log-normal scatter within \cnc{}'s formalism, as \cnc{} can only deal with Gaussian scatter.

Let us now consider some real catalogues. In their cosmological analysis of the \textit{Planck} MMF3 cosmology sample, \citet{Ade2016} assumed a two-layer model for the only mass observable used, the tSZ signal-to-noise $q$, which is consistent with \cnc{}'s formalism. In their first layer, the cluster mass, redshift, and sky location specify the value of the logarithm of cluster's `mean signal-to-noise', $\ln \bar{q}_{\mathrm{m}}$, which, in our notation, is the intermediate variable of the first layer. Then, Gaussian (`intrinsic') scatter is applied, leading to the logarithm of the `true signal-to-noise', $\ln q_{\mathrm{m}}$, which is our output variable of the first layer. Next, in the second layer, the signal-to-noise is simply exponentiated, after which Gaussian scatter is added, obtaining the observed signal-to-noise $q$.

In the \textit{Planck} MMF3 reanalysis with CMB lensing mass calibration, \citet{Zubeldia2019} assumed a two-layer hierarchical model also consistent with our formalism. They considered two mass observables, the tSZ signal-to-noise, $q_{\mathrm{obs}}$ (their selection observable), and the CMB lensing signal-to-noise, $p_{\mathrm{obs}}$, the first of which was modelled using the same model as in \citet{Ade2016}. In their model, first, the cluster mass and redshift specify the values of the logarithms of the mean tSZ and CMB lensing signal-to-noises, $\ln \bar{q}$ and  $\ln \bar{p}$, respectively, which are the intermediate variables of the first layer. Then, correlated Gaussian (`intrinsic') scatter links $\ln \bar{q}$ and $\ln \bar{p}$ to the logarithms of the cluster's true signal-to-noises, $\ln q$ and $\ln p$, which are the output variables of the first layer. In the next (second) layer, a second set of scaling relations simply exponentiates the two variables, leading to $q$ and $p$ (the intermediate variables of the second layer), and, finally, uncorrelated Gaussian (`observational') scatter links $q$ and $p$ with their observed values, $q_{\mathrm{obs}}$ and $p_{\mathrm{obs}}$, respectively. This two-layer model for two mass observables leads to a two-dimensional scatter integral (Eq.\,\ref{eq:scatterintegral}). 

Similar two-layer models, with a first layer of intrinsic scatter and a second one of observational scatter or `noise', have been used in other cosmological analyses, e.g., in \citealt{Bocquet2018,Bocquet2024} to model the tSZ signal-to-noise. In \citet{Bocquet2018}, however, a three-layer model was used to model the weak lensing mass observables, with each layer accounting, respectively, for intrinsic scatter in the lensing mass, uncorrelated LSS scatter in the lensing mass, and noise in the lensing profiles. 

\subsection{Unbinned likelihood}\label{subsec:unbinned}

In the unbinned approach, the likelihood $\mathcal{L}_{\mathrm{unbinned}}$ is simply the pdf of the data in the cluster catalogue at given model parameters $\mathbfit{p}$ (cosmological and scaling relation), $\mathcal{L}_{\mathrm{unbinned}} = P(\mathcal{C} | \mathbfit{p})$. Dropping the explicit dependency on $\mathbfit{p}$ in order to avoid clutter in the notation, the likelihood can be written as

\begin{equation}\label{eq:lik}
    P(\mathcal{C}) = P(N_{\mathrm{tot}}, \underline{\hat{\mathbfit{n}}}, \underline{\mathbfit{D}}) = P ( \underline{\mathbfit{D}} | \underline{\hat{\mathbfit{n}}}, N_{\mathrm{tot}}) P ( \underline{\hat{\mathbfit{n}}} | N_{\mathrm{tot}} ) P(N_{\mathrm{tot}}).
\end{equation}
Here, the first term, $P( \underline{\mathbfit{D}} | N_{\mathrm{tot}}, \underline{\hat{\mathbfit{n}}})$, is the probability of obtaining the cluster data points $ \underline{\mathbfit{D}} $ given that $N_{\mathrm{tot}}$ clusters have been included in the catalogue and that their sky locations have been found to be $\underline{\hat{\mathbfit{n}}}$. The second term, $P ( \underline{\hat{\mathbfit{n}}} | N_{\mathrm{tot}} )$ is the probability that, given $N_{\mathrm{tot}}$ clusters have been included in the catalogue, they are located at the sky locations $\underline{\hat{\mathbfit{n}}}$. Finally, the third term, $P(N_{\mathrm{tot}})$, is the probability of including a total of $N_{\mathrm{tot}}$ clusters in the catalogue.

Assuming that the clusters in the catalogue are statistically independent from each other, $P(N_{\mathrm{tot}})$ is a Poisson distribution, 

\begin{equation}
P(N_{\mathrm{tot}}) = \frac{e^{-\bar{N}_{\mathrm{tot}}}\bar{N}_{\mathrm{tot}}^{N_{\mathrm{tot}}}}{N_{\mathrm{tot}}!},
\end{equation}
where $\bar{N}_{\mathrm{tot}}$ is the expected value of the number of clusters in the catalogue, which can be written as

\begin{equation}\label{eq:nintegral}
\bar{N}_{\mathrm{tot}} = \int \frac{d \bar{N}}{d \Omega} (\hat{\mathbfit{n}}) d \hat{\mathbfit{n}},
\end{equation} 
where $d \bar{N}/d\Omega (\hat{\mathbfit{n}})$ is the mean number of clusters per solid angle in direction $\hat{\mathbfit{n}}$ and where the integral is performed across the survey footprint. \cnc{} assumes that the sky is tessellated into a number of $n_{\mathrm{tile}}$ selection tiles for the purposes of cluster detection. Within each tile, the properties of the selection observable (namely, its scaling relations and scatter) are assumed to be constant. Thus, Eq.\,(\ref{eq:nintegral}) reduces to a sum over the sky tiles, 

\begin{equation}\label{eq:sumn}
\bar{N}_{\mathrm{tot}} = \sum_{k=1}^{n_{\mathrm{tile}}} \frac{d \bar{N}}{d \Omega} (\hat{\mathbfit{n}}_k) \Omega_k,
\end{equation}
where $\hat{\mathbfit{n}}_k$ refers to the centre of tile $k$ and $\Omega_k$ is the solid angle subtended by tile $k$. (Note that since the properties of the selection observable are constant across each selection tile, $\hat{\mathbfit{n}}_k$ can, in fact, refer to any sky direction within tile $k$.) The mean number of clusters per solid angle, $d \bar{N}/d\Omega (\hat{\mathbfit{n}})$, can be, in turn, written as 

\begin{equation}\label{eq:totaln}
\frac{d \bar{N}}{d \Omega} (\hat{\mathbfit{n}})  = \int_{\zeta_{\mathrm{th}}}^{\infty} d \zeta_{\mathrm{obs}} \int_0^{\infty} dz \frac{d^3 N}{d \zeta_{\mathrm{obs}}  dz d\Omega}  (\zeta_{\mathrm{obs}}  ,z, \hat{\mathbfit{n}} ),
\end{equation}
where the integrand is the cluster abundance across the selection observable $\zeta_{\mathrm{obs}}$, true redshift $z$, and sky location $\hat{\mathbfit{n}}$, which is given by

\begin{equation}\label{eq:abundanceobs}
\frac{d^3 N}{d \zeta_{\mathrm{obs}}  dz d\Omega}  (\zeta_{\mathrm{obs}}  ,z, \hat{\mathbfit{n}} ) = \int_0^{\infty} dM   P(\zeta_{\mathrm{obs}}  | M , z,  \hat{\mathbfit{n}} ) \frac{d^3 N}{d M dz d\Omega}  (M,z).  
\end{equation}
Here, $P(\zeta_{\mathrm{obs}} | M , z, \hat{\mathbfit{n}} )$ is the conditional pdf followed by the selection observable at given mass $M$, true redshift $z$, and sky location $\hat{\mathbfit{n}}$, which is given by the scatter integral of Eq\,(\ref{eq:scatterintegral}) if it is restricted to the selection observable. On the other hand,  $d^3 N / (dM dz d\Omega)$ is the cluster abundance across mass and redshift, which does not depend on sky direction $\hat{\mathbfit{n}}$ and is given by the product of the halo mass function, $d^2 N / (dV dM)$, times the comoving volume element, $d^2V/(dz d\Omega)$,

\begin{equation}\label{eq:abundance}
\frac{d^3 N}{dM dz d\Omega} (M,z)  =   \frac{d^2 N}{dV dM} (M,z) \frac{d^2V}{dz d\Omega} (z).
\end{equation}
Taking into account the intermediate variables in the scatter integral (see Eq.\,\ref{eq:scatterintegral}), the integral in Eq.\,(\ref{eq:totaln}) has $n_{\mathrm{layer}} + 2$ dimensions ($n_{\mathrm{layer}} - 1$ dimensions from the scatter integral, in addition to one dimension each from the integration over mass, redshift, and the final layer of the selection observable). We note that we can also calculate the cluster abundance across only $\zeta_{\mathrm{obs}}$ and $z$ by integrating Eq.\,(\ref{eq:abundanceobs}) over the survey area (i.e., summing over the selection tiles), which gives

\begin{equation}\label{eq:sumabunance}
\frac{d^2 N}{d \zeta_{\mathrm{obs}}  dz}  (\zeta_{\mathrm{obs}}  ,z ) = \sum_{k=1}^{n_\mathrm{tile}} \frac{d^3 N}{d \zeta_{\mathrm{obs}}  dz d\Omega} (\zeta_{\mathrm{obs}}  ,z, \hat{\mathbfit{n}}_k ) \Omega_k.
\end{equation}
As in Eq.\,(\ref{eq:sumn}), here $\hat{\mathbfit{n}}_k$ denotes the central sky location of tile $k$, and $\Omega_k$, the solid angle subtended by the tile.


The second term in Eq.\,(\ref{eq:lik}), $P (\underline{\hat{\mathbfit{n}}} | N_{\mathrm{tot}})$, can be written as

\begin{equation}\label{eq:skylocation}
    P (\underline{\hat{\mathbfit{n}}} | N_{\mathrm{tot}}) =  N_{\mathrm{tot}}! \prod_{i=1}^{N_{\mathrm{tot}}} \frac{1}{ \bar{N}_{\mathrm{tot}}} \frac{d \bar{N} (\hat{\mathbfit{n}}_i )}{d \Omega},
\end{equation}
where $d \bar{N}/d\Omega$ is the mean number of clusters per solid angle in direction $\hat{\mathbfit{n}}_i$, and where the product runs over all of the clusters in the sample. We include the $N_{\mathrm{tot}}$ factor in Eq.\,(\ref{eq:skylocation}) to reflect the fact that the ordering of the elements in $\underline{\hat{\mathbfit{n}}}$ does not matter. 

Finally, the first term in Eq.\,(\ref{eq:lik}), $P(\underline{\mathbfit{D}} | \underline{\hat{\mathbfit{n}}}, N_{\mathrm{tot}})$, can be written as

\begin{equation}
P(\underline{\mathbfit{D}} | \underline{\hat{\mathbfit{n}}}, N_{\mathrm{tot}}) = \prod_{i=1}^{N_{\mathrm{tot}}} P ( \mathbfit{D}_i | \hat{\mathbfit{n}},  \mathrm{in}),
\end{equation}
where the product runs over all the clusters in the sample and where the condition that $N_{\mathrm{tot}}$ clusters have been included in the sample is translated into each individual $P ( \mathbfit{D}_i | \hat{\mathbfit{n}},  \mathrm{in})$ as the condition that each of the clusters is included in the sample. Using Bayes' theorem, we can write each $P ( \mathbfit{D}_i | \hat{\mathbfit{n}},  \mathrm{in})$ as

\begin{equation}\label{eq:datain}
    P ( \mathbfit{D}_i | \hat{\mathbfit{n}},  \mathrm{in}) = \frac{   P (\mathrm{in} | \mathbfit{D}_i , \hat{\mathbfit{n}}_i ) P (\mathbfit{D}_i |\hat{\mathbfit{n}}_i) }{ P ( \mathrm{in} | \hat{\mathbfit{n}}_i )}.
\end{equation}
Here, $P(\mathrm{in} | \mathbfit{D}_i , \hat{\mathbfit{n}}_i)$ is the probability for cluster $i$ to be included in the catalogue given the values of its sky location and cluster data vector $\mathbfit{D}_i$. This probability is just $P(\mathrm{in} | \mathbfit{D}_i , \hat{\mathbfit{n}}_i) = \Theta( \zeta_{\mathrm{obs},i} - \zeta_{\mathrm{th}})$, where $\Theta(x)$ denotes the step function and $\zeta_{\mathrm{th}}$, we recall, is the selection threshold. On the other hand, $P (\mathbfit{D}_i |\hat{\mathbfit{n}}_i)$ is the pdf followed by $\mathbfit{D}_i$ conditioned on $\hat{\mathbfit{n}}_i$ but unconditioned on the cluster being included in the catalogue. We refer to $P (\mathbfit{D}_i |\hat{\mathbfit{n}}_i)$ as the `individual cluster likelihood' of cluster $i$. If a redshift measurement is available,  recall that $\mathbfit{D}_i = \{z_{\mathrm{obs},i}, \bmath{\omega}_{\mathrm{obs,i}}\}$. Taking the scatter in the redshift measurements to be uncorrelated with the scatter in the mass observables, we can write

\begin{multline}\label{eq:unconditioneddata}
P (\mathbfit{D}_i |\hat{\mathbfit{n}}_i) = P (z_{\mathrm{obs},i}, \bmath{\omega}_{\mathrm{obs},i} |\hat{\mathbfit{n}}_i) = \\ \int_0^{\infty} dz P (z_{\mathrm{obs},i} | z, \hat{\mathbfit{n}}_i ) P (  \bmath{\omega}_{\mathrm{obs},i} | z, \hat{\mathbfit{n}}_i) P(z | \hat{\mathbfit{n}}_i).
\end{multline}
Here, $P (z_{\mathrm{obs},i} | z, \hat{\mathbfit{n}}_i )$ is the probability for obtaining a redshift measurement $z_{\mathrm{obs},i}$ given a true redshift $z$ and a sky location $\hat{\mathbfit{n}}_i$. We take this pdf to be either (i) a Gaussian centred on $z$ and with some standard deviation $\sigma_{z,i}$, where $\sigma_{z,i}$ can be cluster-dependent, or (ii) a delta function centred on $z$, if redshift measurement uncertainties can be neglected. On the other hand $P(\bmath{\omega}_{\mathrm{obs},i} | z, \hat{\mathbfit{n}}_i) P(z|\hat{\mathbfit{n}}_i)$ can be written as

\begin{multline}\label{eq:pdata}
P(\bmath{\omega}_{\mathrm{obs},i} | z, \hat{\mathbfit{n}}_i) P(z|\hat{\mathbfit{n}}_i) = \\ \int_0^{\infty} dM P(\bmath{\omega}_{\mathrm{obs},i} | M ,z, \hat{\mathbfit{n}}_i) P(M | z, \hat{\mathbfit{n}}_i )P(z|\hat{\mathbfit{n}}_i),
\end{multline}
where the first term in the integrand is the scatter integral, given by Eq.\,(\ref{eq:scatterintegral}), and $ P(M | z, \hat{\mathbfit{n}}_i )P(z|\hat{\mathbfit{n}}_i) =  P(M,z | \hat{\mathbfit{n}}_i )$ is the unconditioned pdf followed by the cluster mass $M$ and its true redshift $z$, which does not depend on sky location and is proportional to $d^3 N / (dM dz d\Omega)$ given by Eq.\,(\ref{eq:abundance}).

Taking into account the scatter integral, in order to compute $P (\mathbfit{D}_i |\hat{\mathbfit{n}}_i)$, as given by Eq.\,(\ref{eq:unconditioneddata}), in principle one needs to compute an integral with $(n_{\mathrm{layer}} - 1 ) \times n_{\mathrm{obs}} + 2$ dimensions ($(n_{\mathrm{layer}} - 1 ) \times n_{\mathrm{obs}}$ dimensions from the scatter integral, in addition to two more dimensions from the integration over mass $M$ and true redshift $z$). 

If no redshift measurement is available, $P (\mathbfit{D}_i |\hat{\mathbfit{n}}_i) = P(\bmath{\omega}_{\mathrm{obs},i} |\hat{\mathbfit{n}}_i)$, which can be written as

\begin{equation}\label{eq:noz}
P (\bmath{\omega}_{\mathrm{obs},i} |\hat{\mathbfit{n}}_i) = \int_0^{\infty} dz \int_0^{\infty} dM P ( \bmath{\omega}_{\mathrm{obs},i} | M, z, \hat{\mathbfit{n}}_i ) P (M, z, \hat{\mathbfit{n}}_i),
\end{equation}
where we note that this integral is the same as that in Eq.\,(\ref{eq:unconditioneddata}) if $P (z_{\mathrm{obs},i} | z, \hat{\mathbfit{n}}_i )$ is set to unity.

Finally, in Eq.\,(\ref{eq:datain}), $P ( \mathrm{in} | \hat{\mathbfit{n}}_i )$ is the probability for a cluster at sky location $\hat{\mathbfit{n}}_i$ to be included in the sample. This probability can be written as

\begin{multline}
P ( \mathrm{in} | \hat{\mathbfit{n}}_i ) = \int_0^{\infty} dz \int_0^{\infty} dM P (  \mathrm{in} | M, z, \hat{\mathbfit{n}}_i) P ( M, z | \hat{\mathbfit{n}}_i) \\
\int_{\zeta_{\mathrm{th}}}^{\infty} d \zeta_{\mathrm{obs}} \int_0^{\infty} dz \int_0^{\infty} dM   P(\zeta_{\mathrm{obs}} | M , z )  P ( M, z | \hat{\mathbfit{n}}_i).
\end{multline}
Since $P ( M, z | \hat{\mathbfit{n}}_i)$ is proportional to $d^3 N / (dM dz d\Omega)$, it follows that $P ( \mathrm{in} | \hat{\mathbfit{n}}_i )$ is proportional to $d\bar{N}/ d\Omega (\hat{\mathbfit{n}}_i)$, which is given by Eq.\,(\ref{eq:totaln}).

Putting the three terms of the likelihood together again, we can then write

\begin{equation}\label{eq:unbinned}
\mathcal{L}_{\mathrm{unbinned}} \propto e^{-\bar{N}_{\mathrm{tot}}} \prod_{i=1}^{N_{\mathrm{tot}}} P ( \mathbfit{D}_i | \hat{\mathbfit{n}}_i),
\end{equation}
where the proportionality factor does not depend on any of the model parameters (cosmological or scaling relation), and where the likelihood is only evaluated at values of the selection observable that are greater than the selection threshold (otherwise, due to the step function accounting for sample selection, $\mathcal{L}_{\mathrm{unbinned}} = 0$). 

Therefore, the unbinned likelihood  consists of two terms: 
\begin{itemize}
    \item a `cluster abundance' term that depends only on the total number of clusters in the sample;
    \item a `cluster data' term that iterates over all the clusters in the sample, multiplying their individual cluster likelihoods.
\end{itemize} We note that, unlike the three factors in Eq.\,(\ref{eq:lik}), these two terms do not individually constitute a likelihood by themselves.

\subsection{Binned likelihood}\label{subsec:binned}

In the binned approach, the data in the catalogue $\mathcal{C}$ is binned into a set of $n_{\mathrm{bin}}$ bins in observable space (selection observable and/or redshift). The likelihood then is the pdf of the number counts across bins $\mathbfit{N}$, $ \mathcal{L}_{\mathrm{binned}} = P(\mathbfit{N} | \mathbfit{p})$, where the vector dimension of $\mathbfit{N}$ is equal to $n_{\mathrm{bin}}$ and where $ \mathbfit{p}$ are the model parameters (cosmological and scaling relation). Hereafter, we drop the explicit dependency of the likelihood on $\mathbfit{p}$ in our notation, as we did in Section\,\ref{subsec:unbinned}. 

We consider three binning schemes: (i) across the selection observable $\zeta_{\mathrm{obs}}$, (ii) across true redshift $z$, and (iii) across both $\zeta_{\mathrm{obs}}$ and $z$. Our binned likelihood therefore cannot accommodate any cluster observables other than $\zeta_{\mathrm{obs}}$ and $z$. In its current implementation in \cnc{}, it also cannot account for redshift measurement uncertainties, unlike its unbinned counterpart.

Consider a rectangular bin in the $\zeta_{\mathrm{obs}}$--$z$ plane, labelled with index $s$ and defined by a boundary given by $\zeta_{\mathrm{low},s}$, $\zeta_{\mathrm{up},s}$, $z_{\mathrm{low},s}$, and $z_{\mathrm{up},s}$. If binning over only $\zeta_{\mathrm{obs}}$ is desired, the same formalism as described here applies by just taking the $z$ limits of the bin to be $z_{\mathrm{low},s}=0$ and $z_{\mathrm{up},s}=\infty$. Similarly, if binning over only $z$ is desired, the $\zeta_{\mathrm{obs}}$ limits of the bin can be taken to be $\zeta_{\mathrm{low},s} = \zeta_{\mathrm{th}}$ and $\zeta_{\mathrm{up},s} = \infty$. The mean number of clusters within this bin, denoted with $\bar{N}_s$, is

\begin{equation}\label{totalnbinned}
\bar{N}_s = \int \frac{d \bar{N}_s}{d \Omega} (\hat{\mathbfit{n}}) d \hat{\mathbfit{n}},
\end{equation}
where $d \bar{N}_s/d\Omega (\hat{\mathbfit{n}})$ is the mean number of clusters in bin $s$ per solid angle in sky direction $\hat{\mathbfit{n}}$ and where the integral is performed across the survey footprint. As in our unbinned likelihood, here we also assume that for cluster detection the sky is tessellated into a set of selection tiles, across which the properties of the selection observable (scaling relations and scatter) are assumed to be constant. Thus, as it was the case in Eq.\,(\ref{eq:totaln}), the integral in Eq.\,(\ref{totalnbinned}) reduces to a sum over $n_{\mathrm{tile}}$ sky tiles,

\begin{equation}\label{eq:meani}
\bar{N}_{s} = \sum_{k=1}^{n_{\mathrm{tile}}} \frac{d \bar{N}_{s}}{d \Omega} (\hat{\mathbfit{n}}_k) \Omega_k,
\end{equation}
where $\hat{\mathbfit{n}}_k$ is the centre of tile $k$ and $\Omega_k$ is the solid angle it subtends. The mean number of clusters in tile $i$ per solid angle in sky direction $\hat{\mathbfit{n}}$, $d \bar{N}_i/d\Omega (\hat{\mathbfit{n}})$, can, in turn, be written as

\begin{multline}\label{eq:meannsolidangle}
\frac{d \bar{N}_{s}}{d \Omega} (\hat{\mathbfit{n}})  = \int_{\zeta_{\mathrm{low},s}}^{\zeta_{\mathrm{up},s}} d \zeta_{\mathrm{obs}} \int_{z_{\mathrm{low},s}}^{z_{\mathrm{up},s}} dz \frac{d^3 N}{d \zeta_{\mathrm{obs}} dz d\Omega}  (\zeta_{\mathrm{obs}} , z, \hat{\mathbfit{n}} ),
\end{multline}
where the integrand is the cluster abundance across the selection observable $\zeta_{\mathrm{obs}}$ and true redshift $z$ at sky location $\hat{\mathbfit{n}}$, which is given by Eq.\,(\ref{eq:abundanceobs}). As expected, Eq.\,(\ref{eq:meannsolidangle}) has the same form as Eq.\,(\ref{eq:totaln}), which gives the total number of clusters per solid angle, the only difference being the integration limits.

As in the unbinned case, in our binned likelihood clusters are assumed to be statistically independent from each other. Therefore, the observed number of clusters within bin $s$, $N_s$, follows a Poisson distribution with an expected value $\bar{N}_{s}$ given by Eq.\,(\ref{eq:meani}),

\begin{equation}\label{eq:poisson}
    P (N_s) = \frac{e^{-\bar{N}_{s}}\bar{N}_{s}^{N_{s}}}{N_{s}!}.
\end{equation}
The total likelihood, $\mathcal{L}_{\mathrm{binned}}$, is simply a product of the individual Poisson likelihoods for each bin,

\begin{equation}\label{eq:binned}
\mathcal{L}_{\mathrm{binned}} = P (\mathbfit{N}) = \prod_{s=1}^{n_{\mathrm{bin}}} P (N_s).
\end{equation}

In the limit of small bin sizes, this binned likelihood reduces to an unbinned likelihood. We refer the reader to Appendix\,B of \citet{Zubeldia2019} for a rigorous derivation of this connection.

\subsection{Extreme value likelihood}\label{subsec:extreme}

It is possible to consider the likelihood of the most extreme object in the catalogue (e.g., \citealt{Bahcall1998,Sahlen2016}). Let us consider the cluster in the catalogue with the largest measured value of the selection observable $\zeta_{\mathrm{obs}}$, $\zeta_{\mathrm{max}}$ (our `most extreme' cluster). The probability for $\zeta_{\mathrm{max}}$ to be the maximum value of the selection observable in the catalogue is equal to the probability that no clusters are found with $\zeta_{\mathrm{obs}} > \zeta_{\mathrm{max}}$. Assuming clusters to be statistically independent from each other, this probability is given by a Poisson distribution with zero occurrences and an expected value given by the mean number of clusters with $\zeta_{\mathrm{obs}}  > \zeta_{\mathrm{max}}$, $\bar{N} (\zeta_{\mathrm{obs}}  > \zeta_{\mathrm{max}})$,

\begin{equation}\label{eq:evdef}
\mathcal{L}_{\mathrm{extreme}} = P (\zeta_{\mathrm{obs}} = \zeta_{\mathrm{max}} | \mathbfit{p}) = e^{-\bar{N} (\zeta_{\mathrm{obs}} > \zeta_{\mathrm{max}})},
\end{equation}
where $\mathbfit{p}$ are the model parameters (cosmological and scaling relation). Here, $\bar{N} (\zeta_{\mathrm{obs}} > \zeta_{\mathrm{max}})$ can be calculated in the same way as the mean number of clusters within each bin is computed for the binned likelihood, namely using Eqs.\,(\ref{totalnbinned}--\ref{eq:meannsolidangle}), setting the bin limits to be  $\zeta_{\mathrm{low}} = \zeta_{\mathrm{max}}$, $\zeta_{\mathrm{up}} = \infty$, $z_{\mathrm{low}} = 0$, and $z_{\mathrm{up}} = \infty$ (note that here we have dropped the bin index from the bin limits as there is only one bin).

Another quantity of interest is the pdf followed by $\zeta_{\mathrm{max}}$, $P(\zeta_{\mathrm{max}} |  \mathbfit{p})$. It is obtained by differentiating $P (\zeta_{\mathrm{obs}} = \zeta_{\mathrm{max}} | \mathbfit{p})$ with respect to $\zeta_{\mathrm{max}}$.

\subsection{Unconfirmed detections}\label{sec:nonval}

So far we have assumed that the objects in the catalogue correspond to real clusters. However, in some scenarios, the catalogue may contain unconfirmed detections, which may be real clusters or false detections. \cnc{} allows for these unconfirmed detections to be present in (i) the unbinned likelihood and (ii) the binned likelihood with binning across selection observable, modelling their abundance in a consistent way.

Let us consider a catalogue with confirmed and unconfirmed detections. In it, each cluster features a boolean validation label $V$, which can be either true, $V=T$, if the cluster is confirmed, or false, $V=F$, otherwise. We assume that unconfirmed detections only have a measurement for the selection observable $\zeta_{\mathrm{obs}}$ (i.e., they do not have measurements for any other mass observables or for redshift). The mean total number of objects in the catalogue is given by $\bar{N}_{\mathrm{tot,all}} =  \bar{N}_{\mathrm{tot}} + \bar{N}_{\mathrm{f}}$, where the mean number of true detections, $\bar{N}_{\mathrm{tot}}$, is given by Eq.\,(\ref{eq:sumn}), and where $\bar{N}_{\mathrm{f}}$ is the mean number of false detections, which is given by

\begin{equation}\label{eq:abundancefalse}
   \bar{N}_{\mathrm{f}} =  \sum_{k=1}^{n_{\mathrm{tile}}} \frac{d \bar{N}_{\mathrm{f}}}{d \Omega} (\hat{\mathbfit{n}}_k) \Omega_k =  \sum_{k=1}^{n_{\mathrm{tile}}} \int d \zeta_{\mathrm{obs}}  \frac{dN_{\mathrm{f}}}{ d \zeta_{\mathrm{obs}} d\Omega } (\zeta_{\mathrm{obs}}, \hat{\mathbfit{n}}_k) \Omega_k.
\end{equation}
Here, $dN_{\mathrm{f}} / (d \zeta_{\mathrm{obs}} d\Omega_k)$ is the abundance of false detections across selection observable per unit solid angle for selection tile $k$. This abundance can be, e.g., estimated by applying the relevant cluster detection pipeline to simulated data lacking the cluster signal.

If unconfirmed detections are present in the catalogue, the cluster abundance part of the unbinned likelihood can be evaluated as in the completely pure case but substituting $\bar{N}_{\mathrm{tot}}$ for $\bar{N}_{\mathrm{tot,all}}$, i.e., it is given by $e^{-\bar{N}_{\mathrm{tot,all}}}$. Similarly, the binned likelihood with selection observable binning can be simply computed by integrating the sum of the abundance of true detections (given by integrating Eq.\,\ref{eq:sumabunance} with respect to redshift) and false detections (given by $dN_{\mathrm{f}} / d \zeta_{\mathrm{obs}} = \sum_{k=1}^{n_\mathrm{tile}} dN_{\mathrm{f}} / (d \zeta_{\mathrm{obs}} d\Omega_k) \Omega_k$) within each bin. Note that the binned likelihood with binning across redshift or across both selection observable and redshift cannot be computed.

On the other hand, in the unbinned likelihood, the individual cluster likelihoods now become a function of whether the detection has been confirmed or not. Assuming that false detections are never confirmed, the individual cluster likelihoods can be calculated by considering three different cases separately: confirmed true detections, unconfirmed true detections, and false detections. We offer a derivation of such a calculation in Appendix\,\ref{sec:appendixa}.

\subsection{Stacked likelihood}\label{subsec:stacked}

Within the unbinned formalism, a stacked data set $\mathcal{S} = \{ \mathbfit{S}_r, r = 1,\dots,n_{\mathrm{stacks}} \}$ can be included in the analysis jointly with the cluster catalogue $\mathcal{C}$. The joint likelihood for $\mathcal{S}$ and $\mathcal{C}$ is

\begin{equation}
    P(\mathcal{C},\mathcal{S} | \mathbfit{p}) = P (\mathcal{S} | \mathcal{C}, \mathbfit{p}) P (\mathcal{C}| \mathbfit{p}).
\end{equation}
Here, $P (\mathcal{C}| \mathbfit{p})$ is the unbinned likelihood for the data in the catalogue, $\mathcal{L}_{\mathrm{unbinned}}$ (see Section\,\ref{subsec:unbinned}), and $P (\mathcal{S} | \mathcal{C}, \mathbfit{p})$ is the probability for the stacked data set given the data in the catalogue and the model parameters $ \mathbfit{p}$. We refer to this term as the stacked likelihood, 

\begin{equation}
 \mathcal{L}_{\mathrm{stacked}} = P (\mathcal{S} | \mathcal{C}, \mathbfit{p}),
\end{equation}
which we calculate in this section. As before, we will drop the explicit dependency on $\mathbfit{p}$.

The stacked data set is formed by $n_{\mathrm{stacks}}$ stacked data vectors, $\mathbfit{S}_i$, with $r = 1,\dots,n_{\mathrm{stacks}}$, each of which corresponds to a given `stacked observable' (e.g., a lensing profile, or a stacked scalar quantity). Each stack corresponds to a given subset of the clusters in the catalogue containing a total of $n_{\mathrm{stack},r}$ clusters. That is,

\begin{equation}\label{eq:stacked_data}
    \mathbfit{S}_r = \sum_{i \,\mathrm{in \, stack }\,r} w_i {\mathbfit{s}}_i,
\end{equation}
where $\mathbfit{s}_i$ is the stacked observable of cluster $i$, $w_i$ is the stacking weight of cluster $i$, and the sum is carried out across all the $n_{\mathrm{stack},r}$ clusters in stack $r$. Assuming that the different stacks are uncorrelated among them, which implies that a given cluster can only appear at most in one of the stacks, the stacked likelihood can be written as

\begin{equation}
\mathcal{L}_{\mathrm{stacked}} = P (\mathcal{S} | \mathcal{C}) = \prod_{r=1}^{n_{\mathrm{stacks}}} P ( \mathbfit{S}_r | \mathcal{C}),
\end{equation}
where we have dropped the explicit dependency on model parameters $\mathbfit{p}$. 

Computing the probability for each of the stacks, $P ( \mathbfit{S}_r | \mathcal{C})$, requires, in principle, jointly marginalising over a number of intermediate variables (e.g., cluster mass) for \textit{each} of the clusters in the stack, which can be computationally expensive. However, if the number of clusters in each stack is large enough, $P ( \mathbfit{S}_r | \mathcal{C})$ is approximately a Gaussian distribution with some mean $\bar{\mathbfit{S}}_r (\mathcal{C})$ and some covariance $\mathbfss{C}_r (\mathcal{C})$. \cnc{} makes such a Gaussian approximation, first computing $\bar{\mathbfit{S}}_r (\mathcal{C})$ and $\mathbfss{C}_r (\mathcal{C})$ assuming a hierarchical model for the stacked observable of each cluster in the stack, and then evaluating $P ( \mathbfit{S}_r | \mathcal{C})$ as a Gaussian distribution centred on $\mathbfit{S}_r - \bar{\mathbfit{S}}_r (\mathcal{C})$ with covariance $\mathbfss{C}_r (\mathcal{C})$. The covariance $\mathbfss{C}_r (\mathcal{C})$ can alternatively be given as an input for each stacked data vector, instead of it being computed within the hierarchical model framework.

Following this approach, the mean of the stacked data vector in stack $r$, $\bar{\mathbfit{S}}_r (\mathcal{C})$, can be written as

\begin{equation}\label{stack_mean_sum}
    \bar{\mathbfit{S}}_r (\mathcal{C}) = \sum_{i \,\mathrm{in \, stack }\,r}  w_i \bar{\mathbfit{s}}_i (\mathcal{C}),
\end{equation}
where $\bar{\mathbfit{s}}_i (\mathcal{C})$ is the expected value of the stacked observable for cluster $i$ (e.g., if the stacked observable is a lensing profile, the mean lensing profile of cluster $i$), and where, as in Eq.\,(\ref{eq:stacked_data}), the sum is carried out over all clusters in stack $r$. Assuming that the clusters in the stack are statistically independent from each other (as it has been assumed in our three catalogue likelihoods), $\bar{\mathbfit{s}}_i (\mathcal{C})$ depends only on the data vector and sky location of cluster $i$, $\mathbfit{D}_i$ and $\hat{\mathbfit{n}}_i$, respectively, $\bar{\mathbfit{s}}_i  (\mathcal{C}) = \bar{\mathbfit{s}} ( \mathbfit{D}_i, \hat{\mathbfit{n}}_i )$, and can be written as

\begin{equation}\label{eq:meanstack}
\bar{\mathbfit{s}} (\mathbfit{D}_i, \hat{\mathbfit{n}}_i ) = \int  d \mathbfit{s}  \mathbfit{s} P ( \mathbfit{s} | \mathbfit{D}_i, \hat{\mathbfit{n}}_i ).
\end{equation}
Here, $ P ( \mathbfit{s} | \mathbfit{D}_i, \hat{\mathbfit{n}}_i )$ is the pdf followed by the stacked observable $\mathbfit{s}$ conditioned on the data for cluster $i$, which can be, in turn, written as

\begin{equation}
P ( \mathbfit{s} | \mathbfit{D}_i, \hat{\mathbfit{n}}_i ) = \int dM P ( \mathbfit{s} | M, \mathbfit{D}_i, \hat{\mathbfit{n}}_i ) P ( M | \mathbfit{D}_i, \hat{\mathbfit{n}}_i ),
\end{equation}
where $M$ is the cluster mass and $P ( M | \mathbfit{D}_i, \hat{\mathbfit{n}}_i )$ is the pdf for the cluster mass given its cluster data. Inserting this expression into Eq.\,(\ref{eq:meanstack}) leads to

\begin{equation}\label{eq:stacked_mean}
\bar{\mathbfit{s}}_i (\mathcal{C}) = \bar{\mathbfit{s}} (\mathbfit{D}_i, \hat{\mathbfit{n}}_i ) = \int dM \bar{\mathbfit{s}} (M, \mathbfit{D}_i, \hat{\mathbfit{n}}_i) P ( M | \mathbfit{D}_i, \hat{\mathbfit{n}}_i ),
\end{equation}
where $\bar{\mathbfit{s}} (M, \mathbfit{D}_i, \hat{\mathbfit{n}}_i)$ is the expected value of the stacked observable $\mathbfit{s}$ for a given cluster with mass $M$, cluster data $\mathbfit{D}_i$ and sky location $\hat{\mathbfit{n}}_i$. \cnc{} uses this expression to compute $\bar{\mathbfit{S}}_r (\mathcal{C})$. $\bar{\mathbfit{s}} (M, \mathbfit{D}_i, \hat{\mathbfit{n}}_i)$ can be calculated for a given hierarchical model describing the stacked data and is given as an input function to the code. On the other hand, $P ( M | \mathbfit{D}_i, \hat{\mathbfit{n}}_i )$ is obtained as a byproduct of the evaluation of the individual cluster likelihoods in the unbinned approach (see Section\,\ref{sec:stacked_implementation}).

Following a similar argument, the second-order moment of the stacked observable for cluster $i$, $\langle s_{i,l} s_{i,m} \rangle (\mathcal{C}) = \langle s_{i,l} s_{i,m} \rangle (\mathbfit{D}_i, \hat{\mathbfit{n}}_i )$, where $l$ and $m$ denote the vector indices of $\mathbfit{s}_i$, can be written as

\begin{equation}\label{stackvar1}
\langle s_{i,l} s_{i,m} \rangle (\mathcal{C})  = \int dM \bar{s}^{(2)}_{lm} (M, \mathbfit{D}_i, \hat{\mathbfit{n}}_i) P ( M | \mathbfit{D}_i, \hat{\mathbfit{n}}_i ).
\end{equation}
Here, $\bar{s}^{(2)}_{lm}$ is the second-order moment of the stacked observable for a given cluster with mass $M$, cluster data $\mathbfit{D}_i$ and sky location $\hat{\mathbfit{n}}_i$, which, as for $\bar{\mathbfit{s}} (M, \mathbfit{D}_i, \hat{\mathbfit{n}}_i)$, can be computed for a given hierarchical model and is given as an input function to \cnc{}. The components of the stacked observable covariance for cluster $i$, $\mathbfss{c}_i$, are then given by

\begin{equation}\label{stackvar2}
    c_{i,lm}  (\mathcal{C}) = \langle s_{i,l} s_{i,m} \rangle (\mathcal{C}) - \bar{s}_{i,l} (\mathcal{C}) \bar{s}_{i,m} (\mathcal{C}),
\end{equation}
and, in turn, the total covariance for the stacked data vector $r$ is given by

\begin{equation}\label{stackvar3}
 \mathbfss{C}_r (\mathcal{C}) = \sum_{i \,\mathrm{in \, stack }\,r}  w_i^2 \mathbfss{c}_i  (\mathcal{C}).
\end{equation}
Note that, as expected, the stacked covariance  $\mathbfss{C}_r (\mathcal{C})$ benefits from the `$1/n$' reduction from stacking. Indeed, if, e.g., $\omega_i = 1/n_{\mathrm{stack},r}$, the stacked covariance $\mathbfss{C}_r (\mathcal{C})$ is a factor of $n_{\mathrm{stack},r}$ smaller than the stacked observable covariance of the individual clusters averaged over the clusters in the stack. \cnc{} follows this approach in order to compute the stacked covariance $\mathbfss{C}_r (\mathcal{C})$, provided that a hierarchical model for the stacked observable is given. Alternatively, $\mathbfss{C}_r (\mathcal{C})$ can be given directly as an input to the code, which can be useful if, e.g., it has been estimated with simulations.

\section{Likelihood implementation}\label{sec:implementation}

In this section we describe how \cnc{} computes the three cluster number count likelihoods presented in Section\,\ref{sec:formalism}, as well as the stacked likelihood, at a given point $\mathbfit{p}$ in input parameter space. The input parameter space is spanned by two types of parameters: cosmological and scaling relation parameters. The former can include the total matter density parameter, $\Omega_{\mathrm{m}}$; the baryonic density parameter, $\Omega_{\mathrm{b}}$; the Hubble constant, $H_0$; the linear amplitude of the scalar perturbations, $A_s$ (or, alternatively, $\sigma_8$); the scalar spectral index, $n_s$; the sum of the neutrino masses, $\sum m_{\nu}$; the effective number of relativistic degrees of freedom, $N_{\mathrm{eff}}$; and the dark energy equation of state parameter, $w$. This list of parameters, however, is not exhaustive, as \cnc{} can work with a broad range of cosmological models via the interface with \texttt{class}\footnote{  \href{https://lesgourg.github.io/class_public/class.html}{ \texttt{lesgourg.github.io/class\_public/class.html}}} \citep{Lesgourgues2011,Blas2011} and \texttt{class\_sz}. The scaling relation parameters (including the mass observable covariance parameters), on the other hand, can be defined in a custom way for each set of observables of choice.

\cnc{} starts with an initialisation step in which the data (the cluster catalogue, the stacked data, if available, and any data needed for the computation of the mass--observable scaling relations) is loaded. If the likelihood is to be evaluated at several points in input parameter space, as it would be the case, e.g., in an MCMC analysis, this initialisation step is only performed once for the sake of efficiency. 

Then, for any of the three available likelihoods, \cnc{} evaluates the cluster abundance across mass and redshift, $d^2 N / (dM dz)$, and, from it, it computes the abundance across the selection observable and true redshift, $d^2 N / (d \zeta_{\mathrm{obs}} dz)$ (defined in Eq.\,\ref{eq:sumabunance}), which is obtained following a `forward convolutional' approach. These `halo mass function' and `cluster abundance' steps are described in Sections\,\ref{sec:hmf} and\,\ref{sec:abundance}, respectively. In the binned and extreme value cases, the likelihood is evaluated in a straightforward way using $d^2 N / (d \zeta_{\mathrm{obs}} dz)$ (see Sections\,\ref{sec:binned_implementation} and\,\ref{sec:ev_implementation}, respectively). In the unbinned case, on the other hand, $d^2 N / (d \zeta_{\mathrm{obs}} dz)$ is used in order to calculate the `cluster abundance' part of the likelihood, whereas the `cluster data' part is evaluated from $d^2 N / (d \zeta_{\mathrm{obs}} dz)$ if the selection observable is the only available mass observable, and following a `backward convolutional' approach otherwise. This is described in detail in Section\,\ref{sec:unbinned_implementation}. Finally, for the unbinned likelihood and if stacked data is provided, the stacked likelihood is computed, as we describe in Section\,\ref{sec:stacked_implementation}.

We note that, for the sake of numerical stability, \cnc{} always computes the logarithm of the likelihood, instead of the likelihood itself. In addition, we note that all the necessary background cosmology quantities can be computed with either the \texttt{astropy.cosmology}\footnote{ \href{https://docs.astropy.org/en/stable/cosmology/index.html}{\texttt{docs.astropy.org}}} package or with \texttt{class\_sz} \citep{Bolliet2023b}, as specified by an input parameter. 
\cnc{} has a number of input parameters controlling the different aspects of its likelihood computations. Some of them are described throughout this section, and an exhaustive list is given in Appendix\,\ref{appendix:c}.

Throughout the rest of this section we illustrate how \cnc{} operates with the specific example of a Simons-Observatory-like catalogue, which we generate with \cnc{}'s synthetic catalogue generator (see Sections\,\ref{subsec:generator} and \ref{subsec:catalogues}). This catalogue features two mass observables: the tSZ signal-to-noise, $q_{\mathrm{obs}}$, and the CMB lensing signal-to-noise, $p_{\mathrm{obs}}$, for which we assume that a measurement is available for every cluster in the sample. We also assume that the sample is selected by imposing a tSZ signal-to-noise threshold at $q_{\mathrm{th}}=5$. We refer the reader to Section\,\ref{sec:features} for further details about this catalogue, in particular for the exact definition of the mass observables, as well as for the values of the input \cnc{} parameters that we use to perform our calculations. Despite illustrating the implementation with this particular catalogue, we stress that the code is implemented in a completely general way and that it can deal with an arbitrary number of custom-defined mass observables.


\subsection{Halo mass function}\label{sec:hmf}

For any of the three likelihoods (unbinned, binned, and extreme value), \cnc{} starts by computing the cluster abundance across mass $M$ and true redshift $z$, $d^3 N / (dM dz d\Omega)$, on a grid in the $M$--$z$ plane. We refer to this first step as the `halo mass function' step. As noted in Section\,\ref{sec:formalism}, $d^3 N / (dM dz d\Omega)$ is just a product of the halo mass function $ d^2 N / (dV dM)$ times the comoving volume element $ d^2V / (dz d\Omega)$. \cnc{} computes the volume element with either \texttt{astropy.cosmology} or \texttt{class\_sz}, and offers four different ways of computing the halo mass function:

\begin{itemize}
    \item A custom implementation of the \citet{Tinker2008} halo mass function in which the linear matter power spectrum is obtained with the \texttt{cosmopower}\footnote{ \href{https://github.com/cosmopower-organization}{\texttt{github.com/cosmopower-organization}}} emulator \citep{Spurio2022,Bolliet2023} and the top-hat linear overdensity variance is efficiently calculated with the \texttt{mcfit}\footnote{ \href{https://github.com/eelregit/mcfit}{\texttt{github.com/eelregit/mcfit}}} package. \texttt{cosmpower} provides a very fast computation of the linear power spectrum ($\sim 0.015$\,s on a laptop). Coupled with the efficiency of \texttt{mcfit}, this option is the fastest way of calculating the \citet{Tinker2008} halo mass function within \cnc{}.
    \item All the halo mass function implemented in \texttt{class\_sz}, which are also evaluated using \texttt{cosmopower} (see \citealt{Bolliet2023b}).
    \item A number of halo mass functions, including that of  \citet{Tinker2008}, as implemented in the \texttt{hmf} package \citep{Murray2013}, which uses \texttt{CAMB}\footnote{\href{https://camb.info}{\texttt{camb.info}}} \citep{Lewis1999} in order to compute the linear matter power spectrum. This option is slower than the two previous ones, but is included for completeness and as a robustness check.
    \item The Mira-Titan halo mass function emulator\footnote{\href{https://miratitanhmfemulator.readthedocs.io/en/latest/}{\texttt{miratitanhmfemulator.readthedocs.io}} }\citep{Bocquet2020}, which, for a given set of input cosmological parameters, emulates the halo mass function. If the halo mass function is evaluated at about $10^2$ redshift values, the Mira-Titan emulator provides a comparable speed to that given by \cnc{}'s custom implementation of the \citet{Tinker2008} halo mass function.

\end{itemize}

The mass and redshift limits of the grid on which $d^3 N / (dV dz d\Omega)$ is computed are specified as input parameters. These limits must span the region in the $M$--$z$ plane that contributes in a non-negligible way to the cluster abundance, which depends on the cluster catalogue to be analysed. For SO, for instance, we find that a minimum and a maximum mass of $M_{500,\mathrm{min}} = 10^{13} M_{\odot}$ and $M_{500,\mathrm{max}} = 10^{16} M_{\odot}$ suffice, as do a minimum and a maximum redshift of $z_{\mathrm{min}}=0.01$ and $z_{\mathrm{max}}=3$. The number of evaluations of the  across mass and redshift are also specified as input parameters. These numbers are equal to the number of evaluations of the cluster abundance in the selection observable--redshift plane, as we describe in Section\,\ref{sec:abundance}.


\subsection{Cluster abundance across selection observable and redshift: a forward convolutional approach}\label{sec:abundance}

After the evaluation of $d^3 N / (dM dz d\Omega)$ on a grid in the $M$--$z$ plane, \cnc{} proceeds by computing the cluster abundance across selection observable $\zeta_{\mathrm{obs}}$ and true redshift $z$, $d^3 N / (d \zeta_{\mathrm{obs}} dz d\Omega)$, for each of the $n_{\mathrm{tile}}$ selection tiles. We refer to this second step as the `cluster abundance' step.

The cluster abundance $d^2 N / (d \zeta_{\mathrm{obs}} dz d\Omega)$ can be calculated in a `brute-force' way by evaluating, for each tile, the $n_{\mathrm{layer}}$-dimensional integral in Eq.\,(\ref{eq:abundanceobs}) on a $\zeta_{\mathrm{obs}}$--$z$ grid. If we denote the number of evaluations across $\zeta_{\mathrm{obs}}$ and $z$ with $n_{\zeta}$ and $n_z$, respectively, and assume for simplicity that the number of evaluations in the selection observable intermediate variables is also $n_{\zeta}$, this brute-force approach leads to a computation time that scales roughly as $n_{\zeta}^{n_{\mathrm{layer}}+1} \times n_z \times n_{\mathrm{tile}}$\footnote{It may be possible to integrate analytically over some of the intermediate variables in the hierarchical model. In that case, the dimensionality of the `brute-force' numerical integral would be reduced.}. 

It is possible to calculate the cluster abundance in a faster way by noting that the hierarchical model describing the mass observables can be thought of as a series of changes of variables and convolutions. \cnc{} follows such an approach, which we refer to as the `forward convolutional' approach and which we now describe.

Let us consider a given true redshift $z$ and a given selection tile $k$. The cluster abundance across mass and redshift, $d^3 N / (dM dz d\Omega)$, becomes a single-variable function of the mass $M$. In addition, within our hierarchical model, for a given redshift and selection tile, there is a one-to-one map between mass $M$ and the intermediate selection observable variable of the the first layer,  $\zeta^{(1)}_{\mathrm{in}} = f_\zeta^{(1)} (M, z, \hat{\mathbfit{n}}_k)$, where $f_\zeta^{(1)}$ denotes the scaling relation of the selection observable in the first layer, and where $\hat{\mathbfit{n}}_k$ denotes the centre of tile $k$. Thus, changing variables, the abundance across $\zeta^{(1)}_{\mathrm{in}}$ and $z$ can be written as

\begin{equation}\label{eq:changev}
\frac{d^3 N}{d \zeta^{(1)}_{\mathrm{in}} dz d\Omega} (\zeta^{(1)}_{\mathrm{in}}, z) = \frac{d^3 N}{dM dz d\Omega} \left( \frac{d f_\zeta^{(1)} }{dM} \right)^{-1},
\end{equation}
where the two factors on the right hand side of the equation are evaluated at the mass $M$ given by $\zeta^{(1)}_{\mathrm{in}} = f_\zeta^{(1)} (M, z, \hat{\mathbfit{n}}_k)$. 

As described in Section\,\ref{sec:hierarchical}, $\zeta^{(1)}_{\mathrm{in}}$ is connected to the output selection observable variable in the first layer, $\zeta^{(1)}$, through Gaussian scatter. We can then write the cluster abundance across $\zeta^{(1)}$ and $z$ as

\begin{equation}\label{eq:convo}
  \frac{d^3 N}{d \zeta^{(1)} dz d\Omega} (\zeta^{(1)},z ) =  \int d \zeta^{(1)}_{\mathrm{in}} P (\zeta^{(1)} | \zeta^{(1)}_{\mathrm{in}}) \frac{d^3 N}{d \zeta^{(1)}_{\mathrm{in}} dz d\Omega}  (\zeta^{(1)}_{\mathrm{in}} , z ),
\end{equation}
where $P (\zeta^{(1)} | \zeta^{(1)}_{\mathrm{in}})$ is a Gaussian centred at $ \zeta^{(1)}_{\mathrm{in}}$ and with some standard deviation $\sigma_\zeta^{(1)}$. This integral is a convolution, namely of $ d^3 N / (d \zeta^{(1)}_{\mathrm{in}} dz d\Omega)$ with a Gaussian kernel with standard deviation $\sigma_\zeta^{(1)}$.

As each layer in the model consists of this same structure, only depending on the output variable of the previous layer, we can iterate this change of variable + convolution procedure over all the layers in the model in order to finally obtain the cluster abundance across the observed selection observable, $\zeta_{\mathrm{obs}}$, and true redshift $z$, $ d^3 N / (d \zeta^{(1)}_{\mathrm{obs}} dz d\Omega)$.

\cnc{} follows this forward convolutional approach in order to compute $ d^3 N / (d \zeta_{\mathrm{obs}} dz d\Omega)$ for each selection tile and each of the redshifts for which $d^3 N / (dM dz d\Omega)$ was evaluated in the halo mass function step. In each layer, the convolution can be carried out as either a real-space or a Fourier space operation (the latter using FFTs and the convolution theorem), as specified by an input parameter. The former is faster if $n_{\zeta}$ is small, whereas the latter becomes faster for larges values of $n_{\zeta}$. 

We note that, in each layer, the change-of-variable step requires evaluating the derivative of the corresponding scaling relation with respect to the output variable of the previous layer. These derivatives can be given as input to the code, which can be useful if they can be calculated analytically. Alternatively \cnc{} can calculate them numerically.

The forward convolutional approach is significantly faster than the brute-force one. Indeed, if the convolutions are performed with FFTs, each change of variable + convolution step scales, approximately, as $n_{\zeta} \log n_{\zeta}$, and the total computation time thus scales as $n_{\mathrm{layer}} \times n_{\zeta} \log n_{\zeta} \times n_z \times n_{\mathrm{tile}}$.

The final output of this cluster abundance step is $ d^3 N / (d \zeta_{\mathrm{obs}} dz d\Omega)$ evaluated on a $n_{\zeta} \times n_z$ rectangular grid on the $\zeta_{\mathrm{obs}}$--$z$ plane for all the selection tiles. \cnc{} also computes the total cluster abundance, $ d^2 N / (d \zeta_{\mathrm{obs}} dz)$, by adding the individual abundances for each tile multiplied by the corresponding tile solid angles (see Eq.\,\ref{eq:sumabunance}). In addition, for each tile, \cnc{} integrates $ d^3 N / (d \zeta_{\mathrm{obs}} dz d\Omega)$ along $\zeta_{\mathrm{obs}}$ and $z$, obtaining, respectively, the one-dimensional cluster distributions across $z$, $ d^2 N / (dz d\Omega)$, and across $ \zeta_{\mathrm{obs}}$, $d^2 N/ (d \zeta_{\mathrm{obs}} d\Omega)$. It also integrates the total abundance across both $\zeta_{\mathrm{obs}}$ and $z$, obtaining the mean total number of clusters in the the catalogue, $\bar{N}_{\mathrm{tot}}$. \cnc{} performs all these integrals using Simpson's rule, as implemented in \texttt{scipy}.


\begin{figure}
\centering
\includegraphics[width=0.5\textwidth,trim={00mm 0mm 0mm 0mm},clip]{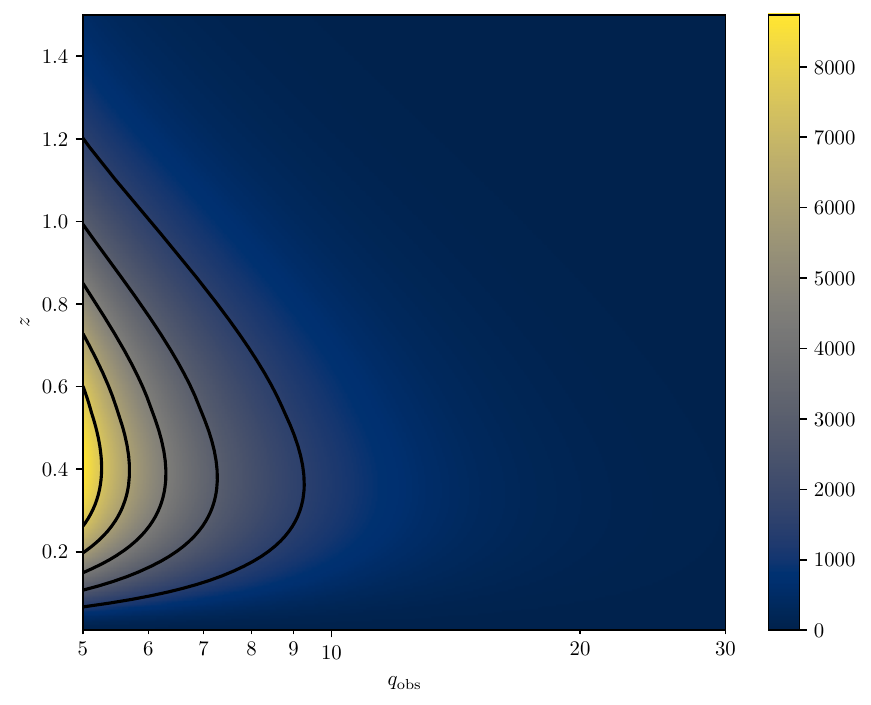}
\caption{Galaxy cluster abundance across tSZ signal-to-noise $q_{\mathrm{obs}}$ and true redshift $z$, $d^2N/(dq_{\mathrm{obs}}dz)$, for our SO-like cluster sample, computed with \cnc{}'s forward convolutional approach.}
\label{fig:abundance_2d}
\end{figure}

\begin{figure}
\centering
\includegraphics[width=0.48\textwidth,trim={30mm 0mm 30mm 0mm},clip]{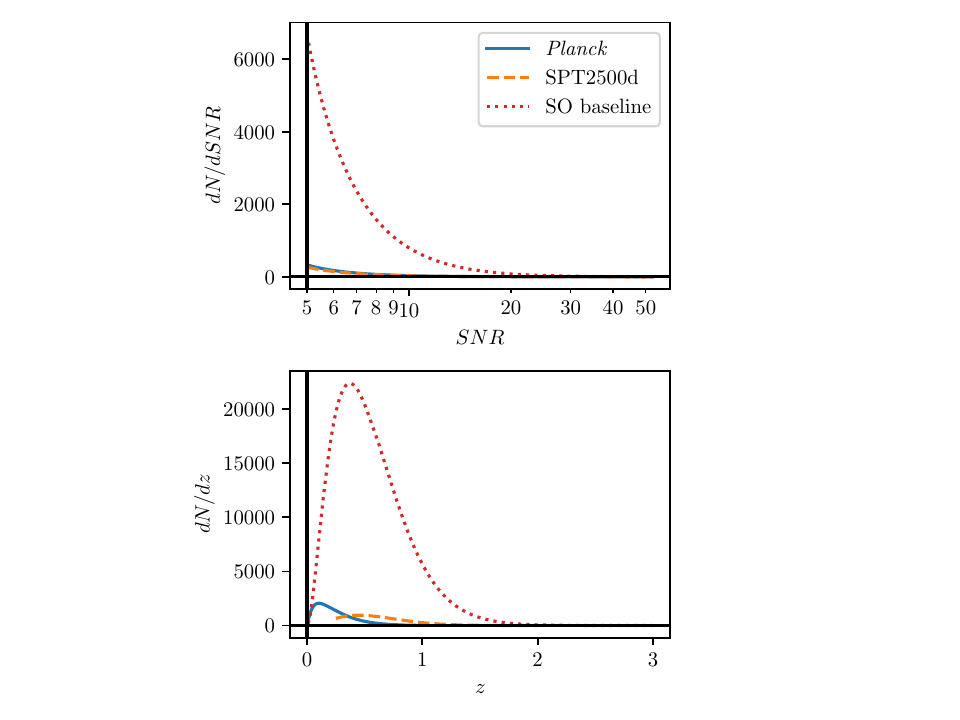}
\caption{Galaxy cluster abundance across tSZ signal-to-noise (top panel) and across true redshift (bottom panel) for our SO-like sample (dotted red curve) and for the \textit{Planck} MMF3 cosmology cluster sample and the SPT2500d sample (solid blue and dashed orange curves, respectively), computed with \cnc{}.}
\label{fig:abundance_1d}
\end{figure}

As an illustration, Figure\,\ref{fig:abundance_2d} shows the two-dimensional cluster abundance across the tSZ signal-to-noise $q_{\mathrm{obs}}$ and true redshift $z$ for our SO-like survey as computed by \cnc{}. The one-dimensional integrated distributions across $q_{\mathrm{obs}}$ and $z$, which are obtained by integrating the two-dimensional distribution of Figure\,\ref{fig:abundance_2d} along the appropriate axis, are shown in Figure\,\ref{fig:abundance_1d} (top and bottom panels, respectively), together with the analogous distributions for the \textit{Planck} MMF3 and the SPT2500d cluster surveys \citep{Ade2016,Planck2016xxvii,Bocquet2018}, which are also computed with \cnc{}.

\subsection{Unbinned likelihood}\label{sec:unbinned_implementation}

As shown in Section\,\ref{sec:formalism}, the unbinned likelihood consists of two terms: a `cluster abundance' term and a `cluster data' term (see, in particular, Eq.\,\ref{eq:unbinned}). The former is simply given by $\exp{(-\bar{N}_{\mathrm{tot}})}$, where $\bar{N}_{\mathrm{tot}}$ is the total number of clusters in the catalogue. As described in Section\,\ref{sec:abundance}, \cnc{} computes $\bar{N}_{\mathrm{tot}}$ in its cluster abundance step, and so the cluster abundance of the unbinned likelihood term is trivially obtained.

The cluster data term, on the other hand, is the product of the individual cluster likelihoods $P( \mathbfit{D}_i | \hat{\mathbfit{n}}_i)$ for all clusters in the catalogue. \cnc{} computes it by iterating over the clusters in the catalogue, computing each individual cluster likelihood in one of the following three different ways, depending on the data available for each cluster:

\begin{itemize}
\item \textbf{Only the selection observable is available}: In this case, $P( \mathbfit{D}_i | \hat{\mathbfit{n}}_i) = P( \zeta_{\mathrm{obs}} | \hat{\mathbfit{n}}_i)$, which is proportional to the cluster abundance across $\zeta_{\mathrm{obs}}$ at the cluster sky location $\hat{\mathbfit{n}}_i$, $d^2 N / (d\zeta_{\mathrm{obs}} d\Omega)$ (as it should be clear from Eq.\,\ref{eq:unconditioneddata}). \cnc{} obtains the individual cluster likelihood by simply interpolating linearly the cluster abundance $d^2 N / (d\zeta_{\mathrm{obs}} d\Omega)$ for the selection tile within which the cluster was selected (which was evaluated in the cluster abundance step) at the measured value of $\zeta_{\mathrm{obs}}$. This is an extremely fast operation.

\item \textbf{Only the selection observable and the redshift are available}: In this case, the individual cluster likelihood can be written as (see Eq.\,\ref{eq:unconditioneddata})

\begin{equation}\label{eq:intzz}
P (z_{\mathrm{obs}}, \zeta_{\mathrm{obs}} |  \hat{\mathbfit{n}}_i ) = \int_0^{\infty} dz P (z_{\mathrm{obs},i} | z, \hat{\mathbfit{n}}_i ) P (  \zeta_{\mathrm{obs}} | z, \hat{\mathbfit{n}}_i) P(z | \hat{\mathbfit{n}}_i).
\end{equation}
Here, $P (z_{\mathrm{obs},i} | z, \hat{\mathbfit{n}}_i )$ accounts for the redshift measurement uncertainty. If it is negligible, $P(z_{\mathrm{obs},i} | z, \hat{\mathbfit{n}}_i ) = \delta ( z_{\mathrm{obs},i} - z)$, and the integral over true redshift $z$ reduces to evaluating the integrand at $z=z_{\mathrm{obs}}$. Otherwise, \cnc{} takes $P (z_{\mathrm{obs},i} | z, \hat{\mathbfit{n}}_i )$ to be a Gaussian centred on $z$ with standard deviation $\sigma_{z,i}$, which can be cluster-dependent. On the other hand, $P (  \zeta_{\mathrm{obs}} | z, \hat{\mathbfit{n}}_i) P(z | \hat{\mathbfit{n}}_i) = P (  \zeta_{\mathrm{obs}} ,z  | \hat{\mathbfit{n}}_i)$, which is proportional to the cluster abundance across the selection observable and true redshift at the cluster's sky location $\hat{\mathbfit{n}}_i$, $d^3N / (d\zeta_{\mathrm{obs}} dz d\Omega)$, computed in the cluster abundance step. If the redshift measurement uncertainty can be neglected, the individual cluster likelihood is obtained by interpolating linearly $d^3N / (d\zeta_{\mathrm{obs}} dz d\Omega)$ at the cluster's measured values of $\zeta_{\mathrm{obs}}$ and $z_{\mathrm{obs}}$. Otherwise, following Eq.\,(\ref{eq:intzz}), the distribution is evaluated at the measured value of $\zeta_{\mathrm{obs}}$ and integrated against the redshift uncertainty Gaussian using Simpson's rule. Both procedures are extremely fast.

\item \textbf{General case}: If a given cluster has measurements available for more than one mass observable (i.e., for other observables in addition to the selection observable), the individual cluster likelihood cannot be calculated from the quantities obtained in the cluster abundance step. In this case, \cnc{} follows a more general approach that we refer to as our `backward convolutional' approach, and which we motivate and describe in detail in the rest of this section. We note that this approach can also be followed in the two previous cases, with \cnc{} offering this possibility as a (slower) option. In Appendix\,\ref{appendix:b} we compare the numerical values of the unbinned likelihood computed with the backward convolutional to those obtained from the cluster abundance in this specific case, finding excellent agreement.

\end{itemize}

\subsubsection{Backward convolutional approach: motivation}

Let us first motivate the backward convolutional approach. Consider a cluster with a set of $n_{\mathrm{obs}}$ mass observables $\bmath{\omega}_{\mathrm{obs}}$ (including the selection observable) and with observed redshift $z_{\mathrm{obs}}$. Its individual cluster likelihood is given by Eq.\,(\ref{eq:unconditioneddata}). Unlike in Section\,\ref{sec:formalism}, here we drop the cluster index from all the variables in order to avoid clutter in the notation. Computing the individual cluster likelihood involves, in principle, carrying out an integral across cluster mass $M$, true redshift $z$ (if redshift measurement uncertainty cannot be neglected), and the $n_{\mathrm{obs}}$ intermediate mass observable variables of $n_{\mathrm{layer}}-1$ layers. Thus, a `brute-force' evaluation of the individual cluster likelihood scales roughly as $n_{\mathrm{eval}}^{(n_{\mathrm{layer}}-1)n_{\mathrm{obs}}+1} \times n_{z,\mathrm{eval}}$, where $n_{\mathrm{eval}}$ is the number of evaluations across the intermediate mass observable variables and the mass (assumed to be the same for all of them), and $n_{z,\mathrm{eval}}$ is the number of evaluations across redshift, with $n_{z,\mathrm{eval}} = 1$ if the redshift measurement uncertainty can be neglected.

As with the selection observable, the hierarchical model linking the cluster mass and true redshift with $\bmath{\omega}_{\mathrm{obs}}$ can also be thought of as a series of changes of variables and convolutions, this time in an $n_{\mathrm{obs}}$-dimensional space. Therefore, it is possible to consider an $n_{\mathrm{obs}}$-dimensional version of the forward convolutional approach in order to compute the individual cluster likelihood for each cluster. This approach would involve evaluating $d^2 N / ( |d \bmath{\omega}_{\mathrm{obs}}| dz d\Omega)$ for each true redshift and relevant sky location on an $n_{\mathrm{eval}}^{n_{\mathrm{obs}}}$ grid and interpolating the distribution at the location of each cluster's mass observable data point, integrating over true redshift if necessary. The evaluation of each individual cluster likelihood with this approach would scale roughly as $n_{\mathrm{layer}} \times (n_{\mathrm{eval}} \log n_{\mathrm{eval}})^{n_{\mathrm{obs}}} \times n_{z,\mathrm{eval}}$, which makes it potentially faster than the brute-force evaluation.

In this approach, the $n_{\mathrm{obs}}$-dimensional distribution  $d^2 N / ( |d \bmath{\omega}_{\mathrm{obs}}| dz d\Omega)$ could be evaluated for the observable ranges spanning all the clusters in the catalogue, as \cnc{} does for the abundance across selection observable. In this case, $n_{\mathrm{eval}}$ ought to be a large number, large enough for the scatter convolutions to be well sampled while ensuring that the entirety of the allowed observable ranges are spanned. In addition, $d^2 N / ( |d \bmath{\omega}_{\mathrm{obs}}| dz d\Omega)$ would have to be evaluated for as many sky location combinations as necessary. Note that the sky dependence of the different mass observables may be different, which would lead to a large number of different combinations. 

Alternatively, $d^2 N / ( |d \bmath{\omega}_{\mathrm{obs}}| dz d\Omega)$ could be computed following an $n_{\mathrm{obs}}$-dimensional forward convolutional approach on a cluster-by-cluster basis, considering only the vicinity of each cluster's data point in cluster data space. This approach would still require $n_{\mathrm{obs}}$-dimensional operations, but $n_{\mathrm{eval}}$ could potentially be a much smaller number. A version of this `cluster-by-cluster' forward convolutional approach was followed, e.g., in \citet{Bocquet2018} in order to evaluate their `mass calibration' likelihood. Note that the brute-force approach could also be restricted to the vicinity of each cluster's data point, but we expect it to be slower than the cluster-by-cluster forward convolutional approach.

This cluster-by-cluster forward convolutional approach is promising, but there is one scenario in which its computational efficiency is suboptimal. This is the case in which there are mass observables with negligible correlation between them, i.e., observables whose scatter can be assumed to be uncorrelated across all the layers in the model. This may be the case, e.g., for a CMB lensing observable, for which the scatter may be dominated by reconstruction noise and thus be mostly uncorrelated with the cluster structure and, therefore, with the scatter in other cluster mass observables such as the tSZ signal (see, e.g., \citealt{Zubeldia2020}). The correlation between tSZ and optical weak lensing observables has also been found to have a negligible impact on cosmological constraints in a recent SPT analysis \citep{Bocquet2024}. In this scenario, the scatter integral, given by Eq.\,(\ref{eq:scatterintegral}), separates into a product of as many integrals as there are sets of mass observables with completely independent scatter, which we refer to as `correlation sets', each integral having a lower number of dimensions. The evaluation time can therefore be greatly reduced if this separability of the scatter integral is taken advantage of. The forward convolutional approach, however, does not take advantage of this fact. Our backward convolutional approach does. Indeed, its evaluation time scales roughly as $n_{\mathrm{layer}} \times (n_{\mathrm{eval}} \log n_{\mathrm{eval}})^{n_{\mathrm{obs,max}}} \times n_{z,\mathrm{eval}}$, where $n_{\mathrm{obs,max}}$ is the number of observables in the largest correlation set, with $n_{\mathrm{obs,max}} \leq n_{\mathrm{obs}}$. 

We note that, as a robustness check, \cnc{} also offers the possibility of brute-force evaluation of the individual cluster likelihoods.

\subsubsection{Backward convolutional approach: description}

For each cluster, our backward convolutional approach starts by dividing the available mass observable measurements $\bmath{\omega}_{\mathrm{obs}}$ into $n_{\mathrm{corr}}$ correlation sets. A mass observable belongs to a given correlation set if its scatter has non-zero correlation with the scatter of any of the other mass observables in the set for any of the layers in the model. Any two observables belong to two different sets if their scatter is uncorrelated across all the layers. 

For example, if a given cluster has measurements available for three mass observables, $A$, $B$, and $C$, and the scatter in $C$ is completely uncorrelated with the scatter in both $A$ and $B$, but the scatters in $A$ and $B$ are correlated in at least one of their layers, there are two correlation sets, $\{ A, B \}$ and $\{ C \}$. The different correlation sets are specified as an input to the code, and we note that there can be at most as many correlation sets as observables, and, at least, just one set, which will contain all the mass observables.

For a given cluster, the pdf followed by the cluster mass observables, $\bmath{\omega}_{\mathrm{obs}}$, conditioned on the cluster mass $M$, true redshift $z$, and sky location $\hat{\mathbfit{n}}$, the `scatter integral', can then be factorised as a product of $n_{\mathrm{corr}}$ scatter integrals, one for each correlation set, i.e.,

\begin{equation}\label{eq:corrset}
P (\bmath{\omega}_{\mathrm{obs}} | M, z, \hat{\mathbfit{n}})  = \prod_{l=1}^{n_{\mathrm{corr}}} P (\bmath{\omega}_{\mathrm{obs},l} | M, z, \hat{\mathbfit{n}}),
\end{equation}
where $\bmath{\omega}_{\mathrm{obs},l}$ denotes the measured values of the mass observables of correlation set $l$. Using Eq.\,(\ref{eq:pdata}), we can then write the pdf followed by $\bmath{\omega}_{\mathrm{obs}}$ at given true redshift $z$ and sky location $\hat{\mathbfit{n}}$ as

\begin{equation}\label{eq:pdatacorr}
P (\bmath{\omega}_{\mathrm{obs}} | z, \hat{\mathbfit{n}}) = \int_0^{\infty} dM \prod_{l=1}^{n_{\mathrm{corr}}} P (\bmath{\omega}_{\mathrm{obs},l} | M, z, \hat{\mathbfit{n}}) P (M, z | \hat{\mathbfit{n}}).
\end{equation}
For each value of $z$, \cnc{} evaluates the scatter integral of each correlation set on a grid of evaluation masses $\bmath{M}_{\mathrm{eval}}$ following our backward convolutional approach, proceeding backwards from mass observable to mass (see below). On the other hand, $P (M, z | \hat{\mathbfit{n}})$ is obtained by interpolating linearly the cluster abundance computed in the halo mass function step at the cluster redshift $z$ and at the mass grid $\bmath{M}_{\mathrm{eval}}$. For efficiency reasons, a downsampled version of the halo function may be interpolated, as specified by an input parameter (see Appendix\,\ref{appendix:c}). The integral over mass is then performed using Simpson's rule. 

The individual cluster likelihood $P (z_{\mathrm{obs}}, \bmath{\omega}_{\mathrm{obs}} |  \hat{\mathbfit{n}} )$ is finally given by (see Eq.\,\ref{eq:unconditioneddata})

\begin{equation}\label{eq:intz}
P (z_{\mathrm{obs}}, \bmath{\omega}_{\mathrm{obs}} |  \hat{\mathbfit{n}} ) = \int_0^{\infty} dz P (z_{\mathrm{obs},i} | z, \hat{\mathbfit{n}}_i ) P (  \bmath{\omega}_{\mathrm{obs}} | z, \hat{\mathbfit{n}}),
\end{equation}
where $P(z_{\mathrm{obs},i} | z, \hat{\mathbfit{n}}_i )$ accounts for the redshift measurement uncertainty. If this uncertainty can be neglected,  $P(z_{\mathrm{obs},i} | z, \hat{\mathbfit{n}}_i ) = \delta ( z_{\mathrm{obs},i} - z)$ and $P (z_{\mathrm{obs}}, \bmath{\omega}_{\mathrm{obs}} |  \hat{\mathbfit{n}} ) $ reduces to $P(\bmath{\omega}_{\mathrm{obs}} | z_{\mathrm{obs}}, \hat{\bmath{n}} )$, i.e., only one redshift evaluation suffices. Otherwise, \cnc{} computes the integral in Eq.\,(\ref{eq:intz}) in a `brute-force' way by evaluating $P(\bmath{\omega}_{\mathrm{obs}} | z, \hat{\bmath{n}})$ on a redshift grid with $n_{z,\mathrm{eval}}$ values around the measured value $z_{\mathrm{obs}}$ and integrating it multiplied by $P(z_{\mathrm{obs}} | z, \hat{\mathbfit{n}})$ using Simpson's rule\footnote{An alternative approach could involve convolving the mass--redshift distribution $P(M,z)$ with the redshift uncertainty kernel $P(z_{\mathrm{obs}} | z, \hat{\mathbfit{n}} )$. We leave an exploration of this idea for further work.}. In \cnc{}, $P(z_{\mathrm{obs}} | z, \hat{\mathbfit{n}} )$ is assumed to be a Gaussian centred at $z_{\mathrm{obs}}$ with standard deviation $\sigma_{z,i}$, which can be different for each cluster and is given as input data to the code. We note that \cnc{} can deal with the possibility of some clusters in the catalogue having negligible redshift measurement uncertainties and others having non-negligible ones.

Let us now consider a given true redshift $z$ and a correlation set $l$. In the rest of this section, we describe how $\bmath{M}_{\mathrm{eval}}$ is chosen and how backward convolutional approach delivering $P (\bmath{\omega}_{\mathrm{obs},l} | M, z, \hat{\mathbfit{n}})$ operates. 

The mass grid $\bmath{M}_{\mathrm{eval}}$ has to be broad enough for the integral in Eq.\,(\ref{eq:pdatacorr}) to converge, but not too broad so that a relatively small number of points $n_{\mathrm{eval}}$ suffices in order to compute the integral with acceptable accuracy. In order to achieve this, $\bmath{M}_{\mathrm{eval}}$ is obtained by estimating the range of masses for which $g (M) \equiv P (\zeta_{\mathrm{obs}} | M, z, \hat{\mathbfit{n}})$ takes non-negligible values. For this, \cnc{} first finds the peak and width of $g(M)$. The location of the peak is estimated by propagating the set of masses used in the halo mass function step through the hierarchical model for the selection observable, assuming no scatter. This procedure assigns a value of the selection observable for each mass. The mass for which the difference with the measured value of the observable, $\zeta_{\mathrm{obs}}$, is the smallest is then chosen as the estimate for the peak of $g(M)$, $M_{\mathrm{peak}}$.

In order to estimate the width of $g(M)$, $\Delta M$, \cnc{} propagates the scatter in the selection observable through the hierarchical model, starting from the measured value $\zeta_{\mathrm{obs}}$ and proceeding backwards. In each layer, the standard deviation of its scatter, $\sigma_{\zeta}^{(j)}$, where $j$ denotes the $j$-th layer, is added in quadrature to the combined scatter due to all the outer layers, which is estimated using standard error propagation. That is, the propagated scatter down to the $j$-th layer, $\Delta \zeta^{(j)}$, is given by

\begin{equation}
    \Delta \zeta^{(j)} = \left[ \left( \frac{d f_{\zeta}^{(j+1)}}{d \zeta^{(j)} } \right)^{-1} \left( \Delta \zeta^{(j+1)} \right)^2 +  \left( \sigma_{\zeta}^{(j)} \right)^2 \right]^{1/2},
\end{equation}
where $j=n_{\mathrm{layer}}-1,\ldots,0$, $\Delta \zeta^{(n_{\mathrm{layer}})} = \sigma_{\zeta}^{(n_{\mathrm{layer}})} $, and $\Delta \zeta^{(0)} = \Delta M$, and where the scaling relation derivative for layer $j$ is evaluated at $f_{\zeta}^{(j)} \circ f_{\zeta}^{(j-1)} \circ \ldots \circ f_{\zeta}^{(1)} (M_{\mathrm{obs}}) $, $\circ$ denoting function composition. 

With $M_{\mathrm{peak}}$ and $\Delta M$ in hand, \cnc{} then takes $\bmath{M}_{\mathrm{eval}}$ to be a set of $n_{\mathrm{eval}}$ linearly-spaced values between $\max \{ M_{\mathrm{est}} - c_M \Delta M  , M_{\mathrm{min}}\} $ and $\min \{ M_{\mathrm{est}} + c_M \Delta M , M_{\mathrm{max}} \}$, where $c_M$ is an input parameter controlling how far away from $M_{\mathrm{peak}}$ is to be considered, and where $M_{\mathrm{min}}$ and $M_{\mathrm{max}}$ are the minimum and maximum masses considered in the halo mass function step.



Once the mass grid $\bmath{M}_{\mathrm{eval}}$ is obtained, \cnc{} propagates it forward through the scaling relations up to the last layer of the model for all the mass observables in the correlation set. More precisely, for each layer, \cnc{} produces an $n_{\mathrm{eval}}^{n_{\mathrm{obs},l}}$ grid in the space spanned by the layer's input variables. In general, the points in each of these grids will not be equally spaced. Because of this, for each layer \cnc{} also produces an $n_{\mathrm{eval}}^{n_{\mathrm{obs},l}}$ grid with equally-spaced points across each dimension, covering the same region in the layer's input variable space.

The code then proceeds to propagate backwards the scatter of \emph{all} the mass observables in the correlation set, starting at the observed data point $\bmath{\omega}_{\mathrm{obs},l}$, in order to evaluate $P (\bmath{\omega}_{\mathrm{obs},l} | M, z, \hat{\mathbfit{n}})$ on the mass grid $\bmath{M}_{\mathrm{eval}}$. Following the hierarchical model described in Section\,\ref{sec:hierarchical}, the pdf of $\bmath{\omega}_{\mathrm{obs},l}$ conditioned on the input variables of the last layer, $\bmath{\omega}^{(n_{\mathrm{layer}}-1)}_l$, and on $z$ and $\hat{\bmath{n}}$, $P( \bmath{\omega}_{\mathrm{obs},l} | \bmath{\omega}^{(n_{\mathrm{layer}}-1)}_l, z, \hat{\bmath{n}})$, is an $n_{\mathrm{obs},l}$-dimensional Gaussian centred on $\bmath{f}^{(n_{\mathrm{layer}})} (\bmath{\omega}^{(n_{\mathrm{layer}}-1)}_l, z, \hat{\mathbf{n}})$, where $\bmath{f}^{(n_{\mathrm{layer}})}$ is the set of scaling relations of the last layer. \cnc{} evaluates $P( \bmath{\omega}_{\mathrm{obs}} | \bmath{\omega}^{(n_{\mathrm{layer},l}-1)}, z, \hat{\bmath{n}})$ on its equally-spaced $\bmath{\omega}^{(n_{\mathrm{layer}}-1)}$-space grid for the measured values of the mass observables $\bmath{\omega}_{\mathrm{obs},l}$, obtaining a distribution in $\bmath{\omega}_l^{(n_{\mathrm{layer}}-1)}$-space. This distribution is then convolved with an $n_{\mathrm{obs},l}$-dimensional Gaussian accounting for the scatter in the penultimate layer. This way, $P( \bmath{\omega}_{\mathrm{obs},l} | \bmath{\omega}^{(n_{\mathrm{layer}}-1)}_{\mathrm{in},l},  z, \hat{\bmath{n}})$ is evaluated on an equally-spaced grid in $\bmath{\omega}_{\mathrm{in},l}^{(n_{\mathrm{layer}}-1)}$-space, where we recall that  $\bmath{\omega}^{(n_{\mathrm{layer}}-1)}_{\mathrm{in},l}$ is the intermediate variables of the $n_{\mathrm{layer}}-1$-th layer. 

In effect, we can write

\begin{multline}
    P(\bmath{\omega}_{\mathrm{obs},l} | \bmath{\omega}^{(n_{\mathrm{layer}}-1)}_{\mathrm{in},l}) = \\ \int  |d \bmath{\omega}^{(n_{\mathrm{layer}}-1)}_{l}| P( \bmath{\omega}_{\mathrm{obs},l} | \bmath{\omega}^{(n_{\mathrm{layer}}-1)}_l,  z, \hat{\bmath{n}}) \\ P( \bmath{\omega}^{(n_{\mathrm{layer}}-1)}_l | \bmath{\omega}^{(n_{\mathrm{layer}}-1)}_{\mathrm{in},l},  z, \hat{\bmath{n}}),
\end{multline}
and so $P( \bmath{\omega}_{\mathrm{obs},l} | \bmath{\omega}^{(n_{\mathrm{layer}}-1)}_{\mathrm{in},l},  z, \hat{\bmath{n}})$ is indeed given by a convolution of $P( \bmath{\omega}_{\mathrm{obs},l} | \bmath{\omega}^{(n_{\mathrm{layer}}-1)}_l,  z, \hat{\bmath{n}})$ with the Gaussian kernel describing the scatter in the penultimate layer, $P( \bmath{\omega}^{(n_{\mathrm{layer}}-1)}_l | \bmath{\omega}^{(n_{\mathrm{layer}}-1)}_{\mathrm{in},l},  z, \hat{\bmath{n}})$. \cnc{} can evaluate this $n_{\mathrm{obs},l}$-dimensional convolution in either real or Fourier space, the latter using FFTs and the convolution theorem.

This evaluation of $P( \bmath{\omega}_{\mathrm{obs},l} | \bmath{\omega}^{(n_{\mathrm{layer}}-1)}_{\mathrm{in},l},  z, \hat{\bmath{n}})$ on an equally-spaced $\bmath{\omega}^{(n_{\mathrm{layer}}-1)}$-space grid is then interpolated at the nonlinear grid of the same layer, allowing for the pdf to be linked to the previous (the $n_{\mathrm{layer}}-2$-th) layer. 

This procedure is repeated until the first layer of the model is reached, with one convolution per layer and a total of $n_{\mathrm{layer}}$-1 convolutions. The final output is $P( \bmath{\omega}_{\mathrm{obs},l} | M,  z, \hat{\bmath{n}} )$ evaluated on an $n_{\mathrm{obs},l}$-dimensional grid in which all the axes correspond to $\bmath{M}_{\mathrm{eval}}$, as the mass is the input variable to the first layer for all mass observables. Simply extracting the diagonal of this distribution delivers a one-dimensional evaluation of $P( \bmath{\omega}_{\mathrm{obs},l} | M, z, \hat{\mathbfit{n}} )$ at $\bmath{M}_{\mathrm{eval}}$. This algorithm constitutes our backward convolutional approach. To the authors' knowledge, this is the first time this approach has been proposed in the literature and \cnc{} constitutes its first implementation.

In the case in which the hierarchical model consists of only one layer, no convolutions are required, and the scatter integral reduces to a Gaussian evaluated at the observed values of the mass observables. This case is incorporated as a possibility in \cnc{}.

Finally, we note that, unlike in the forward convolutional approach used to calculate the cluster abundance, the backward convolutional approach does not require to evaluate the derivative of the scaling relations of the mass observables, except for the selection observable, its derivatives being used to compute $\Delta M$.

\begin{figure}
\centering
\includegraphics[width=0.3\textwidth,trim={00mm 0mm 0mm 0mm},clip]{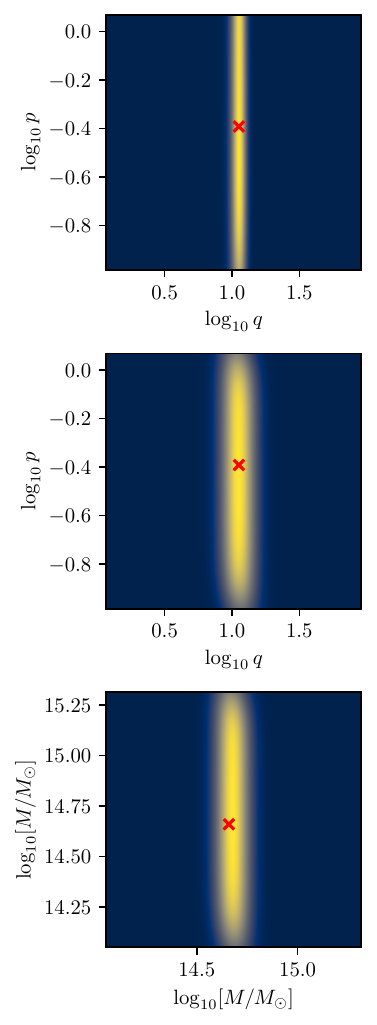}
\caption{\textit{Top panel}:  $P( q_{\mathrm{obs}}, p_{\mathrm{obs}} | \ln q, \ln p, z_{\mathrm{obs}}, \hat{\bmath{n}})$ evaluated by \cnc{} on a grid on the $\ln q$--$\ln p$ plane for a reference cluster in our reference SO-like catalogue. We recall that $q$ and $p$ are, respectively, the true tSZ and CMB lensing signal-to-noise, and $q_{\mathrm{obs}}$ and $p_{\mathrm{obs}}$ are their respective observed values. The observed data point is shown as a red cross. Note that the distribution is unbounded along the $ \ln p$ direction, as the CMB lensing measurement has a very small signal-to-noise. \textit{Middle panel}: $P( q_{\mathrm{obs}}, p_{\mathrm{obs}} | \ln \bar{q}, \ln \bar{p}, z_{\mathrm{obs}}, \hat{\bmath{n}})$,  for the same cluster, evaluated on the same grid on the $\ln \bar{q}$--$\ln \bar{p}$ plane, obtained by \cnc{} by convolving the distribution in the top panel with the Gaussian describing the scatter in the first layer of the model. $\bar{q}$ and $\bar{p}$ are, respectively, the mean tSZ and CMB lensing signal-to-noise, which, for a given redshift, are single-valued functions of the mass. \textit{Bottom panel}: Same distribution as in the middle panel, but recast in terms of two mass variables, with the cluster true mass shown as a red cross. Note that there are edge effects at the two ends of mass variable associated to $\bar{p}$. These are are caused by the distribution in the last layer (the one at the top panel) being unbounded in that direction and are of no concern, as only the diagonal of the map is of interest.}
\label{fig:backward_2d}
\end{figure}

 \begin{figure}
\centering
\includegraphics[width=0.4\textwidth,trim={00mm 0mm 0mm 0mm},clip]{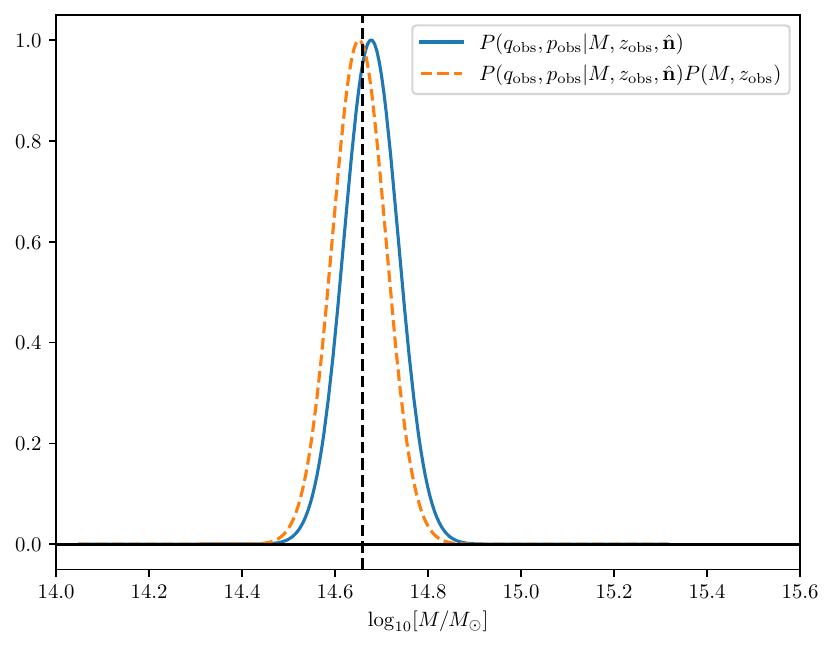}
\caption{$P( q_{\mathrm{obs}}, p_{\mathrm{obs}} | M, z_{\mathrm{obs}}, \hat{\bmath{n}})$ as a function of the mass $M$ for the same reference cluster as the one considered in Figure\,\ref{fig:backward_2d} (blue curve), as computed by \cnc{}. This one-dimensional distribution is obtained as the diagonal of the two-dimensional distribution in the bottom panel of Figure\,\ref{fig:backward_2d}. Its product with the cluster abundance across mass and redshift, $P(M,z_{\mathrm{obs}})$, is also shown (orange curve). Both curves have been rescaled by their maximum values in order to aid visualisation. If normalised to unity, the orange curve can be seen as a posterior for the cluster mass given the cluster mass observables, its redshift, and its sky location (see Section\,\ref{subsec:masstheory}). For comparison, the vertical dashed line shows the cluster's true mass.}
\label{fig:backward_1d}
\end{figure}

\subsubsection{Backward convolutional approach: illustration}\label{sec:illustration}

Figure\,\ref{fig:backward_2d} illustrates how the backward convolutional approach operates with the specific example of one cluster of our reference SO synthetic catalogue. The chosen cluster has $q_{\mathrm{obs}} = 11.25$, $p_{\mathrm{obs}} = 0.41$, a measured redshift $z_{\mathrm{obs}} = 0.65$ (with negligible measurement uncertainty), and a mass of $M = 4.56 \times 10^{14} M_{\odot}$. 

The top panel shows $P( \bmath{\omega}_{\mathrm{obs}} | \bmath{\omega}^{(n_{\mathrm{layer}}-1)},  z_{\mathrm{obs}}, \hat{\bmath{n}}) = P( q_{\mathrm{obs}}, p_{\mathrm{obs}} | \ln q, \ln p, z_{\mathrm{obs}}, \hat{\bmath{n}})$ evaluated on a $n_{\mathrm{eval}} \times n_{\mathrm{eval}}$ grid in the $\ln q$--$\ln p$ plane, with $n_{\mathrm{eval}} = 128$. A cross marks the logarithm of the coordinates of the observed data point, $(q_{\mathrm{obs}},p_{\mathrm{obs}})$. Here, $\ln q$ and $\ln p$ are the output variables of the penultimate (i.e., first) layer in the model, and $P( q_{\mathrm{obs}}, p_{\mathrm{obs}} | \ln q, \ln p, z_{\mathrm{obs}}, \hat{\bmath{n}})$ is a two-dimensional Gaussian centred at $(q-q_{\mathrm{obs}},p-p_{\mathrm{obs}})$ and with an identity covariance matrix. Note that the distribution is unbounded along the $\ln p$ axis for the chosen observable range, as the CMB lensing measurement is noise dominated, with a typical signal-to-noise per cluster of a fraction of unity. 

The middle panel shows the result of convolving the distribution in the top panel with the Gaussian describing the scatter in the first layer, which we take to be a two-dimensional Gaussian with variance along the $\ln q$ and $\ln p$ directions equal to $\sigma_q = 0.173$ and $\sigma_p = 0.22$, respectively, and a correlation coefficient $r = 0.77$. The convolved distribution therefore corresponds to $P( \bmath{\omega}_{\mathrm{obs}} | \bmath{\omega}_{\mathrm{in}}^{(n_{\mathrm{layer}}-1)},  z_{\mathrm{obs}}, \hat{\bmath{n}}) = P( q_{\mathrm{obs}}, p_{\mathrm{obs}} | \ln \bar{q}, \ln \bar{p},  z_{\mathrm{obs}}, \hat{\bmath{n}})$ evaluated on a grid in the $\ln \bar{q}$--$\ln  \bar{p}$ plane, where $\ln \bar{q}$ and $\ln \bar{p}$ are the mean tSZ and CMB lensing signal-to-noise, respectively. Note that, as expected, the distribution across the $\ln \bar{q}$ direction is noticeably broader than in the top panel, as an additional layer of scatter has now been taken into account. 

Since the first layer in the model has already been reached, this two-dimensional distribution can be recast in terms of two mass variables, there being a one-to-one map between mass and both $\ln \bar{q}$ and $\ln \bar{p}$ at fixed redshift and sky location. This distribution is shown in the bottom panel, with both axes corresponding to the mass $M$, and with the cluster true mass being shown as the red cross. Extracting the diagonal of this two-dimensional distribution leads to a one-dimensional evaluation of $P( q_{\mathrm{obs}}, p_{\mathrm{obs}} | M, z_{\mathrm{obs}}, \hat{\bmath{n}})$, which is shown in Figure\,\ref{fig:backward_1d} (solid blue curve) along with its product with $P(M,z)$ (dashed orange curve). We note that we have rescaled both curves by their maximum values to aid visualisation. The integral of $P( q_{\mathrm{obs}}, p_{\mathrm{obs}} | M, z_{\mathrm{obs}}, \hat{\bmath{n}})P(M,z)$ (the dashed orange curve) with respect to the mass gives the value of the individual cluster likelihood for this reference cluster.

\subsection{Cluster mass inference}\label{subsec:masstheory}

Since the backward convolutional approach evaluates $P( \bmath{\omega}_{\mathrm{obs}} | M,  z, \hat{\bmath{n}})$ for each cluster to which it is applied, it naturally provides a way of obtaining cluster mass estimates for a given point in (cosmological and scaling relation) parameter space. Indeed, using Bayes' theorem, the pdf followed the cluster mass $M$ given the observed mass observable data point $\bmath{\omega}_{\mathrm{obs}}$, its true redshift $z$, and its sky location $ \hat{\bmath{n}}$ is given by

\begin{equation}\label{eq:mass}
    P (M | \bmath{\omega}_{\mathrm{obs}},  z, \hat{\bmath{n}} ) \propto  P( \bmath{\omega}_{\mathrm{obs},l} | M,  z, \hat{\bmath{n}}) P (M, z | \hat{\bmath{n}}),
\end{equation}
where the proportionality factor can be obtained by normalising the distribution, and where we have used that $P(M,z | \hat{\bmath{n}}) = P(M |z, \hat{\bmath{n}}) P (z | \hat{\bmath{n}})$.

In its implementation of the backward convolutional approach, \cnc{} evaluates the right-hand side of Eq.\,(\ref{eq:mass}) on the mass grid $\mathbfit{M}_{\mathrm{eval}}$. If requested to do so (through an input parameter, see Appendix\,\ref{appendix:c}), \cnc{} can also trivially obtain the cluster mass pdf $ P (M | \bmath{\omega}_{\mathrm{obs}},  z, \hat{\bmath{n}} ) $ evaluated at $\mathbfit{M}_{\mathrm{eval}}$ and at the grid of evaluation redshifts (which we recall reduces to just the observed redshift $z_{\mathrm{obs}}$ if the redshift measurement uncertainty can be neglected). It can also deliver the mean estimated mass and its standard deviation, which are obtained by numerically computing the moments of $P (M | \bmath{\omega}_{\mathrm{obs}},  z, \hat{\bmath{n}})$.

Figure\,\ref{fig:backward_1d} illustrates \cnc{}'s mass inference process, with the dashed orange curve corresponding (up to a normalisation) to the posterior on the mass given the cluster mass observables, cluster redshift, and sky location for our reference cluster in our synthetic SO-like catalogue.

Note that these mass estimates are not Eddington-biased, as this bias is naturally accounted for by the second factor on the right-hand side of Eq.\,(\ref{eq:mass}). They do, however, suffer from Malmquist bias, as we illustrate in Section\,\ref{subsec:masses}.


\subsection{Binned likelihood}\label{sec:binned_implementation}

\cnc{} evaluates the binned likelihood making use of the cluster abundances that it computes in its cluster abundance step (see Section\,\ref{sec:abundance}). As noted in Section\,\ref{subsec:binned}, \cnc{} supports three binning schemes: across the selection observable $\zeta_{\mathrm{obs}}$, across redshift, and across both $\zeta_{\mathrm{obs}}$ and redshift. In each of them, the bin edges are given as input parameters. Unlike its unbinned counterpart, the binned likelihood implementation does not incorporate the possibility of redshift measurement uncertainties, and so assumes that $z_{\mathrm{obs}}=z$, where $z$ is true redshift. It also cannot take as data other observables different from the selection observable and/or redshift, and cannot deal with missing data (e.g., missing redshift measurements in the cases of binning across redshift or across both $\zeta_{\mathrm{obs}}$ and redshift).

If binning across both $\zeta_{\mathrm{obs}}$ and redshift is desired, \cnc{} integrates the cluster abundance across selection observable and redshift, $d^2 N / (d \zeta_{\mathrm{obs}} dz)$, which was evaluated on a grid on the $\zeta_{\mathrm{obs}}$--$z$ plane in the cluster abundance step, within each bin using Simpson's rule. This procedure gives the expected number of clusters within each bin, $\bar{N}_s$, from which the binned likelihood is trivially evaluated (see Eqs.\,\ref{eq:poisson} and \ref{eq:binned}). Similarly, if binning across only $\zeta_{\mathrm{obs}}$ or redshift is desired, the relevant distributions, $d N / d \zeta_{\mathrm{obs}}$ and $d N / dz$, respectively, both of which were also evaluated in the cluster abundance step, are integrated instead.

\subsection{Extreme value likelihood}\label{sec:ev_implementation}

As with the binned likelihood, \cnc{} evaluates the extreme value likelihood using the cluster abundances computed in the cluster abundance step. Namely, it obtains the mean number of clusters with $\zeta_{\mathrm{obs}} > \zeta_{\mathrm{max}}$, $\bar{N}(\zeta_{\mathrm{obs}} > \zeta_{\mathrm{max}})$, by integrating, using Simpson's rule, $d N / d \zeta_{\mathrm{obs}}$ from $\zeta_{\mathrm{obs}} = \zeta_{\mathrm{max}}$ to the maximum $\zeta_{\mathrm{obs}}$ value for which it was evaluated, which we denote with $\zeta_{\mathrm{\infty}}$. Note that $d N / d \zeta_{\mathrm{obs}}$ has to be negligible at $\zeta_{\infty}$ so that its integral with this upper integration limit constitutes a good approximation of the improper integral to $\zeta_{\mathrm{obs}}=\infty$. The likelihood is then given by $\mathrm{exp}[- \bar{N}(\zeta_{\mathrm{obs}} > \zeta_{\mathrm{max}})]$ (see Section\,\ref{subsec:extreme}).

\cnc{} can also calculate the pdf followed by $\zeta_{\mathrm{max}}$, $P(\zeta_{\mathrm{max}})$ which, as noted in Section\,\ref{subsec:extreme}, is simply given by the derivative of the extreme value likelihood with respect to $\zeta_{\mathrm{max}}$. \cnc{} computes this derivative numerically, delivering $P(\zeta_{\mathrm{max}})$ evaluated on a $\zeta_{\mathrm{max}}$ grid. Using this pdf, \cnc{} can also compute the expected value of $\zeta_{\mathrm{max}}$, $\bar{\zeta}_{\mathrm{max}}$.

\subsection{Unconfirmed detections}\label{sec:nonvalimplementation}

As noted in Section\,\ref{sec:nonval}, \cnc{} allows for the presence of unconfirmed detections, i.e., possible false detections, in the cluster catalogue for two of its likelihoods: the unbinned likelihood and the binned likelihood with selection observable binning.

Both likelihoods require as input the abundance of false detections as a function of the selection observable and, if relevant, selection tile, $dN_{\mathrm{f}} / (d \zeta_{\mathrm{obs}} d\Omega)$. In addition, as explained in Appendix\,\ref{sec:appendixa}, the unbinned likelihood also needs as input the probability of validation as a function of the selection observable and, if relevant, selection tile. These additional pieces of information are loaded by \cnc{} and used in order to account for unconfirmed detections in a consistent way, following the formalism developed in Section\,\ref{sec:nonval} and Appendix\,\ref{sec:appendixa}. The implementation details are described in Appendix\,\ref{sec:appendixa}.

\subsection{Stacked likelihood}\label{sec:stacked_implementation}

\subsubsection{Likelihood computation}

As detailed in Section\,\ref{subsec:stacked}, \cnc{} allows to combine the unbinned likelihood for the catalogue $\mathcal{C}$ with a stacked data set $\mathcal{S}$ comprising $n_{\mathrm{stack}}$ stacked data vectors. This is done by calculating the stacked likelihood $\mathcal{L}_{\mathrm{stacked}} (\mathbfit{p}) = P ( \mathcal{S} | \mathcal{C}, \mathbfit{p})$, which is then multiplied by the unbinned likelihood, $\mathcal{L}_{\mathrm{unbinned}} (\mathbfit{p})  = P(\mathcal{C} | \mathbfit{p})$. As explained in Section\,\ref{subsec:stacked}, \cnc{} approximates the pdf of each stacked data vector with a Gaussian with mean  $\bar{\mathbfit{S}}_r (\mathcal{C})$ and covariance  $\mathbfss{C}_r (\mathcal{C})$, where the index $i$ denotes stack $r$. The mean $\bar{\mathbfit{S}}_r (\mathcal{C})$ is calculated assuming a model for the stacked observable for each cluster in the stack. The covariance, on the other hand, can be either calculated assuming the same model for each cluster in the stack, or, alternatively, can be given as an input.

More specifically, $\bar{\mathbfit{S}}_r (\mathcal{C})$ is given by the sum of the expected values of the stacked observable for all the clusters in stack $r$ (Eq.\,\ref{stack_mean_sum}), each of which is given by Eq.\,(\ref{eq:stacked_mean}), which we repeat here for convenience, dropping the cluster index to avoid clutter in the notation:

\begin{equation}\label{eq:stacked_mean2}
\bar{\mathbfit{s}} (\mathcal{C}) = \bar{\mathbfit{s}} (\mathbfit{D}, \hat{\mathbfit{n}} ) = \int dM \bar{\mathbfit{s}} (M, \mathbfit{D}, \hat{\mathbfit{n}}) P ( M | \mathbfit{D}, \hat{\mathbfit{n}} ).
\end{equation}
Here, $\bar{\mathbfit{s}} (M, \mathbfit{D}, \hat{\mathbfit{n}})$ is the expected value of the stacked observable for a cluster with mass $M$, cluster data $\mathbfit{D}$, and sky location $\hat{\mathbfit{n}}$. This can be calculated for a given hierarchical model and is given as an input function to \cnc{}. On the other hand, $P ( M | \mathbfit{D}, \hat{\mathbfit{n}})$ is the pdf followed by the cluster mass given the cluster data $\mathbfit{D}$ and sky location $\hat{\mathbfit{n}}$. Recalling that $\mathbfit{D} = \{ z_{\mathrm{obs}},  \bmath{\omega}_{\mathrm{obs}} \}$, and assuming the redshift measurement uncertainty to be negligible, this pdf is given by Eq.\,(\ref{eq:mass}) and is obtained as a by-product of the individual cluster likelihood evaluation with the backward convolutional approach (see Section\,\ref{sec:unbinned_implementation}).

In order to evaluate $\bar{\mathbfit{s}} (\mathbfit{D}, \hat{\mathbfit{n}} )$ for each cluster in the stack, \cnc{} first evaluates its individual cluster likelihood with the backward convolutional approach, obtaining $P ( M | \mathbfit{D}, \hat{\mathbfit{n}})$ evaluated at a set of masses $\mathbfit{M}_{\mathrm{eval}}$. Note that this calculation may have already taken place if the individual cluster likelihood for that cluster was computed with the backward convolutional approach; in that instance, \cnc{} does not repeat the calculation. \cnc{} then evaluates $\bar{\mathbfit{s}} (M, \mathbfit{D}, \hat{\mathbfit{n}})$ at $\mathbfit{M}_{\mathrm{eval}}$, and finally computes the integral in Eq.\,(\ref{eq:stacked_mean2}) with Simpson's rule.

If the stacked observable covariance is also to be calculated assuming a model for the stacked observable for each cluster, \cnc{} does so following an analogous approach, similarly using the mass pdf delivered by the backward convolutional approach (see Eqs.\,\ref{stackvar1},\,\ref{stackvar2}, and\,\ref{stackvar3}).

Note that, in its current implementation, \cnc{} computes the stacked likelihood taking the redshift measurement uncertainties to be negligible for all the clusters in the stack. 

\subsubsection{Illustration}

As an illustration, consider our reference synthetic SO catalogue, from which we can construct a stacked data set $\mathcal{S}$ consisting of one one-dimensional stacked data vector $\mathbfit{S} = p_{\mathrm{stacked}}$ resulting from averaging the CMB lensing signal-to-noise $p_{\mathrm{obs}}$ across the cluster sample, 

\begin{equation}\label{eq:stackedmean}
    p_{\mathrm{stacked}} = \frac{1}{n_{\mathrm{tot}}} \sum_{i=1}^{n_{\mathrm{tot}}} p_{\mathrm{obs},i}.
\end{equation}
Neglecting the correlation in the scatter between the CMB lensing and the tSZ signal-to-noise observables and following the notation of Section\,\ref{sec:illustration}, $\bar{\mathbfit{s}} (M, \mathbfit{D}, \hat{\mathbfit{n}})$ can be written as $\bar{\mathbfit{s}} (M, \mathbfit{D}, \hat{\mathbfit{n}}) = \bar{p}_{\mathrm{obs}} (M,z_{\mathrm{obs}},q_{\mathrm{obs}}, \hat{\mathbfit{n}}) = \exp[ \ln \bar{p} (M,z_{\mathrm{obs}},\hat{\mathbfit{n}}) + \sigma_p^2/2] $, where we recall that $\bar{p}$ is the mean CMB lensing signal-to-noise, which is specified by the cluster mass, redshift and sky location, and where $\sigma_p$ is the intrinsic scatter. On the other hand, the second moment of $p_{\mathrm{obs}}$ at given mass, cluster data and sky location can be written as $\langle s^2 \rangle (M, \mathbfit{D}, \hat{\mathbfit{n}}) = \langle p_{\mathrm{obs}}^2 \rangle (M,z_{\mathrm{obs}},q_{\mathrm{obs}}, \hat{\mathbfit{n}}) = \sigma_{\mathrm{intrinsic}}^2 (M,z_{\mathrm{obs}}, \hat{\mathbfit{n}}) + \sigma_{\mathrm{observational}}^2$,  where $ \sigma_{\mathrm{intrinsic}}^2 = \exp(\sigma_p^2 - 1) \exp[2 \ln \bar{p} (M,z_{\mathrm{obs}},\hat{\mathbfit{n}}) + \sigma_p^2 ] + \bar{p}_{\mathrm{obs}} (M,z_{\mathrm{obs}},q_{\mathrm{obs}}, \hat{\mathbfit{n}})^2 $, and $\sigma_{\mathrm{observational}}^2 = 1$. With these expressions in hand, \cnc{} can evaluate the mean value and the covariance (in this case, just variance) of $p_{\mathrm{stacked}}$. We note that since the CMB lensing signal-to-noise per cluster is very small, the total scatter is dominated by the unit-variance observational scatter, and therefore the variance of $p_{\mathrm{stacked}}$ will be slightly larger than $1/n_\mathrm{tot}$ (see Section\,\ref{sec:validation}).

\subsection{Other features}\label{sec:features}

\subsubsection{Synthetic cluster catalogue generation}\label{subsec:generator}

In addition to computing three different types of cluster number count likelihoods, \cnc{} can generate synthetic cluster catalogues consistent with the the assumptions underlying the likelihoods. 

For a given observational set-up (set of observables with their corresponding hierarchical model, and selection criterion), this is done as follows. First, \cnc{} runs its halo mass function step, evaluating $d^3 N / (dM dz d\Omega)$ on a mass--redshift grid, with its boundaries being specified as input parameters (see Section\,\ref{sec:hmf}). The total mean number of clusters \emph{in the Universe}, $\bar{N}_{\mathrm{Universe}}$ is then computed by integrating $d^3 N / (dM dz d\Omega)$ over mass and redshift and across the survey footprint. The total number of clusters in the Universe $N_{\mathrm{Universe}}$ is then obtained as a random draw from a Poisson distribution with expected value equal to $\bar{N}_{\mathrm{Universe}}$. Next, $N_{\mathrm{Universe}}$ clusters are drawn from the pdf given by $P (M,z,\hat{\mathbfit{n}} ) = P (M,z ) = d^3 N / (dM dz d\Omega) / \bar{N}_{\mathrm{Universe}}$, which generates a catalogue of all the clusters in the Universe (within the survey footprint), each with a mass $M$, redshift $z$ and sky location $\hat{\mathbfit{n}}$. As $d^3 N / (dM dz d\Omega)$ does not depend on sky location, the cluster's sky location is assigned randomly within the survey footprint. This is consistent with the assumption underlying the three \cnc{} likelihoods that the clusters are statistically independent from each other (i.e., the covariance due to cluster clustering is neglected). On the other hand, each $M$--$z$ pair is drawn from $P(M,z)$ by first drawing a redshift from $P(z) = \int P(M,z) dM$ using the cumulative distribution function random number generation algorithm, and then drawing a mass from $P(M|z) = P(M,z)/P(z)$ using the same algorithm.

Next, each point in the $M$--$z$--$\hat{\mathbfit{n}}$ space is propagated through the hierarchical model for all the mass observables in the survey, accounting for the scatter in each layer through the addition of Gaussian random noise with the appropriate covariance. The cluster catalogue is then constructed by selecting all the clusters for which the value of the selection observable is greater than the selection threshold $\zeta_{\mathrm{th}}$.

Synthetic catalogues generated this way can be very useful for testing the accuracy of the \cnc{} likelihoods for a given survey, as they are much easier to generate than it is to compute the likelihood and they are generated making exactly the same assumptions underlying the likelihoods (see, e.g., \citealt{Zubeldia2019,Bocquet2024}). In Section\,\ref{sec:validation} we use a set of these synthetic catalogues in order to test \cnc{} in the context of the upcoming Simons Observatory.

\subsubsection{Goodness of fit}\label{subsec:goodness}

\cnc{} can compute the modified Cash goodness-of-fit statistic of \citet{Kaastra2017}, $C$ (see also \citealt{Cash1979}), for any of the three binning schemes supported by the binned likelihood (rectangular bins across the selection observable $\zeta_{\mathrm{obs}}$, redshift $z$, or both $\zeta_{\mathrm{obs}}$ and $z$). That is, \cnc{} can assess the goodness of fit across the selection observable and/or redshift. The $C$ statistic is given by

\begin{equation}
    C = 2 \sum_{s=1}^{n_{\mathrm{bin}}} \bar{N}_s - N_s + N_s \ln (N_s/\bar{N}_s),
\end{equation}
where $\bar{N}_s$ and $N_s$ are, respectively, the expected and observed number of clusters in bin $s$, the former being computed by \cnc{} as detailed in Section\,\ref{sec:binned_implementation}. \cnc{} can also compute its expected value, $\bar{C}$, and standard deviation, $\sigma_C$, using the formulae of \citet{Kaastra2017} (in particular, their Eqs.\,8--22).  $\bar{C}$ can then be compared to the observed value $C$ for a given point in parameter space (e.g., for the `best-fit' point; see, e.g., \citealt{Bocquet2018}).

\subsubsection{Parallel computing}

If desired, \cnc{} can parallelise several of its computations using Python's \texttt{multiprocessing} module\footnote{\href{https://docs.python.org/3/library/multiprocessing.html}{\texttt{docs.python.org/3/library/multiprocessing.html}}}, leading to enhanced performance through the use of more than one core. \cnc{} can parallelise:

\begin{itemize}
    \item The evaluation of the halo mass function on a grid of redshifts in the halo mass function step. This can be done for all the halo mass functions except for the Mira-Titan emulator. The grid of redshifts is divided into $n_{\mathrm{core,hmf}}$ smaller batches of similar sizes, each of which is assigned to a different core.
    \item The forward convolutional approach in the cluster abundance step, splitting either the redshift grid or the set of selection tiles into $n_{\mathrm{core,abundance}}$ smaller batches, each of which is assigned to a different core. The parallelisation scheme (redshifts or tiles) can be specified through an input parameter.
    \item The computation of the cluster data part of the unbinned likelihood, dividing the set of clusters into $n_{\mathrm{core,data}}$ smaller batches, each of which is assigned to a different core.
    \item The computation of the stacked likelihood, dividing the set of clusters in each stack into $n_{\mathrm{core,stacked}}$ batches, each of which is assigned to a different core.
\end{itemize}
The number of cores to be used in each task ($n_{\mathrm{core,hmf}}$, $n_{\mathrm{core,abundance}}$, $n_{\mathrm{core,data}}$, and $n_{\mathrm{core,stacked}}$) are specified as input parameters. The optimal number of cores for each task will depend on the catalogue, the values of \cnc{}'s precision parameters, and the machine on which the code is run.

\subsubsection{Interface with MCMC codes}

\cnc{} is interfaced with the MCMC sampling code \texttt{Cobaya} \citep{Torrado2019,Torrado2021}, allowing for easy-to-run MCMC parameter estimation (see the online documentation). It can also be easily interfaced with other MCMC codes, such as \texttt{emcee} \citep{ForemanMackey2013} and \texttt{CosmoSIS} \citep{Zuntz2015}, something we leave for future work.

\section{Likelihood validation}\label{sec:validation}

In this section we assess the accuracy of the implementation of the likelihoods in \cnc{} using a set of 100 synthetic SO-like cluster catalogues. We first describe our synthetic catalogues in Section\,\ref{subsec:catalogues} and the values for the input parameters that we set in Section\,\ref{sec:parametervalues}. We then analyse the catalogues in observable space in Section\,\ref{susbec:agremeent}, comparing their properties with \cnc{}'s predictions. Next, we use subsets of them for parameter inference, considering four cases:

\begin{itemize}
    \item \textbf{SZ unbinned}: the unbinned likelihood with redshift and tSZ data (56 catalogues).
    \item \textbf{SZ binned}: the binned likelihood with redshift and tSZ data (56 catalogues).
    \item \textbf{SZ+CMBlens}: the unbinned likelihood with redshift, tSZ and CMB lensing data (28 catalogues).    
     \item \textbf{SZ+CMBlens stacked}: the unbinned likelihood with redshift and tSZ data jointly with the stacked likelihood for the CMB lensing data (28 catalogues).
\end{itemize}

We analyse the derived constraints in Section\,\ref{subsec:biases}, quantifying the biases in the inferred parameter values and assessing the goodness of fit. Then, in Section\,\ref{subsec:masses} we illustrate \cnc{}'s cluster mass estimation capabilities, in Section\,\ref{subsec:efficiency} we offer a quantification of \cnc{}'s execution speed, and in Section\,\ref{subsec:classsz} we benchmark the code against \texttt{class\_sz}.

\subsection{Synthetic catalogues}\label{subsec:catalogues}

We generate 100 SO-like cluster catalogues using \cnc{}'s synthetic cluster catalogue generator. The catalogues are all generated assuming the same cosmology and the same scaling relation parameter values, i.e., they are statistically identical. We assume a spatially flat Lambda cold dark matter ($\Lambda$CDM) cosmology with $\Omega_{\mathrm{m}} = 0.315$, $\Omega_{\mathrm{b}} = 0.04897$, $h = 0.674$, $\sigma_8=0.811$, $n_{\mathrm{s}} = 0.96$, and $\sum m_\mathrm{\nu} = 0.06$\,eV \citep{Planck2018VI}, and use \cnc{}'s own implementation of the \citet{Tinker2008} halo mass function, which makes use of the \texttt{cosmopower} power spectrum emulator. We consider two mass observables: the tSZ signal-to-noise $q_{\mathrm{obs}}$ (our selection observable) and the CMB lensing signal-to-noise  $p_{\mathrm{obs}}$. Each cluster has a measurement for both mass observables, and the catalogues are constructed by imposing a selection threshold of $q_{\mathrm{th}}=5$. In addition, each cluster comes with a redshift measurement, which we assume to have negligible measurement uncertainty. We assume a survey footprint covering 40\,\% of the sky \citep{SO2019}, and assume that all the objects in the catalogue are confirmed clusters, not adding any false detections (the addition of false detections is studied in Appendix\,\ref{sec:appendixa}).

The two mass observables are linked to mass and redshift with a two-layer hierarchical model. In the first layer, the \emph{mean} tSZ signal-to-noise  $\bar{q} (M_{500},z)$ is given by

\begin{equation}
    \bar{q} (M_{500},z) = \frac{y_0 ( \beta_{\mathrm{SZ}} M_{500},z)}{\sigma_{y_0} (\theta_{500}  (\beta_{\mathrm{SZ}} M_{500},z))},
\end{equation}
where $y_0$ is the cluster's central Compton-$y$ value, $\sigma_{y_0}$ is the cluster detection multifrequency matched filter (MMF) noise evaluated at the cluster's angular scale $\theta_{500}$, and the tSZ mass bias is $\beta_{\mathrm{SZ}} = 0.8$. $y_0$ is given by

\begin{equation}\label{eq:scalrelsz}
    y_0 = 10^{A_{\mathrm{SZ}}} \left( \frac{ M_{500}}{ 3 \times 10^{14} h_{70}^{-1} M_{\odot}} \right)^{\alpha_{\mathrm{SZ}}} E^2(z) h_{70}^{-1/2},
\end{equation}
where $A_{\mathrm{SZ}}=-4.3054$, $\alpha_{\mathrm{SZ}}=1.1233$, and $h_{70} = h/0.7$. This scaling relation is consistent with the universal pressure profile of \citet{Arnaud2010} and its form follows the tSZ scaling relation of  \citet{Hilton2018}. 

The MMF noise $\sigma_{y_0}$, on the other hand, is computed by applying the tSZ cluster finder \texttt{SZiFi}\footnote{\href{https://github.com/inigozubeldia/szifi}{\texttt{github.com/inigozubeldia/szifi}}} \citep{Zubeldia2021,Zubeldia2022} to SO-like maps from the Websky simulation \citep{Stein2019,Stein2020}, assuming just one selection tile. More specifically, we consider observations at the six frequency channels of the SO Large Aperture Telescope (27, 39, 93, 145, 225, and 278\,GHz), using as input the Websky tSZ, kinetic SZ (kSZ), Cosmic Infrared Background (CIB) and CMB maps, appropriately convolved by the corresponding beams (Gaussian beams with a FWHM of 7.4, 5.1, 2.2, 1.4, 1., and 0.9\,arcmin, respectively; \citealt{SO2019}). We then add white noise, with noise levels, for each channel, of 71, 36, 8, 10, 22, and 54\,$\mu$K\,arcmin, respectively (SO baseline noise levels; \citealt{SO2019}). We tessellate the sky into 768 HEALPix pixels ($N_{\mathrm{side}} = 8$, see \citealt{Gorski2005}) and apply \texttt{SZiFi} to the first 10 tiles, using the standard MMF, i.e., without spectral foreground deprojection, computing the MMF noise for each tile at 15 angular scales, with the angular scale $\theta_{500}$ logarithmically spaced between $\theta_{500}=0.5$\,arcmin and $\theta_{500}=15$\,arcmin. Finally, for each angular scale, we take the average of the MMF noise across the 10 tiles.

Still in the first layer, the \emph{mean} CMB lensing signal-to-noise, $\bar{p} (M_{500},z)$, is given by

\begin{equation}
    \bar{p} (M_{500},z) = \frac{\kappa_0 (\beta_{\mathrm{CMBlens}} M_{500},z)}{\sigma_{\kappa_0} (\theta_{500} (\beta_{\mathrm{CMBlens}} M_{500},z))},
\end{equation}
where $\kappa_0 (M_{500},z)$ is the central value of the cluster's CMB lensing convergence, $\sigma_{\kappa_0} (\theta_{500} (\beta_{\mathrm{CMBlens}}M_{500},z))$ is the CMB lensing matched filter noise, and the CMB lensing mass bias is $\beta_{\mathrm{CMBlens}}=0.92$. This expression assumes that the cluster CMB lensing signal has been extracted with a matched filter approach, as first proposed in \citet{Melin2015} and applied, e.g., in \citet{Ade2016,Zubeldia2019,Zubeldia2020,Huchet2024}. Following \citet{Zubeldia2019,Zubeldia2020}, in order to compute both $\kappa_0$ and the matched filter noise, we assume that the cluster convergence profile is that of a truncated Navarro-Frenk-White profile (NFW; \citealt{Navarro1997}) with a concentration $c_{\mathrm{500}}=3$ and a truncation radius of $5 R_{500}$. We compute the matched filter noise using the publicly-available SO minimum-variance (temperature+polarisation) quadratic estimator reconstruction noise curve\footnote{\href{https://github.com/simonsobs/so_noise_models/tree/master/LAT_lensing_noise/lensing_v3_1_1/nlkk_v3_1_0_deproj0_SENS1_fsky0p4_qe_lT30-3000_lP30-5000.dat}{\texttt{github.com/simonsobs/so\_noise\_models/blob/master/ \\ LAT\_lensing\_noise/lensing\_v3\_1\_1/nlkk\_v3\_1\_0\_deproj0\_SENS1 \\ \_fsky0p4\_qe\_lT30-3000\_lP30-5000.dat }}}, taking it to be the same for all the clusters in the sample.

The mean tSZ and CMB lensing signal-to-noises are then linked to the logarithms of the \emph{true} tSZ and CMB lensing signal-to-noises through Gaussian intrinsic scatter, with a covariance matrix given by $\sigma_{\ln q} = 0.173$, $\sigma_{\ln p} = 0.22$, and a correlation coefficient $r = 0$ (i.e., no intrinsic correlation). In the second layer of the model, the set of scaling relations simply exponentiates $\ln q$ and $\ln p$, which are then linked to the observed values ($q_{\mathrm{obs}}$ and $p_{\mathrm{obs}}$, respectively) through uncorrelated Gaussian scatter with unit variance for both observables. Note that this model is very similar to that used in \citet{Zubeldia2019} and, for the tSZ observable, in \citet{Ade2016}, the only difference being the form of the tSZ scaling relation (Eq.\,\ref{eq:scalrelsz}).

In order to validate the implementation of the stacked likelihood we also consider, for each catalogue, a stacked data set consisting of a single one-dimensional stacked data vector given by the mean CMB lensing signal-to-noise across the cluster sample, $p_{\mathrm{stacked}}$ (see Eq.\,\ref{eq:stackedmean}).

\subsection{\cnc{} parameter values}\label{sec:parametervalues}

For both the generation of our synthetic catalogues and the evaluation of the \cnc{} likelihoods, we set a minimum mass $M_{\mathrm{min}} = 10^{13} M_{\odot}$, a maximum mass $M_{\mathrm{max}} = 10^{16} M_{\odot}$, a minimum redshift $z_{\mathrm{min}} = 0.01$, and a maximum redshift $z_{\mathrm{max}} = 3$. The cluster halo mass function and abundance steps are computed setting $n_{\zeta} = 2^{14}$ in the SZ unbinned and binned cases, and $n_{\zeta} = 2^{17}$ in the SZ+CMBlens and SZ+CMBlens stacked cases, and $n_z = 100$ in all the cases. In the SZ+CMBlens and SZ+CMBlens stacked cases, the backward convolutional approach is followed to compute the individual cluster likelihoods, for which we set $n_{\mathrm{eval}} = 2^{11}$ and $n_{\mathrm{eval},z} = 1$. Since there is no correlation between the scatter in the tSZ and CMB lensing signal-to-noise observables, we consider them to be part of two different correlation sets. In order to obtain the evaluation mass grid $\mathbfit{M}_{\mathrm{eval}}$, we set $c_M = 10$. We compute the derivatives of the scaling relations of the selection observable numerically.

In the SZ binned case, we consider binning across both the selection observable and redshift. Across redshift, we consider 9 equally-spaced bins from $z=0.01$ to $z=3$, whereas across the selection observable we consider 9 logarithmically-spaced bins from $q_{\mathrm{obs}} = 5$ to $q_{\mathrm{obs}} = 200$.

\subsection{Consistency in data space}\label{susbec:agremeent}

We first assess the agreement between our synthetic catalogues and \cnc{}'s predictions in data space, considering the number counts as a function of tSZ signal-to-noise and redshift (Section\,\ref{subsec:nc}), the stacked CMB lensing observable (Section\,\ref{subsec:stobs}), and the most extreme cluster (Section\,\ref{subsec:evv}). In this section, all of the \cnc{}'s calculations are carried out at the true input parameter values. 

\subsubsection{Number counts}\label{subsec:nc}

\begin{figure*}
\centering
\includegraphics[width=0.8\textwidth,trim={00mm 0mm 0mm 0mm},clip]{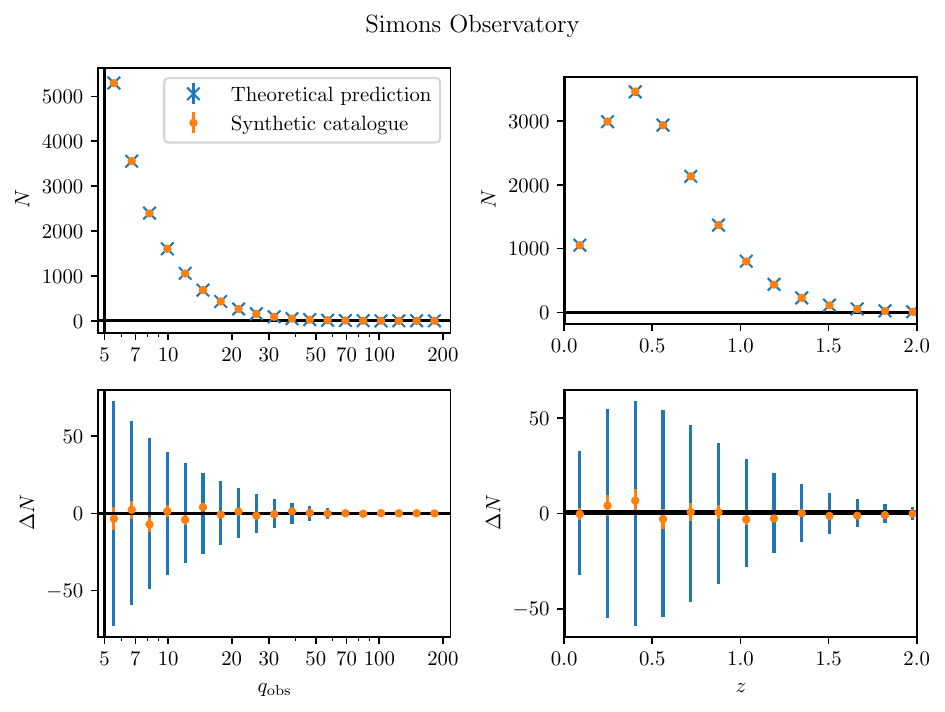}
\caption{\textit{Top panels}: Mean number counts across signal-to-noise and redshift (left and right panels, respectively) across our 100 synthetic SO-like catalogues (blue data points), shown along \cnc{}'s theoretical prediction (orange points). \textit{Bottom panels}: Difference between the mean number counts for our synthetic catalogues and \cnc{}'s theoretical prediction (orange data points), shown along the Poisson errors for one catalogue (blue error bars). No evidence for any disagreement between the synthetic catalogues and the theoretical prediction can be seen, with any biases in the number counts being constrained to be significantly smaller than the Poisson errors.}
\label{fig:number_counts}
\end{figure*}

The top panels of Figure\,\ref{fig:number_counts} show the mean number counts across our 100 synthetic catalogues as a function of tSZ signal-to-noise and redshift, binned in 19 and 13 bins across signal-to-noise and redshift, respectively (orange data points in the left and right panels, respectively). \cnc{}'s theoretical prediction is shown in blue. The agreement between the synthetic catalogues and the theoretical prediction is excellent. This can be seen more clearly in the bottom panels, which depict the difference between mean number counts in the synthetic catalogues and the theoretical prediction (orange data points). For each bin, the Poisson error for one catalogue is shown as the blue error bar. No evidence for a bias in the theoretical calculation can be seen, with the bias being constrained to be significantly smaller than the Poisson error for every bin.

We find a predicted mean total number of clusters of 15634.38 objects and an empirical mean across the 100 synthetic catalogues of $15626.99 \pm 12.50$ objects, both numbers being fully consistent and constituting a 0.047\,\% agreement.

\subsubsection{Stacked observable}\label{subsec:stobs}


As a test in data space of the backward convolutional approach and of the stacked likelihood, we calculate the mean of the stacked CMB lensing observable $p_{\mathrm{stacked}}$ for each of our synthetic catalogues and compare it to its respective observed value. This can be easily done using \cnc{}'s stacked likelihood machinery. We find a mean stacked CMB lensing signal-to-noise across all the synthetic catalogues of $ \langle p_{\mathrm{stacked}}\rangle = 0.30859 \pm 0.00078$, where angular brackets denote averaging over all the catalogues, and where the standard deviation is obtained empirically as the sample standard deviation. The corresponding theoretical prediction is $\langle \bar{p}_{\mathrm{stacked}} \rangle = 0.307442 \pm 0.000080$, where here we have computed the standard deviation of the stacked observable following its hierarchical model. The empirical and theoretical values are in excellent agreement. We also find the mean of the stacked observable residuals to be $\langle p_{\mathrm{stacked}} - \bar{p}_{\mathrm{stacked}}\rangle =  0.00114 \pm 0.00079$, with the standard deviation being empirically obtained, there being no evidence for a bias and constituting a $0.26$\,\% agreement. 

\subsubsection{Most extreme cluster}\label{subsec:evv}

\begin{figure}
\centering
\includegraphics[width=0.42\textwidth,trim={00mm 0mm 0mm 0mm},clip]{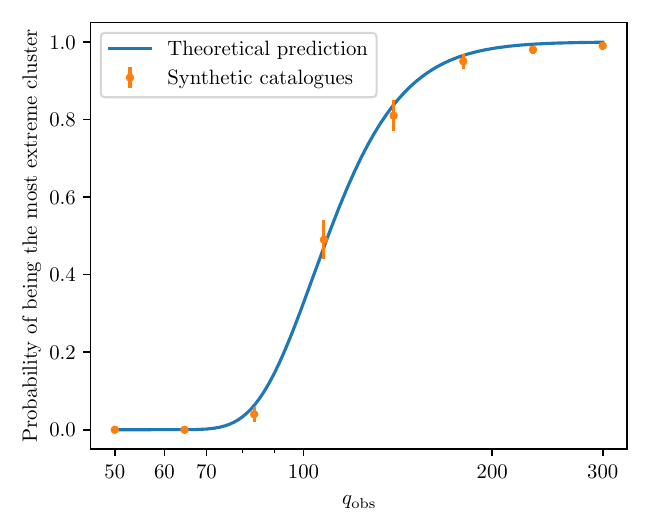}
\caption{Probability for being the cluster with the largest value of $q_{\mathrm{obs}}$ in the catalogue for our reference SO-like survey, as computed with \cnc{} (blue curve), shown along the corresponding estimate from our synthetic catalogues (orange data points). There is excellent agreement between the synthetic catalogues and the theoretical prediction.}
\label{fig:extreme}
\end{figure}

Figure\,\ref{fig:extreme} shows, in blue, the probability for $q_{\mathrm{obs}}$ to be the largest value of the selection observable in the sample (the `most extreme' cluster), as computed with \cnc{}'s extreme value likelihood machinery. In addition, the orange data points show the fraction of the synthetic catalogues for which this is true, with the error bars obtained with bootstrapping. There is complete agreement between the theoretical prediction and the synthetic catalogues.

Furthermore, we find the empirical mean of the largest value of the tSZ signal-to-noise across the 100 synthetic catalogues to be $\langle \zeta_{\mathrm{max}} \rangle = 117.8 \pm 3.6$, where angular brackets denote averaging over the catalogues. Its predicted value, as computed with \cnc{}, is $\bar{\zeta}_{\mathrm{max}} = 115.61$, there being full agreement with the empirical value, to $3.1$\,\%. Together with Figure\,\ref{fig:extreme}, this result demonstrates that \cnc{} can successfully compute the extremes of the cluster distribution.

\subsection{Parameter constraints: biases and goodness of fit}\label{subsec:biases}

\subsubsection{Parameters and priors}

We derive parameter constraints from our synthetic catalogues using the CosmoMC MCMC sampler \citep{Lewis2002,Lewis2013} as implemented in \texttt{Cobaya} \citep{Torrado2019,Torrado2021}. In every analysis we vary the following cosmological parameters: $\Omega_{\mathrm{m}}$, $\Omega_{\mathrm{b}}$, $h$, $\sigma_8$, and $n_{\mathrm{s}}$. We impose \textit{Planck}-derived Gaussian priors \citep{Planck2018I} on parameters that are poorly constrained by the cluster number counts, namely on $\Omega_{\mathrm{b}} h^2$, setting $\Omega_{\mathrm{b}} h^2 =  0.02224 \pm 0.00015$, and on $n_{\mathrm{s}}$, setting $n_{\mathrm{s}} = 0.96 \pm 0.0042$ (mean and standard deviation in both cases). In addition, since the Hubble constant is also poorly constrained by cluster data, we impose a further prior on the CMB acoustic scale parameter, $\theta_{\mathrm{MC}}$ \citep{Kosowsky2002}, which at fixed baryon density depends only on $\Omega_{\mathrm{m}}$ and $h$, effectively fixing $h$ for given values of $\Omega_{\mathrm{m}}$ and $\Omega_{\mathrm{b}}$ (see also \citealt{Zubeldia2019}, where the same prior was imposed). We choose a Gaussian prior with \textit{Planck}'s value, $100 \theta_{\mathrm{MC}} = 1.04093 \pm 0.00030$ (mean and standard deviation; \citealt{Planck2016xiii}), computing $\theta_{\mathrm{MC}}$ within \cnc{}.

In addition, we vary the following scaling relation parameters: the tSZ signal-to-noise amplitude parameter $A_{\mathrm{SZ}}$, the tSZ mass slope parameter $\alpha_{\mathrm{SZ}}$, the tSZ intrinsic scatter $\sigma_{\mathrm{SZ}}$, and, in the two sets of analyses including CMB lensing measurements, the CMB lensing mass bias, $\beta_{\mathrm{CMBlens}}$, and the CMB lensing intrinsic scatter, $\sigma_{\mathrm{CMBlens}}$. We impose Gaussian priors on the two scatter parameters, namely $\sigma_{\mathrm{SZ}} = 0.173 \pm 0.05$ and $\sigma_{\mathrm{CMBlens}} = 0.22 \pm 0.05$ (mean and standard deviation), motivated, respectively, by the \textit{Planck} SZ counts analysis \citep{Ade2016} and by the analysis of CMB lensing simulations \citep{Zubeldia2020}. In addition, in the SZ unbinned and binned analyses, we further impose priors on the scaling relation parameters $A_{\mathrm{SZ}}$ and $\alpha_{\mathrm{SZ}}$, setting $A_{\mathrm{SZ}} = -4.3054 \pm 0.033$ and $\alpha_{\mathrm{SZ}} = 1.233 \pm 0.014$ (mean and standard deviation). The standard deviations of these priors are chosen to be identical to those of the corresponding posteriors in the SZ+CMBlens case, in which these two parameters are constrained by the CMB lensing data. On the other hand, in the SZ+CMBlens and SZ+CMBlens stacked analyses, we impose a Gaussian prior on $\beta_{\mathrm{CMBlens}}$, setting $\beta_{\mathrm{CMBlens}} = 0.092 \pm 0.02$ (mean and standard deviation), effectively setting a $\sim\,2\,\%$ systematic floor on the calibration of the CMB lensing mass observable, which we believe to be a reasonable assumption for SO.

Taking all the priors into account, in the SZ unbinned and binned cases, the parameters that are left free, with broad, uniform priors, are $\Omega_{\mathrm{m}}$ and $\sigma_8$. In the SZ+CMBlens and SZ+CMBlens cases, these are $\Omega_{\mathrm{m}}$, $\sigma_8$, $A_{\mathrm{SZ}}$, and $\alpha_{\mathrm{SZ}}$.

\subsubsection{Parameter constraints and biases}\label{subsec:constraints}

We obtain parameter posteriors for each MCMC analysis and compute the mean and standard deviation of each parameter, which, for parameter $p$, we denote with $\hat{p}$ and $\sigma_{p}$, respectively. We then define the bias on parameter $p$ as $b_p \equiv \langle \hat{p} \rangle - p_{\mathrm{true}} $, where $ p_{\mathrm{true}}$ is its true, input value, and where angular brackets denote ensemble averaging over data realisations, i.e., over our catalogues. Table\,\ref{table:significance} shows the parameter biases that we obtain for the four sets of analyses. These are shown in `1\,$\sigma$' units, i.e., dividing each $b_p$ estimate by the standard deviation of parameter $p$, $\sigma_p$, averaged over all the catalogues that are analysed, and also as a percentage, i.e., $100  b_p/p_{\mathrm{true}}$. The errors on the parameter biases are obtained by bootstrapping over the catalogues for each set of analyses.

\begin{table*}
\centering
\begin{tabular}{lllll}
\thickhline

\textbf{Parameter} & \textbf{SZ unbinned} & \textbf{SZ binned} & \textbf{SZ+CMBlens} & \textbf{SZ+CMBlens stacked}
\\
\thickhline
\multicolumn{5}{c}{\textit{Parameter biases (in units of 1\,$\sigma$)}} \\

\\

$\sigma_8$ &  $-0.023 \pm 0.082$  &  $-0.131 \pm 0.081$  &  $0.15 \pm 0.19$  &  $0.08 \pm 0.19$  \\ $\Omega_{\mathrm{m}}$ &  $0.146 \pm 0.049$  &  $0.110 \pm 0.045$  &  $0.08 \pm 0.15$  &  $0.03 \pm 0.14$  \\ $\Omega_{\mathrm{b}}$ &  $0.136 \pm 0.046$  &  $0.096 \pm 0.043$  &  $0.07 \pm 0.14$  &  $0.02 \pm 0.14$  \\ $H_0$ &  $-0.126 \pm 0.048$  &  $-0.092 \pm 0.044$  &  $-0.05 \pm 0.15$  &  $-0.00 \pm 0.14$  \\ $n_{\mathrm{s}}$ &  $-0.03 \pm 0.01$  &  $-0.010 \pm 0.012$  &  $-0.01 \pm 0.01$  &  $-0.0049 \pm 0.0086$  \\ $A_{\mathrm{SZ}}$ &  $-0.113 \pm 0.039$  &  $-0.039 \pm 0.032$  &  $-0.20 \pm 0.15$  &  $-0.13 \pm 0.15$  \\ $\alpha_{\mathrm{SZ}}$ &  $-0.052 \pm 0.078$  &  $-0.092 \pm 0.071$  &  $-0.31 \pm 0.17$  &  $-0.32 \pm 0.17$  \\ $\sigma_{\mathrm{SZ}}$ &  $0.062 \pm 0.048$  &  $0.067 \pm 0.049$  &  $0.194 \pm 0.084$  &  $0.212 \pm 0.068$  \\ $1-b_{\mathrm{CMBlens}}$ &  &  &  $0.114 \pm 0.038$  &  $0.106 \pm 0.032$  \\ $\sigma_{\mathrm{CMBlens}}$ &  &  &  $-0.064 \pm 0.067$  &  $0.064 \pm 0.021$  \\  \hline \multicolumn{5}{c}{\textit{Parameter biases (in $\%$)}} \\ \\ $\sigma_8$ &  $-0.016 \pm 0.057$  &  $-0.10 \pm 0.06$  &  $0.09 \pm 0.11$  &  $0.05 \pm 0.12$  \\ $\Omega_{\mathrm{m}}$ &  $0.36 \pm 0.12$  &  $0.30 \pm 0.12$  &  $0.22 \pm 0.39$  &  $0.1 \pm 0.4$  \\ $\Omega_{\mathrm{b}}$ &  $0.21 \pm 0.07$  &  $0.164 \pm 0.073$  &  $0.12 \pm 0.23$  &  $0.03 \pm 0.24$  \\ $H_0$ &  $-0.092 \pm 0.036$  &  $-0.076 \pm 0.037$  &  $-0.04 \pm 0.12$  &  $-0.00 \pm 0.12$  \\ $n_{\mathrm{s}}$ &  $-0.0128 \pm 0.0044$  &  $-0.0045 \pm 0.0052$  &  $-0.0059 \pm 0.0044$  &  $-0.0021 \pm 0.0037$  \\ $A_{\mathrm{SZ}}$ &  $0.072 \pm 0.025$  &  $0.028 \pm 0.022$  &  $0.1 \pm 0.1$  &  $0.09 \pm 0.11$  \\ $\alpha_{\mathrm{SZ}}$ &  $-0.045 \pm 0.068$  &  $-0.083 \pm 0.064$  &  $-0.35 \pm 0.19$  &  $-0.37 \pm 0.19$  \\ $\sigma_{\mathrm{SZ}}$ &  $1.5 \pm 1.2$  &  $1.6 \pm 1.2$  &  $4.9 \pm 2.1$  &  $5.5 \pm 1.8$  \\ $1-b_{\mathrm{CMBlens}}$ &  &  &  $0.24 \pm 0.08$  &  $0.226 \pm 0.067$  \\ $\sigma_{\mathrm{CMBlens}}$ &  &  &  $-1.4 \pm 1.4$  &  $1.44 \pm 0.47$  \\ 

\hline
\multicolumn{5}{c}{\textit{Goodness of fit}} \\ \\

$\langle C \rangle$ & $15.68 \pm 0.82$ & $17.3 \pm 1.4$ & $18.5 \pm 1.2$ & $18.5 \pm 1.2 $\\
$\langle \bar{C} \rangle$ & $16.94 \pm 0.14 $ & $16.95 \pm 0.14$ & $16.61 \pm 0.15$ & $16.61 \pm 0.15$ \\
$\langle C - \bar{C} \rangle / \langle \sigma_C \rangle $ & $-0.22 \pm 0.14$ & $0.07 \pm 0.25$ & $0.33 \pm 0.21$ & $0.33 \pm 0.21$ \\
\thickhline
\end{tabular}
\caption{Biases on the parameters constrained in the analyses of our synthetic SO-like catalogues for the four cases considered, shown both in units of $1\,\sigma$ and as a percentage. The errors on all the parameters correspond to their standard deviation across our analysis sets, estimated with bootstrapping. In addition, we show the value of the modified Cash goodness-of-fit statistic $C$, ensemble-averaged over each set of analyses, as well as the ensemble average of its expected value $\bar{C}$ and of its residuals $C-\bar{C}$, the latter in units of $1\,\sigma$.}
\label{table:significance}
\end{table*}

\begin{figure*}
\centering
\includegraphics[width=0.9\textwidth,trim={00mm 0mm 0mm 0mm},clip]{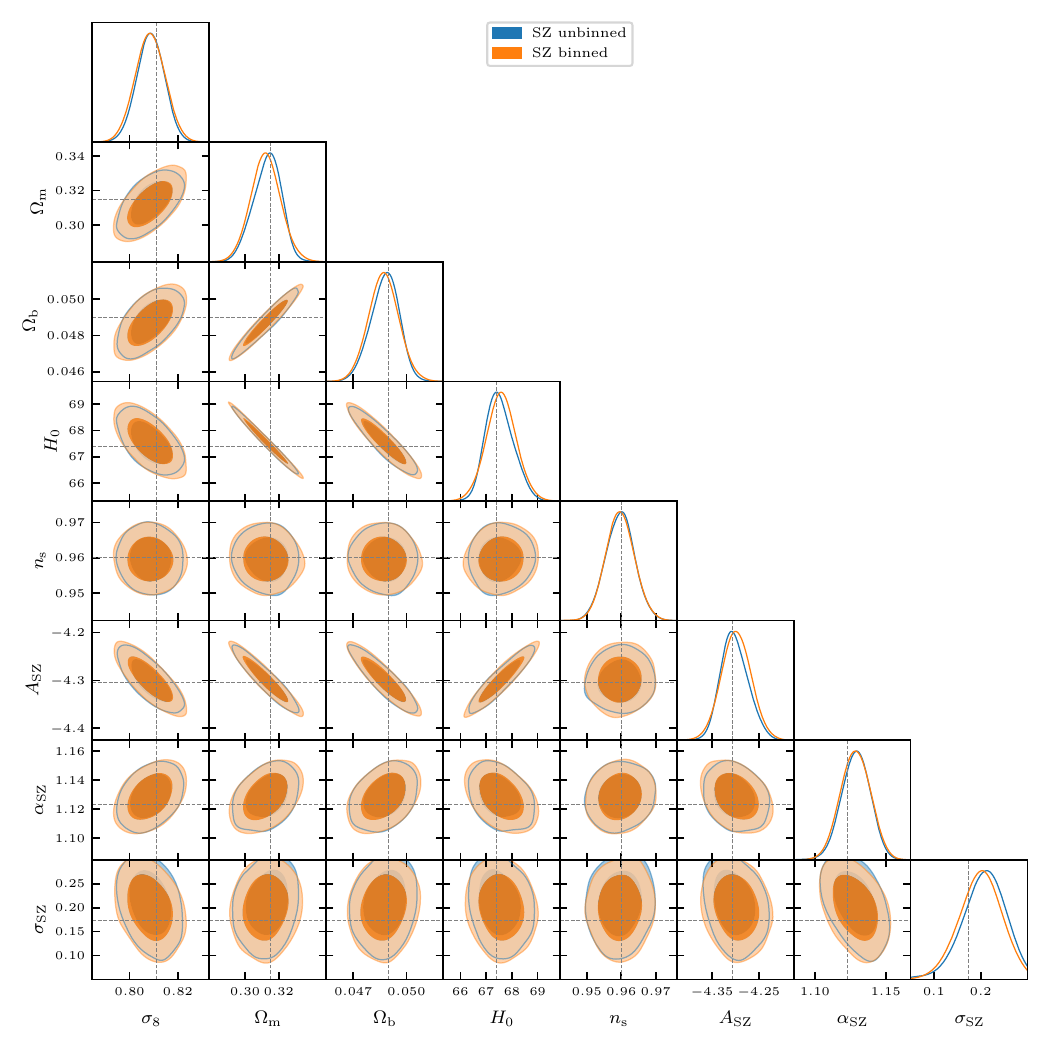}
\caption{Two-dimensional marginalised parameter constraints for our SZ unbinned and binned analyses for one of our synthetic SO-like catalogues, with the true parameter values shown as the dashed lines.}
\label{fig:mcmc_sz}
\end{figure*}

\begin{figure*}
\centering
\includegraphics[width=0.9\textwidth,trim={00mm 0mm 0mm 0mm},clip]{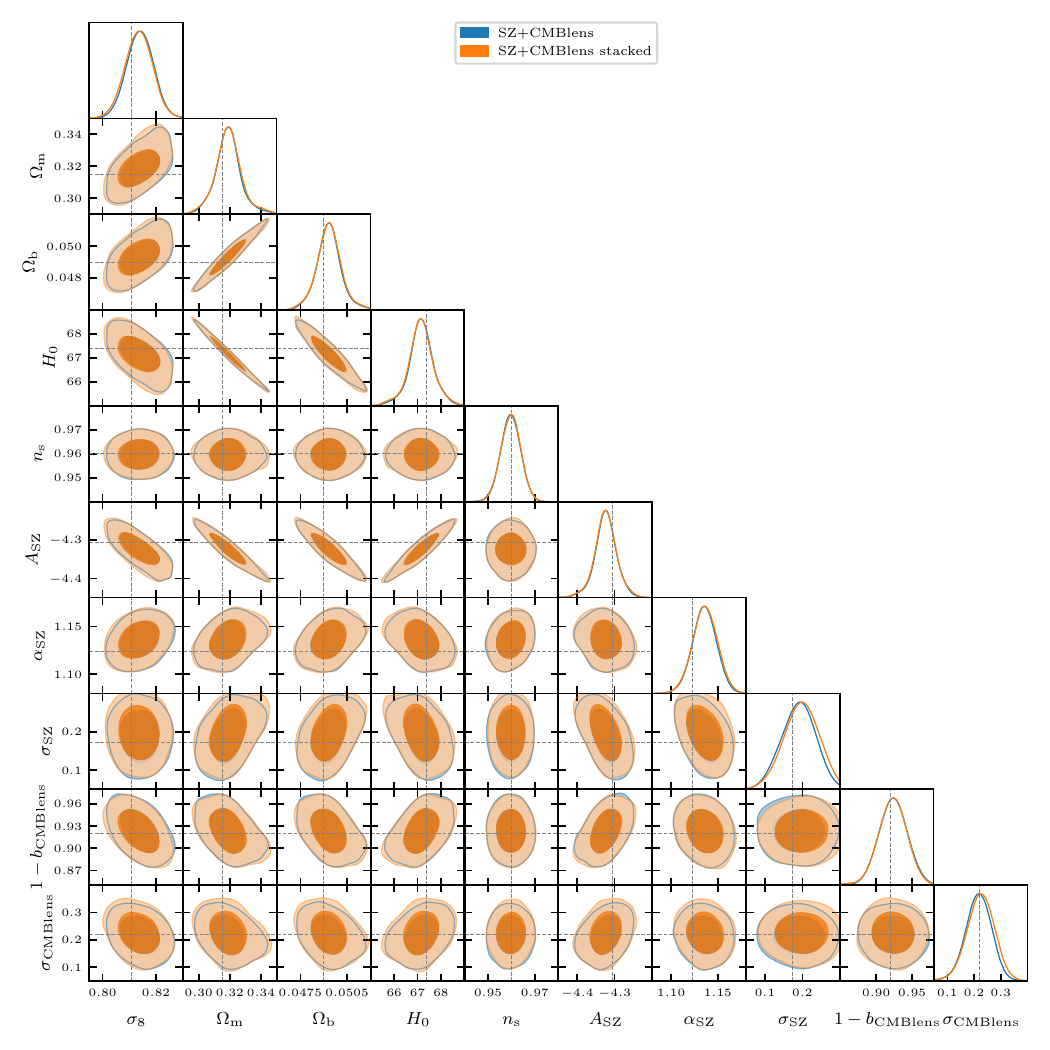}
\caption{As Figure\,\ref{fig:mcmc_sz}, but for our SZ+CMBlens and SZ+CMBlens stacked analyses of the same synthetic cluster catalogue.}
\label{fig:mcmc_szlens}
\end{figure*}

As it can be seen in Table\,\ref{table:significance}, the biases in all the parameters are constrained to be a small fraction of one standard deviation for all sets of analyses. The parameter featuring the largest bias is $\alpha_{\mathrm{SZ}}$ for the SZ+CMBlens and SZ+CMBlens cases, for which the bias is constrained to be $b_{\alpha_{SZ}} = -0.31 \pm 0.17$ and $b_{\alpha_{SZ}} = -0.32 \pm 0.17$, respectively, in $1\,\sigma$ units. All the cosmological parameters, on the other hand, have biases of $0.15\,\sigma$ or less, and of less than $1\,\%$, with no bias being detected in most cases.


For illustration, Figure\,\ref{fig:mcmc_sz} shows the two-dimensional marginalised parameter constraints on all the model parameters for the SZ unbinned and SZ binned analyses of one of our catalogues, with the true parameter values shown as the dashed lines. As it can be seen, the constraints obtained with the unbinned and the binned likelihoods are almost identical. Figure\,\ref{fig:mcmc_szlens} is an analogous plot for our SZ+CMBlens and SZ+CMBlens stacked analyses of the same catalogue, with the constraints between them being also virtually identical. This indicates that the $\sim 2\,\%$ systematic uncertainty in the calibration of the CMB lensing signal-to-noise, as accounted for by the CMB lensing bias parameter $\beta_{\mathrm{CMBlens}}$, contributes significantly to the overall mass calibration uncertainty in the analysis. Indeed, stacking the CMB lensing signal-to-noise measurements entails a certain signal-to-noise loss relative to the cluster-by-cluster case (see Appendix\,\ref{appendix:b}), which here gets washed away.

There are some parameter degeneracies that can be observed in Figures\,\ref{fig:mcmc_sz} and\,\ref{fig:mcmc_szlens} that are worth commenting on. There are strong degeneracies between $H_0$ and $\Omega_{\mathrm{m}}$ and between $H_0$ and $\Omega_{\mathrm{b}}$. These are caused by the priors on $\theta_{\mathrm{MC}}$ and $\Omega_{\mathrm{b}} h^2$, respectively. More interesting, regarding the cluster counts, are the negative degeneracies between $A_{\mathrm{SZ}}$ and both $\sigma_8$ and $\Omega_{\mathrm{m}}$. They indicate that a stronger tSZ signal per cluster (which, in turn, increases the number of clusters in the catalogue) can be partially compensated by lower values of $\sigma_8$ or $\Omega_{\mathrm{m}}$. The degeneracy between $A_{\mathrm{SZ}}$ and $\Omega_{\mathrm{m}}$ also translates into degeneracies between $A_{\mathrm{SZ}}$ and $H_0$ (through the prior on $\theta_{\mathrm{MC}}$) and, in turn, between $A_{\mathrm{SZ}}$ and $\Omega_{\mathrm{b}}$ (through the prior on $\Omega_{\mathrm{b}} h^2$). 

\begin{figure}
\centering
\includegraphics[width=0.4\textwidth,trim={00mm 0mm 0mm 0mm},clip]{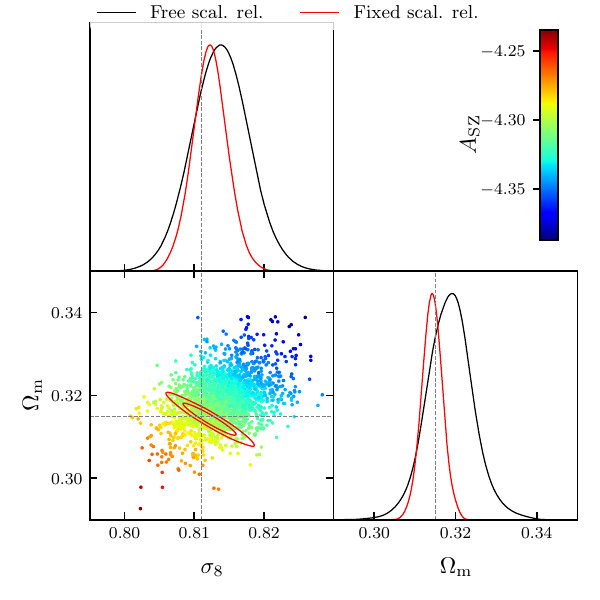}
\caption{Parameter constraints on the $\Omega_{\mathrm{m}}$--$\sigma_8$ plane for the SZ+CMBlens analysis for one of our synthetic SO-like catalogues, both if all parameters, including $A_{\mathrm{SZ}}$, are allowed to vary (as in Figure\,\ref{fig:mcmc_szlens}; coloured points), and if only the cosmological parameters are varied, the scaling relation parameters being fixed to their true values (red contours). The true values of $\Omega_{\mathrm{m}}$ and $\sigma_8$ as shown as the dashed lines. This plot clearly illustrates the impact of the tSZ amplitude parameter $A_{\mathrm{SZ}}$ on the degeneracy between $\Omega_{\mathrm{m}}$ and $\sigma_8$ (see the discussion in Section\,\ref{subsec:constraints}).}
\label{fig:mcmc_onlycosmo}
\end{figure}

The most interesting degeneracy regarding the cluster counts, however, is that between $\Omega_{\mathrm{m}}$ and $\sigma_8$, as these are the cosmological parameters that are most tightly constrained by them within the $\Lambda$CDM model. $\Omega_{\mathrm{m}}$ and $\sigma_8$ are found to be positively correlated in all four analyses. This is a result of the strong negative correlation between $A_{\mathrm{SZ}}$ and both $\Omega_{\mathrm{m}}$ and $\sigma_8$, which causes both $\Omega_{\mathrm{m}}$ and $\sigma_8$ to be positively correlated between them. This effect can be seen more clearly in Figure\,\ref{fig:mcmc_onlycosmo}, which shows the marginalised constraints on the $\Omega_{\mathrm{m}}$--$\sigma_8$ plane for the SZ+CMBlens case for the same catalogue as shown in Figures\,\ref{fig:mcmc_sz} and \ref{fig:mcmc_szlens}, showing also the values taken by $A_{\mathrm{SZ}}$ (coloured scatter map). It can be clearly seen that the value of $A_{\mathrm{SZ}}$ runs along the long degeneracy axis, causing $\Omega_{\mathrm{m}}$ and $\sigma_8$ to be positively correlated. For comparison, the analogous constraints obtained by varying only the cosmological parameters, setting all the scaling relation parameters to their true values, are also shown (red contours). In this case, $\Omega_{\mathrm{m}}$ and $\sigma_8$ are negatively correlated, as would be expected, with this correlation being reverted as $A_{\mathrm{SZ}}$ is allowed to vary. We note that this effect has already been observed in the analysis of real data sets, e.g., in that of the \textit{Planck} MMF3 cosmology sample (see, in particular Figure\,7 of \citealt{Ade2016}, where the impact of the prior on the `hydrostatic mass bias' parameter, $1-b$, which controls the amplitude of the cluster tSZ signal, on the constraints on the $\Omega_{\mathrm{m}}$--$\sigma_8$ plane is similar to what we observe here for $A_{\mathrm{SZ}}$). We also note that, as Figure\,\ref{fig:mcmc_onlycosmo} clearly illustrates, the degeneracy direction in the $\Omega_{\mathrm{m}}$--$\sigma_8$ plane will strongly depend on how tightly $A_{\mathrm{SZ}}$ can be constrained, which, in turn, depends on the precision of the data used for mass calibration. In the particular context of SO, at low redshifts weak lensing observations from \textit{Euclid} and Rubin/LSST are expected to provide a much higher signal-to-noise than SO CMB lensing measurements for the same clusters (e.g., \citealt{Giocoli2024}). We leave the exploration of these synergies to further work.

\subsubsection{Goodness of fit}

We assess the goodness of fit of our derived parameter constraints using the modified Cash statistic $C$, as implemented in \cnc{} (see Section\,\ref{subsec:goodness}). In particular, for each catalogue in each set of analyses, we evaluate $C$ at the derived parameter means. We also evaluate its theoretically-predicted expected value, $\bar{C}$, and its standard deviation, $\sigma_C$. We do this using the same bins across tSZ signal-to-noise and redshift that were used in the binned likelihood analyses. Then, for each set of analyses, we compute the means of $C$, $\bar{C}$, and $\sigma_C$ across the corresponding catalogues, as well as that of the statistic residuals, $C-\bar{C}$. The means of $C$, $\bar{C}$, and of the residuals $C-\bar{C}$ are shown in Table\,\ref{table:significance}, the latter in units of $\sigma_C$, with the quoted uncertainties being empirically obtained. For all four sets of analyses, the means of $C$ and $\bar{C}$ are found to be in good agreement with each other, and those of the residuals are found to be consistent with zero to a fraction of one standard deviation, indicating a good fit to the synthetic data.

\subsection{Cluster masses}\label{subsec:masses}

\begin{figure}
\centering
\includegraphics[width=0.45\textwidth,trim={00mm 0mm 0mm 0mm},clip]{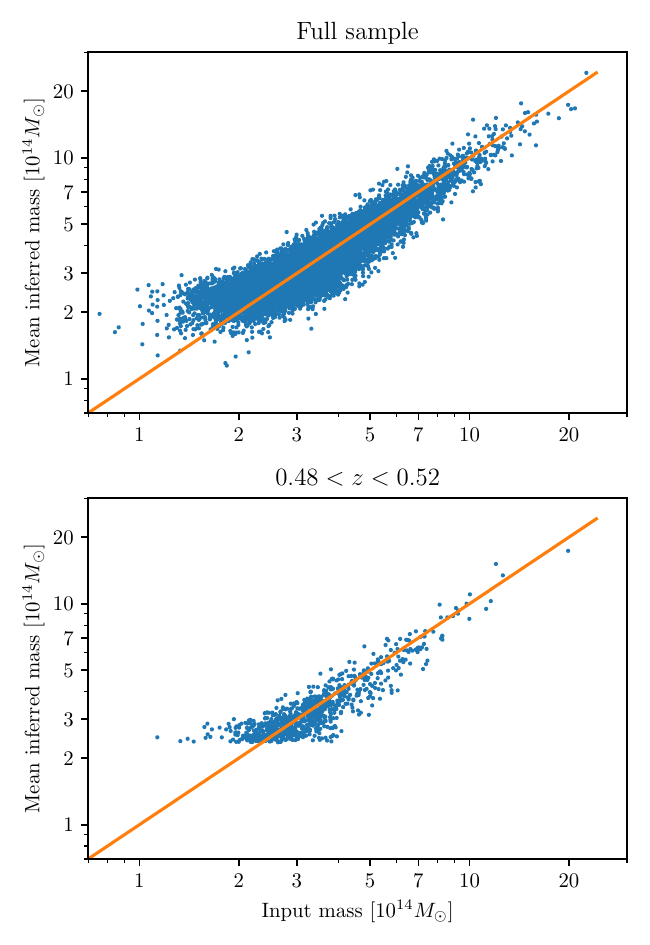}
\caption{Cluster masses inferred by \cnc{} with the SZ+CMBlens likelihood for one of our synthetic SO-like catalogues, plotted against the true cluster masses. The top panel shows the full cluster sample, whereas the bottom panel shows the clusters within a small redshift range, illustrating the impact of Malmquist bias (see Section\,\ref{subsec:masses}).}
\label{fig:masses}
\end{figure}

The top panel of Figure\,\ref{fig:masses} shows the inferred cluster masses obtained with the SZ+CMBlens likelihood for one of our synthetic SO catalogues at the true parameter values, plotted against their corresponding true values, illustrating \cnc{}'s cluster mass inference capabilities. In particular, each inferred mass corresponds to the mean of the mass posterior for each cluster (see Section\,\ref{subsec:masstheory}). 

As expected, the inferred masses suffer from Malmquist bias (see, e.g., \citealt{Mantz2010b,Sereno2016}), which is responsible for the shape of the distribution in the inferred mass--true mass plane. The effect of Malmquist bias can be seen more clearly in the bottom panel of Figure\,\ref{fig:masses}, which shows the clusters in the catalogue lying within a small redshift range, $0.48 < z < 0.52$. In our synthetic catalogues, at fixed redshift there is a one-to-one mapping between the data for each cluster and its mean inferred mass. Therefore, within a small redshift bin, a sample that is selected by thresholding on one of the mass observables (which, incidentally, provides most of the signal-to-noise) translates into a sample that is approximately selected on the mean inferred mass, as is apparent in the bottom panel of Figure\,\ref{fig:masses}. The distribution for the full cluster sample (top panel) can then be understood as the superposition of the different distributions across redshift. As a result of this very significant Malmquist bias, care should be taken when making use of the cluster masses inferred with \cnc{}. Note that, as explained in Section\,\ref{subsec:masstheory}, the mass estimates are, however, Eddington-bias corrected.

We also note that although here we inferred the cluster masses for a given point in parameter space, these can be obtained at every point explored in an MCMC analysis at almost no additional computational cost. Doing this delivers cluster mass estimates for which the uncertainty due to the model parameters is properly marginalised over.

\subsection{Computational efficiency}\label{subsec:efficiency}

In this section we provide a brief account of \cnc{}'s execution speed in several scenarios. All the execution times that we provide are averaged over 10 identical executions. Note that these are illustrative, as they can depend strongly on the machine on which the code is run.

We start with one of our synthetic SO-like catalogues, setting $n_{\mathrm{core,hmf}}=1$, $n_{\mathrm{core,abundance}}=8$, and all the other input parameters set to the values chosen to obtain the results in Section\,\ref{sec:validation}. Let us first consider only one mass observable, the tSZ signal-to-noise. In this case, evaluating the unbinned, binned and extreme value likelihoods takes 0.66\,s, 0.64\,s, and 0.65\,s respectively. As argued in Section\,\ref{sec:implementation}, the evaluation time of the unbinned likelihood barely scales with the number of clusters in the catalogue. Indeed, if we artificially double the number of clusters in the catalogue, we obtain an evaluation time of 0.66\,s. The evaluation time of the binned likelihood also scales very mildly with the number of bins. Indeed, doubling the number of bins across both selection observable and redshift leads to an evaluation time of 0.70\,s.

Let us now consider two mass observables, the tSZ signal-to-noise and the CMB lensing signal-to-noise, setting $n_{\mathrm{core,data}}=16$. In this case, the only likelihood that can be evaluated is the unbinned likelihood, which is computed with the backward convolutional approach. Assuming that the two mass observables belong to two different correlation sets leads to an evaluation time of 5.74\,s. Note that, in this case, the evaluation time scales approximately linearly with the number of clusters. If the mass observables are taken to belong to the same correlation set, the execution time is 8.79\,s for $n_{\mathrm{eval}} = 128$. Note that this value of $n_{\mathrm{eval}}$ is significantly smaller than that used in Section\,\ref{sec:validation}, for which the computation becomes too expensive. This evidences the power of splitting the observables into correlation sets, as carried out in the backward convolutional approach. The minimum value of $n_{\mathrm{eval}}$ for which the likelihood computation is accurate enough will depend significantly on the catalogue to be analysed; we leave a detailed study of this issue to further work. 

For comparison, for $n_{\mathrm{eval}} = 128$ and for the two mass observables, brute-force computation of the individual cluster likelihoods leads to an evaluation time of 181.14\,s. 

Finally, in the case in which the number count likelihood for the tSZ signal-to-noise is combined with the stacked CMB lensing observable, we find an evaluation time of 6.91\,s.

Let us now consider the likelihood for the real \textit{Planck} data considered in Zubeldia \& Bolliet (in prep.). We find that the \textit{Planck} binned likelihood across tSZ signal-to-noise and redshift, which is identical to that used in the official \textit{Planck} analysis \citep{Ade2016}, takes 1.54\,s to evaluate. On the other hand, the \textit{Planck} unbinned likelihood also taking CMB lensing data (the CMB lensing signal-to-noise, which constitutes one single correlation set together with the tSZ signal-to-noise), which is identical to that used in \citet{Zubeldia2019}, has an evaluation time of 2.21\,s. 

\subsection{Benchmarking against \texttt{class\_sz}}\label{subsec:classsz}

As an additional validation step, we benchmark the \cnc{} cluster abundance calculations against \texttt{class\_sz} for our SO-like set-up. 

\texttt{class\_sz} is code in C and Python based on the Boltzmann solver \texttt{class} \citep{Lesgourgues2011,Blas2011}. It allows to calculate a wide range of CMB and LSS observables (beyond the matter and CMB power spectra computed by \texttt{class}) in a fast and accurate way, making use of the \texttt{cosmopower} neural network emulators. We refer to \cite{Bolliet2023b} and to the code repository (see footnote\,\ref{footnote}) for further details about the code.

For a given tSZ observational set-up, \texttt{class\_sz} can calculate the binned cluster abundance across signal-to-noise and redshift in a brute-force way. From this binned abundance, \texttt{class\_sz} can evaluate the number count binned likelihood. This was demonstrated in \citet{Bolliet2019}, where a cosmological analysis of the \textit{Planck} cluster sample was carried out, improving upon \cite{Ade2016} by consistently treating massive neutrinos in the halo mass function. In \citet{Bolliet2019}, the binned abundance calculation followed the same formalism and implementation as that of \cite{Ade2016}. We refer to \cite{Ade2016} for details about the model. We stress, in particular, that the binned abundance in both \citet{Bolliet2019} and \cite{Ade2016} is evaluated in a very different way from \cnc{}'s approach. Indeed, in \cnc{} it is computed from the unbinned abundance, which is, in turn, obtained following with the forward convolutional approach.

\texttt{class\_sz} can also compute the cluster abundance across the selection observable and redshift following the same forward convolutional approach (see Eq.\,\ref{eq:convo}), as \cnc{} does. Although similar to \cnc{}'s (see Section\,\ref{sec:abundance}), the implementation of this approach is fully independent in terms of numerical libraries and parallelisation strategies. In particular, the \texttt{class\_sz} implementation is in C, with the convolutions being performed explicitly with the \texttt{FFTW3} library \citep{Frigo2005}.

\begin{figure*}
\centering
\includegraphics[width=0.90\textwidth,trim={00mm 0mm 0mm 0mm},clip]{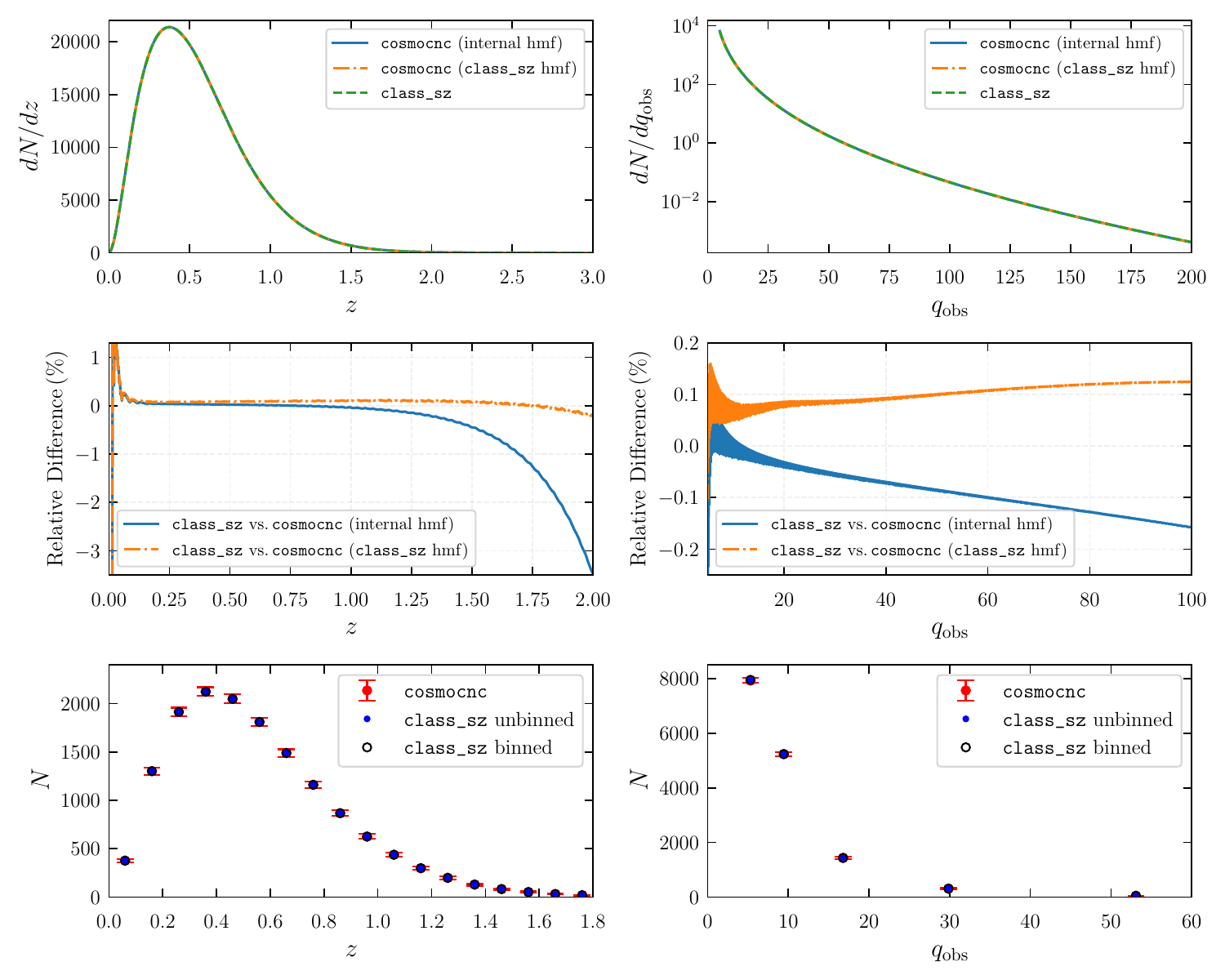}
\caption{\textit{Top panels}: SO cluster abundance as a function of redshift (left) and SZ signal-to-noise (right) computed by \cnc{} with its internal halo mass function implementation (solid blue curves), by \cnc{} with the \texttt{class\_sz} halo mass function implementation (dotted-dashed orange curves), and by \texttt{class\_sz} with the forward convolutional approach (see Section \ref{sec:abundance}). \textit{Middle panels}: Relative difference between the calculations in the top panels in percentage. \textit{Bottom panels}: Binned number counts computed by integrating the abundance from the forward convolutional approach (top panels), both by \cnc{} (red points with Poisson error bars) and \texttt{class\_sz} (plain blue points), and also computed by \texttt{class\_sz}'s brute-force implementation, following \citealt{Ade2016} (black circles). For further details on the settings (e.g., cosmological and scaling relation parameters, precision settings), we refer to the online notebook (see footnote \ref{fn:class_szbm}).
}
\label{fig:class_sz_benchmark}
\end{figure*}

In Figure\,\ref{fig:class_sz_benchmark} we compare the outputs of \cnc{} and \texttt{class\_sz}. In particular, the top panels show the cluster abundance as a function of redshift (left) and signal-to-noise $q_\mathrm{obs}$ (right) computed with the forward convolutional approach by both codes. The relative differences between them are shown in the middle panels. When \cnc{} is called with the \texttt{class\_sz} halo mass function (as can be done by setting \texttt{hmf\_calc: classy\_sz}), the difference between both codes remains below 0.2\,\% for all values of $z$ and $q_\mathrm{obs}$ of interest. This is also the case when \cnc{} is called with its internal halo mass function (\texttt{hmf\_calc: cnc}), except for $z > 1.5$, where \texttt{class\_sz} predicts a slightly larger abundance than \cnc{}, with a relative difference of a few percent. Since the predicted number of SO clusters at such high redshifts is very small, we expect this level of discrepancy to have a negligible impact on the likelihood. 

The bottom panels of Figure\,\ref{fig:class_sz_benchmark} show the cluster abundance binned in $z$ (left) and $q_\mathrm{obs}$ (right) bins, as obtained by \cnc{} with its internal halo mass function (red points) and by \texttt{class\_sz}, the latter both following the forward convolutional approach (blue points, labelled `\texttt{class\_sz} unbinned') and the brute-force calculation (black circles, labelled `\texttt{class\_sz} binned'). The agreement is excellent, the differences being much smaller than the associated Poisson errors (shown in red).

Figure\,\ref{fig:class_sz_benchmark} figure can be reproduced using a notebook available online\footnote{See \href{https://github.com/inigozubeldia/cosmocnc/blob/main/tutorials/cosmocnc_so_benchmark_class_sz.ipynb}{cosmocnc\_so\_benchmark\_class\_sz.ipynb} in the \cnc{} GitHub repository.\label{fn:class_szbm}}.

Finally, we stress that \texttt{class\_sz}'s cluster abundance tools are restricted to tSZ surveys and can only deal with a single mass observable. \cnc{}, on the other hand, is much more flexible, as it can be applied to any cluster survey (X-ray, optical, or tSZ) and can deal with an arbitrary number of mass observables in a consistent way.

\section{Summary}\label{sec:conclusion}

In this work we have introduced \cnc{}, a fast, flexible, and accurate Python package for cluster number count likelihood computation. It was designed with the goal of facilitating cluster number count cosmological inference, hoping that it can be used in order to perform a cosmological analysis with most cluster catalogues with little to no modification. In it, little is hard-coded, but the analysis specifics (cluster catalogue, scaling relations) are passed at a high level, and the same core machinery is always used.

\cnc{} features three types of likelihoods: an unbinned, a binned, and an extreme value likelihood. Its unbinned likelihood, which is the main focus of the code, can deal with an arbitrary of mass observables, missing data, redshift measurement uncertainties, and the presence of unconfirmed detections, amongst other complexities in the data (see Section\,\ref{sec:mainfeatures}). \cnc{} can also take stacked measurements as input, which are modelled consistently with the cluster catalogue data, and can produce mass estimates for the clusters in the catalogue. 

In \cnc{}, the cluster abundance is evaluated following a forward convolutional approach, which has proven to be very efficient. If there are more than one mass observables, the individual cluster likelihoods within the unbinned likelihood are computed following a backward convolutional approach, an approach that has been developed in this work and of which \cnc{} constitutes the first implementation. This approach has also proven to be efficient, particularly in the case in which the set of mass observables can be split into smaller correlation sets, an instance that it takes full advantage of.

After deriving the formalism underlying \cnc{}'s likelihoods and describing in detail the implementation of the likelihoods (Sections\,\ref{sec:formalism} and \ref{sec:implementation}, respectively), we have validated the code for the particular case of the upcoming Simons Observatory, which will detect about 16\,000 clusters in its baseline configuration (Section\,\ref{sec:validation}). In particular, we have produced 100 synthetic SO-like catalogues and compared their properties (number counts as a function of selection observable and redshift, mean of the stacked lensing observable, and observable value for the most extreme cluster) with \cnc{}'s predictions, finding excellent agreement. We have then carried out cosmological parameter inference with subsets of these catalogues in four different likelihood cases (unbinned and binned likelihood with redshift and tSZ data, and unbinned likelihood also adding CMB lensing data for mass calibration, both on a cluster-by-cluster basis and stacked across the catalogue). We have obtained constraints on cosmological and scaling relation parameters featuring biases that constrained to be at most a small fraction of one standard deviation for all parameters and all analysis sets. In particular, the cosmological parameter with the largest observed bias is $\Omega_{\mathrm{m}}$ in the SZ unbinned case, for which we measure a bias of $0.146 \pm 0.049$, in 1\,$\sigma$ units, equivalent to a $0.36 \pm 0.12$\,\% bias, being therefore completely negligible. We have also found good fits to the (synthetic) data in all four analysis sets.

In addition, we have benchmarked the SO cluster abundance computed with \cnc{} against two independent implementations in \texttt{class\_sz}, finding sub-percent level agreement. We can therefore conclude that \cnc{} is, in terms of speed and accuracy, Stage-3-ready.

In an upcoming paper, Zubeldia \& Bolliet (in prep.), we will demonstrate the application of \cnc{} to real data by using it to perform cosmological analyses with the \textit{Planck} MMF3 and the SPT2500d cluster catalogues. We stress that using it in order to analyse other cluster samples with different mass observables ought to be straightforward, including, e.g., the Atacama Cosmology Telescope (ACT) cluster sample \citep{Hilton2020}, and lensing measurements from, e.g., HSC \citep{Miyatake2019}, KiDS \citep{Robertson2024}, DES \citep{Gatti2021}, and ACT \citep{Thornton2016}.

Currently, the main limitation to \cnc{}'s capabilities is the assumption that all the clusters in the catalogue are statistically independent, neglecting sample variance due to cluster clustering. While this assumption has been a good approximation for most of the past cluster samples, in particular X-ray and SZ ones (e.g., \citealt{Mantz2015,Ade2016,Bocquet2023, Ghirardini2024}), it may no longer be so for some of the cluster catalogues due to be delivered by upcoming experiments (see, e.g., \citealt{Payerne2023}), as it is already the case for optical samples (e.g., \citealt{Costanzi2019,Fumagalli2024}). Sample variance will be included in a later release of the code. In addition, in its current form, \cnc{} assumes a simple selection function consisting of a threshold on the selection observable (as well as, optionally, a minimum and a maximum redshift). We note, however, that modifying the code in order to incorporate more complicated selection functions (such as, e.g., that of \citealt{Ghirardini2024}) is expected to be a simple task.

Finally, we note that the general formalism developed in this work for cluster number counts could be easily adapted to other scenarios, such as void number counts, whose potential as a cosmological probe has recently been recognised (e.g., \citealt{Pisani2015,Contarini2022,Contarini2023, Contarini2024}). We leave the exploration of this idea to further work.

\section*{Data Availability}

The code at the centre of this work, \cnc{}, is publicly available at \href{github.com/inigozubeldia/cosmocnc}{\texttt{github.com/inigozubeldia/cosmocnc}}, including a selection of the synthetic data analysed here. The rest of the data underlying this article will be shared on reasonable request to the corresponding author.

\section*{Acknowledgements}

The authors would like to thank Anthony Challinor, Nicholas Battaglia, Sebastian Bocquet,  Will Handley, Steven Gratton, Colin Hill, and  Eunseong Lee for useful discussions, Antony Lewis for his help with \texttt{getdist}, and Yin Li for his help with \texttt{mcfit}. The authors would also like to thanks the Aspen Center for Physics, where part of this work was completed.

\'{I}Z acknowledges support from the STFC (grant numbers ST/W000977/1). BB acknowledges  support from the European Research Council (ERC) under the European Union's Horizon 2020 research and innovation programme (Grant agreement No. 851274).

This work was performed using resources provided by the Cambridge Service for Data Driven Discovery (CSD3) operated by the University of Cambridge Research Computing Service (\href{www.csd3.cam.ac.uk}{\texttt{csd3.cam.ac.uk}}), provided by Dell EMC and Intel using Tier-2 funding from the Engineering and Physical Sciences Research Council (capital grant EP/T022159/1), and DiRAC funding from the Science and Technology Facilities Council (\href{www.dirac.ac.uk}{\texttt{dirac.ac.uk}}), within the DiRAC Cosmos dp002 project.



\bibliographystyle{mnras}
\bibliography{references} 




\appendix

\section{Unconfirmed detections in the unbinned likelihood}\label{sec:appendixa}

In this appendix we derive an expression for the unbinned likelihood in the scenario in which there are unconfirmed detections in the cluster catalogue (see Section\,\ref{sec:nonval}). We then validate its implementation in \cnc{} with a synthetic SO-like catalogue containing false detections.

\subsection{Likelihood formalism}\label{subsec:validationformalism}

Consider a cluster catalogue where each object has a boolean validation variable $V$, which can be either $V=T$ if the object is a confirmed detection, or $V=F$ otherwise. The unbinned likelihood then becomes

\begin{multline}
\mathcal{L}_{\mathrm{unbinned}} = P(N_{\mathrm{tot}}, \underline{\hat{\mathbfit{n}}}, \underline{\mathbfit{D}}, \underline{V}) = \\ P ( \underline{\mathbfit{D}}, \underline{V}) | \underline{\hat{\mathbfit{n}}}, N_{\mathrm{tot}}) P ( \underline{\hat{\mathbfit{n}}} | N_{\mathrm{tot}} ) P(N_{\mathrm{tot}}),
\end{multline}
where $\underline{V}$ is a vector containing all the validation labels, and the other variables the same as in Eq.\,(\ref{eq:lik}). As noted in Section\,\ref{sec:validation}, the third term in the likelihood is a Poisson distribution for which the expected value is given by the sum of the mean total number of true and false detections, $\bar{N}_{\mathrm{tot,all}}$. Similarly, the second term can be written as

\begin{equation}
    P (\underline{\hat{\mathbfit{n}}} | N_{\mathrm{tot}}) =  N_{\mathrm{tot}}! \prod_{i=1}^{N_{\mathrm{tot}}} \frac{1}{ \bar{N}_{\mathrm{tot,all}}} \frac{d \bar{N}_{\mathrm{all}} (\hat{\mathbfit{n}}_i )}{d \Omega},
\end{equation}
where $d\bar{N}_{\mathrm{all}} / d \Omega ( \hat{\mathbfit{n}}_i)$ is the total mean number of objects in the catalogue, both true and false detections, per solid angle at sky location $\hat{\mathbfit{n}}_i$. Finally, the first term can be written as a product over the data likelihoods, $P ( \mathbfit{D}_i, V_i | \hat{\mathbfit{n}}, \mathrm{in})$, for all the objects in the catalogue. In the rest of this appendix we derive an expression for $P ( \mathbfit{D}_i, V_i | \hat{\mathbfit{n}}, \mathrm{in})$.

Let us first introduce another boolean variable, the `truthness' label $C_i$, which is $C_i=T$ if the object is indeed a true cluster, and $C_i=F$ otherwise, i.e., if it is a false detection. We can then write $P ( \mathbfit{D}_i, V_i | \hat{\mathbfit{n}},  \mathrm{in})$ as

\begin{multline}\label{eq:validated}
P ( \mathbfit{D}_i, V_i | \hat{\mathbfit{n}}_i,  \mathrm{in}) =  \\ \sum_{C_i = T,F} P( \mathbfit{D}_i | V_i, C_i, \hat{\mathbfit{n}}_i, \mathrm{in}) P (V_i | C_i, \hat{\mathbfit{n}}_i, \mathrm{in}) P(C_i | \hat{\mathbfit{n}}_i, \mathrm{in}).
\end{multline}
Here, $P( \mathbfit{D}_i | V_i, C_i, \hat{\mathbfit{n}}_i, \mathrm{in})$ is the cluster likelihood for the four different cases spanned by the boolean variables $V_i$ and $C_i$: confirmed true clusters, unconfirmed true clusters, unconfirmed false detections, and confirmed false detections. In the following, we will assume that the last case cannot take place, i.e., that all confirmed detections correspond to true clusters. On the other hand,  $P (V_i | C_i, \hat{\mathbfit{n}}_i, \mathrm{in})$ is the probability that an object in the catalogue at sky location $\hat{\mathbfit{n}}_i$ is confirmed given its status as true or false detection, and $P(C_i | \hat{\mathbfit{n}}_i,\mathrm{in})$ is the probability for an object in the catalogue at sky location $\hat{\mathbfit{n}}_i$ to be either a true or a false detection.

Using Bayes' theorem, the first term in the sum in Eq.\,(\ref{eq:validated}), $P( \mathbfit{D}_i | V_i, C_i, \mathrm{in}_i)$, can be written as
\begin{equation}\label{eq:datainv}
    P ( \mathbfit{D}_i | V_i, C_i, \hat{\mathbfit{n}},  \mathrm{in}) = \frac{   P (\mathrm{in} | \mathbfit{D}_i , V_i, C_i, \hat{\mathbfit{n}}_i ) P (\mathbfit{D}_i | V_i, C_i, \hat{\mathbfit{n}}_i) }{ P ( \mathrm{in} | V_i, C_i, \hat{\mathbfit{n}}_i )}.
\end{equation}
Note that this expression is the same as that given in Eq.\,(\ref{eq:datain}), simply with the additional $V_i$ and $C_i$ labels. As in Eq.\,(\ref{eq:datain}), $P (\mathrm{in} | \mathbfit{D}_i , V_i, C_i \hat{\mathbfit{n}}_i )$ is a step function centred at the selection observable threshold $\zeta_{\mathrm{th}}$, $P (\mathbfit{D}_i | V_i, C_i, \hat{\mathbfit{n}}_i)$ is the unconditioned pdf followed by the object data vector $\mathbfit{D}_i$ given its validation and truthness status and its sky location (the `individual cluster likelihood'), and $P ( \mathrm{in} | V_i, C_i, \hat{\mathbfit{n}}_i )$ is the probability for the object to be included in the catalogue given its validation and truthness status and its sky location, which is proportional to the mean number of objects per solid angle with such validation and truthness status at sky location $\hat{\mathbfit{n}}_i$, $ d\bar{N} / d \Omega ( V_i, C_i, \hat{\mathbfit{n}}_i) $.

On the other hand, the second term in the sum in Eq.\,(\ref{eq:validated}) can be written as

\begin{equation}
P (V_i | C_i, \hat{\mathbfit{n}_i}, \mathrm{in}) = \frac{d\bar{N} / d \Omega (V_i,C_i ,\hat{\mathbfit{n}}_i)}{ d\bar{N} / d \Omega (C_i, \hat{\mathbfit{n}}_i)},
\end{equation}
where $d\bar{N} / d \Omega (C_i, \hat{\mathbfit{n}}_i)$ is the mean number of objects in the catalogue per solid angle with truthness status $C_i$ at sky location $\hat{\mathbfit{n}}_i$. Finally, the third term can be written as 
\begin{equation}
P(C_i | \hat{\mathbfit{n}_i}, \mathrm{in}) = \frac{d\bar{N} / d \Omega (C_i ,\hat{\mathbfit{n}}_i)}{ d\bar{N}_{\mathrm{all}} / d \Omega ( \hat{\mathbfit{n}}_i)}.
\end{equation}

Putting all the factors together, for a confirmed object ($V_i=T$) we can write
\begin{equation}
P ( \mathbfit{D}_i, V_i = T | \hat{\mathbfit{n}}_i),  \mathrm{in}) \propto P (\mathbfit{D}_i | V_i=T, C_i=T, \hat{\mathbfit{n}}_i) \left[ \frac{d\bar{N}_{\mathrm{all}}}{d \Omega ( \hat{\mathbfit{n}}_i)} \right]^{-1}.
\end{equation}
On the other hand, for an unconfirmed object ($V_i=F$) we can write

\begin{multline}\label{eq:nonvaleq}
P ( \mathbfit{D}_i, V_i = F | \hat{\mathbfit{n}}_i,  \mathrm{in}) \propto [ P (\mathbfit{D}_i | V_i=F, C_i=T, \hat{\mathbfit{n}}_i) + \\ P (\mathbfit{D}_i | V_i=F, C_i=F, \hat{\mathbfit{n}}_i) ]  \left[ \frac{d\bar{N}_{\mathrm{all}}}{d \Omega ( \hat{\mathbfit{n}}_i)} \right]^{-1} ,
\end{multline}
where the first and second terms in the sum correspond, respectively, to unconfirmed true and false detections. We can then write the total unbinned likelihood as

\begin{multline}
    \mathcal{L}_{\mathrm{unbinned}} \propto e^{-\bar{N}_{\mathrm{tot,all}}} \prod_{i=1}^{N_{\mathrm{tot,val}}} P ( \mathbfit{D}_i | V_i=T, C_i=T, \hat{\mathbfit{n}}_i) \times \\ \prod_{i=1}^{N_{\textrm{tot,non-val}}} \sum_{C_i=T,F} P ( \mathbfit{D}_i | V_i=F, C_i , \hat{\mathbfit{n}}_i),
\end{multline}
where the first product is over all the confirmed objects in the catalogue, and the second one is over all the unconfirmed ones. Note that, as expected, this expression reduces to that given in Eq.\,(\ref{eq:unbinned}) if all the objects in the catalogue have been validated and no unconfirmed clusters are theoretically expected, so that $\bar{N}_{\mathrm{tot,all}} = \bar{N}_{\mathrm{tot}}$.

Using Bayes' theorem, for true detections ($C_i=T$) we can write $P (\mathbfit{D}_i | V_i, C_i=T, \hat{\mathbfit{n}}_i)$ as

\begin{equation}
P (\mathbfit{D}_i | V_i, C=T, \hat{\mathbfit{n}}_i) \propto P(V_i |  \mathbfit{D}_i, \hat{\mathbfit{n}}_i) P (\mathbfit{D}_i | C_i=T, \hat{\mathbfit{n}}_i),
\end{equation}
where $P(V_i |  \mathbfit{D}_i, \hat{\mathbfit{n}}_i)$ is the probability for a true cluster to have validation status $V_i$ given its cluster data $\mathbfit{D}_i$ and sky location $\hat{\mathbfit{n}}_i$, and $P (\mathbfit{D}_i | C_i=T, \hat{\mathbfit{n}}_i)$ is the true cluster data likelihood regardless of its validation status. \cnc{} computes $P (\mathbfit{D}_i | C_i=T, \hat{\mathbfit{n}}_i)$ as in the standard case with no unconfirmed detections, which is explained in detail in Section\,\ref{sec:unbinned_implementation}. $P(V_i |  \mathbfit{D}_i, \hat{\mathbfit{n}}_i)$, on the other hand, is taken as input data, except if $\mathbfit{D}_i$ contains a redshift measurement or any mass observable other than the selection observable, in which case \cnc{} assumes that the object is automatically confirmed, i.e., $P(V_i |  \mathbfit{D}_i, \hat{\mathbfit{n}}_i)=1$. Note that, in general, $P (\mathbfit{D}_i | V_i, C_i=T, \hat{\mathbfit{n}}_i) $ is \emph{not} proportional to $P (\mathbfit{D}_i | C_i=T, \hat{\mathbfit{n}}_i)$, as the probability of validation can depend on the value of the selection observable, with objects detected with high significance typically having a higher probability of being validated that objects near the selection threshold. Indeed, the empirical purity of a catalogue often increases significantly with the selection observable threshold (see, e.g., \citealt{Planck2016xxvii,Hilton2020,Bleem2020}).

For false detections ($C_i=F$ and $V_i=F$), \cnc{} takes $P (\mathbfit{D}_i | V_i=F, C_i=F, \hat{\mathbfit{n}}_i)$ as input data. We recall that, in this case, $\mathbfit{D}_i = \zeta_{\mathrm{obs},i}$ and, therefore, $P (\mathbfit{D}_i | V_i=F, C_i=F, \hat{\mathbfit{n}}_i)$ is simply proportional to the abundance of false detections $d N_{\mathrm{f}} / (d \zeta_{\mathrm{obs}} d\Omega_k)$, which can be, e.g., estimated from simulations.

\subsection{Likelihood validation}\label{subsec:falsevalidation}

We validate \cnc{}'s features dealing with unconfirmed objects with one of the synthetic SO-like catalogues generated for the general validation of \cnc{} discussed in Section\,\ref{sec:validation}, where the catalogues analysed contained only confirmed objects. In particular, we inject 150 false detections to the catalogue (which contains a total of 15683 true objects), drawn as samples from the following distribution, which is defined for $q_{\mathrm{obs}} > 5$ and is designed to quickly decrease with signal-to-noise:

\begin{equation}
\frac{dN_{\mathrm{f}}}{dq_{\mathrm{obs}}} \propto e^{-\left( \frac{q_{\mathrm{obs}}-3}{1.5} \right)^2}.
\end{equation}

We then consider three likelihood cases, each of them considering only one mass observable, the tSZ signal-to-noise. These are: (1) the binned likelihood with binning across the selection observable (the tSZ signal-to-noise), which is the only binning scheme allowed by our confirmation formalism; (2) the unbinned likelihood with no objects in the catalogue considered to be confirmed; and (3) the unbinned likelihood with all but 150 randomly-chosen true detections being confirmed. We note that the last scenario assumes that the distribution of confirmed true objects is proportional to the distribution of unconfirmed true objects. As noted in Section\,\ref{subsec:validationformalism}, this will not be the case in general, but here we make such a choice for the sake of simplicity.

\begin{figure}
\centering
\includegraphics[width=0.4\textwidth,trim={00mm 0mm 0mm 0mm},clip]{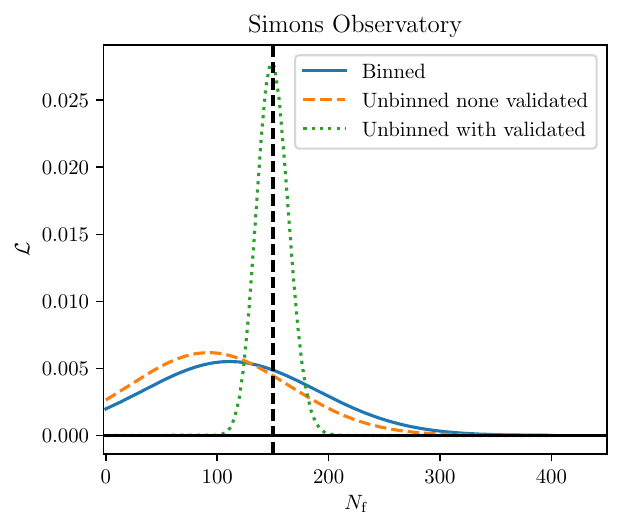}
\caption{Posteriors on the number of false detections for one of our synthetic SO-like catalogues to which 150 false detections have been injected for the three likelihood cases that we have considered (see Section\,\ref{subsec:falsevalidation}): the binned likelihood (solid blue curve), the unbinned likelihood with all objects in the catalogue assumed to be unconfirmed (dashed orange curve), and the unbinned likelihood with only 150 unconfirmed true objects (dotted green curve). The true number of false detections is shown as the vertical dashed line.}
\label{fig:false}
\end{figure}

Figure\,\ref{fig:false} shows our three likelihoods evaluated as a function of the number of false detections, $N_{\mathrm{f}}$, which can be thought of as an input parameter, with the other model parameters fixed to their true values. The true number of false detections is shown as the vertical dashed line. The curves in Figure\,\ref{fig:false} can be thought of as posteriors for $N_{\mathrm{f}}$ (assuming flat priors on $N_{\mathrm{f}}$). The three of them are consistent with the true number of false detections. As expected, the unbinned likelihood in which only 150 true detections are unconfirmed (dotted green curve) delivers much tighter constraints on $N_{\mathrm{f}}$ than the other two cases, for which no confirmation information is available.

\section{Consistency tests}\label{appendix:b}

In some scenarios, \cnc{} allows to evaluate the same likelihood in following different paths, offering the possibility of assessing the consistency between these different avenues as a further check of the code. Here we consider two consistency tests, which we perform with one of our synthetic SO-like catalogues.

\begin{figure}
\centering
\includegraphics[width=0.4\textwidth,trim={00mm 0mm 0mm 0mm},clip]{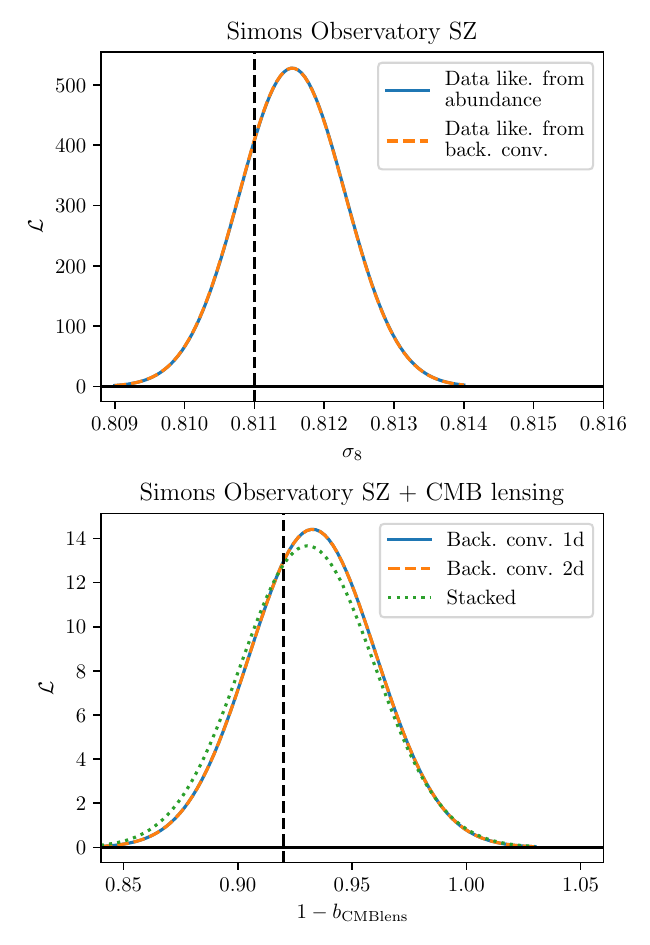}
\caption{\textit{Top panel}: Unbinned likelihood as a function of $\sigma_8$ for one of our synthetic SO-like catalogues with one mass observable (the tSZ signal-to-noise), with the individual cluster likelihoods evaluated both by interpolating over the cluster abundance (blue solid curve) and with the backward convolutional approach (dashed orange curve). The agreement between the two curves is excellent. The true value of $\sigma_8$ is shown as the dashed vertical line. \textit{Lower panel}: Unbinned likelihood also including the CMB lensing mass measurements as a function of the CMB lensing mass bias parameter $\beta_{\mathrm{CMBlens}}$, evaluated assuming two correlation sets (solid blue curve) and one correlation set (dashed orange curve). The agreement between the curves is also excellent. In addition, the likelihood for the case in which the CMB lensing signal-to-noise measurements are stacked across the whole sample is also shown (dotted green curve).}
\label{fig:consistency}
\end{figure}

We first consider the unbinned likelihood for just one mass observable, the tSZ signal-to-noise (the selection observable). As noted in Section\,\ref{sec:unbinned_implementation}, in this case the individual cluster likelihoods can be computed by interpolating the cluster abundance across tSZ signal-to-noise and redshift at the data points for each cluster in the catalogue. Alternatively, they can also be computed with the backward convolutional approach. Both approaches are mathematically identical, meaning that the numerical value of the likelihood at a given point in parameter space ought to be the same for both of them. The top panel of Figure\,\ref{fig:consistency} shows the unbinned likelihood as a function of $\sigma_{8}$, with all other parameters set to their true values, evaluated following these two different approaches. The agreement between the two is excellent.

A further consistency test can be performed when also considering an additional mass observable, in our case the CMB lensing signal-to-noise. If there is no correlation between its scatter and that of the tSZ signal-to-noise, as it is the case, the individual cluster likelihoods can be calculated with the backward convolutional approach assuming that either there is just one correlation set including both mass observables or that each mass observable constitutes a correlation set by itself. In the former case, the various operations (e.g., the convolutions) are two-dimensional, whereas in the latter case they are one-dimensional. Both cases, however, are mathematically equivalent, and should therefore lead to identical numerical values for the likelihood. Note that this would not be the case if there was correlation between the mass observables. The bottom panel of Figure\,\ref{fig:consistency} shows the unbinned likelihood evaluated in these two scenarios as a function of the CMB lensing mass bias, $\beta_{\mathrm{CMBlens}}$ (solid blue curve for two correlation sets and dashed orange curve for one correlation set). The agreement between the two is also excellent. For comparison, Figure\,\ref{fig:consistency} also shows the likelihood for the case in which the CMB lensing signal-to-noise measurements are stacked across the whole sample (the SZ+CMBlens stacked case of Section\,\ref{sec:validation}). This is not equivalent to the other two cases, and therefore the likelihood value is expected to be different (if highly correlated). Note, in particular, that if we interpret the curves as posteriors on $\beta_{\mathrm{CMBlens}}$, the constraint on $\beta_{\mathrm{CMBlens}}$ is slightly wider in the stacked case, indicating that, as expected, some information is lost upon stacking.

\section{Input parameters}\label{appendix:c}

In Table\,\ref{table:parameters} we offer a brief description of all \cnc{}'s input parameters. We note that, as the code continues to be developed, these may change, referring the reader to the online documentation for up-to-date information.
\onecolumn
\centering
\begin{longtable}{p{.20\textwidth} p{.2\textwidth} p{.5\textwidth}}
\hline
\textbf{Parameter name} & \textbf{Allowed values} & \textbf{Description} \\
\hline
\\
\texttt{number\_cores\_hmf} & \texttt{int} & Number of cores to be used in the halo mass function step; $n_{\mathrm{core,hmf}}$ in the text. \\
\texttt{number\_cores\_abundance} & \texttt{int} & Number of cores to be used in the cluster abundance step; $n_{\mathrm{core,abundance}}$ in the text. \\
\texttt{number\_cores\_data} & \texttt{int} & Number of cores to be used in the evaluation of the individual cluster likelihoods with the backward convolutional approach; $n_{\mathrm{core,data}}$ in the text.\\
\texttt{number\_cores\_stacked} & \texttt{int} & Number of cores to be used in the evaluation of the stacked likelihood; $n_{\mathrm{core,stacked}}$ in the text.\\
\texttt{parallelise\_type} & \{\texttt{"patch"}, \texttt{"redshift"}\} & If \texttt{number\_cores\_abundance} > 1, whether the parallelisation is to be carried out across selection tiles (\texttt{"patch"}) or across redshift (\texttt{"redshift"}).\\
\texttt{n\_points} & \texttt{int} & Number of mass and selection observable points at which the halo mass function and the cluster abundance, respectively, are evaluated for each redshift; $n_{\zeta}$ in the text.\\
\texttt{n\_z} & \texttt{int} & Number of redshift evaluations in the halo mass function and cluster abundance steps; $n_z$ in the text.\\
\texttt{n\_points\_data\_lik} & \texttt{int} & Number of points in the operations of the backward convolutional approach; $n_{\mathrm{eval}}$ in the text.\\
\texttt{sigma\_mass\_prior} & \texttt{float} & Width of the evaluation mass range in the backward convolutional approach; $c_M$ in the text.\\
\texttt{downsample\_hmf\_bc} & \texttt{int} & Factor by which the halo mass function is downsampled prior to interpolation in the backward convolutional approach.\\
\texttt{load\_catalogue} & \{\texttt{True}, \texttt{False}\} & Whether a cluster catalogue is to be loaded.\\
\texttt{likelihood\_type} &  \{\texttt{"unbinned"},\texttt{"binned}, \texttt{"extreme\_value"}\} & Likelihood type.\\
\texttt{observable\_select} & \texttt{string} & Name of the selection observable.\\
\texttt{observables} & \texttt{list} & Mass observables to be considered in the likelihood, grouped by correlation set. For example, if there are three mass observables, \texttt{"A"}, \texttt{"B"}, and \texttt{"C"}, with \texttt{"A"} and \texttt{"B"} belonging to the same correlation set and \texttt{"C"} belonging to a different correlation set, the parameter must be set to \texttt{[["A","B"],["C"]]}. Note that if there is only one mass observable, the correct syntax is \texttt{[["A"]]}.\\
\texttt{cluster\_catalogue} & \texttt{string} & Name of the cluster catalogue.\\
\texttt{data\_lik\_from\_abundance} & \{\texttt{True}, \texttt{False}\} & Whether the individual cluster likelihoods are to be computed by interpolating over the cluster abundance (\texttt{True}) or with the backward convolutional approach (\texttt{False}). Only relevant if there is one single mass observable.\\
\texttt{data\_lik\_type} & \{\texttt{"backward\_convolutional"},  \texttt{"direct\_integral"}\} & \,\,\,\,\,\,Whether the individual cluster likelihoods are to be computed with the backward convolutional approach or with brute-force integration.\\
\texttt{abundance\_integral\_type} & \{\texttt{"fft"},  \texttt{"direct"}\} & Whether the convolutions in the cluster abundance step are to be computed in real or in Fourier space.\\
\texttt{compute\_abundance\_matrix} & \{\texttt{True}, \texttt{False}\} & Whether the sum of the cluster abundance across selection observable and redshift across all the selection tiles is to be computed.\\
\texttt{catalogue\_params} & \texttt{dict} & Custom parameters for the cluster catalogue.\\
\texttt{apply\_obs\_cutoff} & \{\texttt{True}, \texttt{False}\} & Whether a threshold in the selection observable abundance is to be applied before the convolution in the last layer.\\
\texttt{get\_masses} & \{\texttt{True}, \texttt{False}\}  & Whether the cluster masses are to be computed.\\
\texttt{delta\_m\_with\_ref} & \{\texttt{True}, \texttt{False}\}  & Whether the evaluation mass set $\mathbfit{M}_{\mathrm{eval}}$ is to be calculated at a fixed set of input parameters (see online documentation).\\
\texttt{obs\_select\_min} & \texttt{float} & Selection observable threshold; $\zeta_{\mathrm{th}}$ in the text.\\
\texttt{obs\_select\_max} & \texttt{float} & Maximum value of the selection observable for which the cluster abundance is computed.\\
\texttt{z\_min} & \texttt{float} & Minimum redshift in the halo mass function and cluster abundance steps; $z_{\mathrm{min}}$ in the text.\\
\texttt{z\_max} & \texttt{float} & Maximum redshift in the halo mass function and cluster abundance steps; $z_{\mathrm{max}}$ in the text.\\
\texttt{M\_min} & \texttt{float} & Minimum mass in the halo mass function and cluster abundance steps; $M_{\mathrm{min}}$ in the text.\\
\texttt{M\_max} & \texttt{float} & Maximum mass in the halo mass function and cluster abundance steps; $M_{\mathrm{max}}$ in the text.\\
\texttt{cosmo\_model} & \{"\texttt{lcdm}", "\texttt{mnu}",  "\texttt{neff}", "\texttt{wcdm}", "\texttt{ede}"\} & Cosmological model.\\ 
\texttt{cosmology\_tool} & \{"\texttt{astropy}", "\texttt{classy\_sz}"\} & Package with which to compute basic cosmology quantities.\\
\texttt{hmf\_calc} &  \{"\texttt{cnc}", "\texttt{hmf}", "\texttt{MiraTitan}","\texttt{classy\_sz}"\} & Package with which to compute the halo mass function.\\
\texttt{hmf\_type} & \texttt{float} & Halo mass function to be used; default is \texttt{"Tinker08"}.\\
\texttt{mass\_definition} & \texttt{float} & Mass definition; default is \texttt{"500c"}.\\
\texttt{hmf\_type\_deriv} & \{\texttt{"numerical"}, \texttt{"analytical"}\} & Whether the derivative of the power spectrum for the halo mass function is computed analytically or numerically. \\
\texttt{power\_spectrum\_type} & \{\texttt{"cosmopower"}\} & Package with which the matter power spectrum is computed; for now, only \texttt{"cosmopower"} is supported.\\
\texttt{cosmo\_amplitude\_parameter} & \,\,\,\,\, \{\texttt{"sigma\_8"}, \texttt{"A\_s"}\} & Cosmological perturbations amplitude parameter to be considered as an input to the code.\\
\texttt{cosmo\_param\_density} & \{\texttt{"physical"}, \texttt{"critical"}\} & Whether the input cosmological densities are $\Omega_{\mathrm{c}} h^2$ and $\Omega_{\mathrm{b}} h^2$ (\texttt{"physical"}), or $\Omega_{\mathrm{m}}$ and $\Omega_{\mathrm{b}}$ (\texttt{"critical"}). \\
\texttt{scalrel\_type\_deriv} & \{\texttt{"numerical"}, \texttt{"analytical"}\} & Whether to compute the derivative of the mass observable scaling relations numerically or analytically.\\
\texttt{sigma\_scatter\_min} & \texttt{float} & Minimum scatter in the selection observable for which the convolution is performed (otherwise it is ignored).  \\
\texttt{z\_errors} &  \{\texttt{True}, \texttt{False}\} & Whether there are non-zero redshift measurement uncertainties.\\
\texttt{n\_z\_error\_integral} & \texttt{int} & Number of redshift points along which the redshift integral is carried out in the individual cluster likelihoods; $n_{\mathrm{eval},z}$ in the text.\\
\texttt{z\_error\_sigma\_integral\_range} &  \,\,\,\,\,\,\,\,\,\,\,\,\,\,\,\,\,\,\texttt{float} & Half-span of the redshift range in the redshift integral of the individual cluster likelihoods, in units of the redshift measurement uncertainty standard deviation. \\
\texttt{z\_error\_min} &\texttt{float} & Minimum value of the redshift measurement uncertainty for which the redshift integral is performed; otherwise, the individual cluster likelihood is only evaluated at the measured redshift value. \\
\texttt{z\_bounds} & \{\texttt{True}, \texttt{False}\} & Whether clusters without a redshift measurement have bounds on their redshift instead.\\
\texttt{non\_validated\_clusters} & \{\texttt{True}, \texttt{False}\} & Whether there are unconfirmed detections in the catalogue. \\
\texttt{binned\_lik\_type} & \{\texttt{"z\_and\_obs\_select"}, \texttt{"obs\_select"}, \texttt{"z"}\} & Binning scheme in the binned likelihood. \\
\texttt{bins\_edges\_z} & \texttt{list} & Edges of the redshift bins. \\
\texttt{bins\_edges\_obs\_select} & \texttt{list} & Edges of the selection observable bins. \\
\texttt{stacked\_likelihood} & \{\texttt{True}, \texttt{False}\}  & Whether to include the stacked likelihood. \\
\texttt{stacked\_data} & \texttt{list} & List containing the names of the stacked data sets as strings. \\
\texttt{compute\_stacked\_cov} & \{\texttt{True}, \texttt{False}\} & Whether the covariance of the stacked data sets is computed within a hierarchical model for the stacked observable or given as input.\\
\texttt{path\_to\_cosmopower\_organization} & \,\,\,\,\,\,\,\,\,\,\,\,\,\,\,\,\,\,\,\,\,\,\,\,\,\,\, \texttt{string} & Path to \texttt{cosmopower}.\\
\texttt{class\_sz\_ndim\_redshifts}& \texttt{int} & Number of redshift points for tabulation of \texttt{class\_sz} HMF.\\
\texttt{class\_sz\_concentration\_parameter}& \,\,\,\,\,\,\,\,\,\,\,\,\,\,\,\,\,\,\,\,\,\,\,\,\,\,\,\,\,\texttt{string} & Concentration-mass relation for converting between overdensity mass definitions. \\
\texttt{class\_sz\_output} & \texttt{string} & Output to be collected from \texttt{class\_sz}. \\
\texttt{class\_sz\_hmf} & \texttt{string} & Halo mass function to be used.\\

\\

\caption{\cnc{}'s input parameters. See the online documentation of up-to-date information.}
\label{table:parameters}
\end{longtable}


\bsp	
\label{lastpage}
\end{document}